\documentclass[final,reqno]{amsart}

\usepackage[colorlinks=true,allcolors=blue]{hyperref}
\usepackage{amsaddr}
\usepackage{amssymb,amsfonts,amsmath,graphicx,amsthm}
\usepackage[square,numbers]{natbib}
\usepackage{bm}
\usepackage{enumerate}
\usepackage[dvipsnames]{xcolor}
\usepackage{physics}
\usepackage{bbold}

\usepackage{mathtools}

\usepackage[capitalise]{cleveref}

\undef\div
\DeclareMathOperator{\div}{div}
\DeclareMathOperator{\supp}{supp}
\DeclareMathOperator{\spec}{spec}

\crefname{equation}{}{}

\newtheorem{theorem}{Theorem}[section]
\newtheorem{lemma}[theorem]{Lemma}
\crefname{lemma}{Lemma}{Lemmas}
\Crefname{lemma}{Lemma}{Lemmas}
\newtheorem{proposition}[theorem]{Proposition}
\crefname{proposition}{Proposition}{Propositions}
\Crefname{proposition}{Proposition}{Propositions}
\newtheorem*{claim}{Claim}

\theoremstyle{remark}
\newtheorem{remark}{Remark}
\crefname{remark}{Remark}{Remarks}
\Crefname{remark}{Remark}{Remarks}

\newcommand{\RR}{\mathbb{R}}
\newcommand{\CC}{\mathbb{C}}
\newcommand{\ZZ}{\mathbb{Z}}

\newcommand{\iden}{\mathbb{1}}
\renewcommand{\i}{\mathrm{i}}
\newcommand{\e}{\mathrm{e}}

\newcommand{\dua}[2]{\langle #1, #2 \rangle}

\newcommand{\ol}[1]{\overline{#1}}

\newcommand{\HC}{\mathcal{H}}
\newcommand{\DC}{D}
\newcommand{\Cc}{\mathcal{C}}

\newcommand{\AF}{\bm{\mathfrak{A}}}
\newcommand{\BF}{\bm{\mathfrak{B}}}
\newcommand{\Af}{\mathfrak{A}}
\newcommand{\Bf}{\mathfrak{B}}
\newcommand{\vbf}{\bm{\mathfrak{b}}}
\newcommand{\vjp}{\mathbf{j}^{\mathrm{p}}}
\newcommand{\vjph}{\mathbf{j}^{\mathrm{ph}}}
\newcommand{\vjc}{\mathbf{j}^{\mathrm{c}}}

\newcommand{\PF}{\mathrm{PF}}

\newcommand{\ph}{\mathrm{ph}}
\newcommand{\rad}{\mathrm{rad}}
\newcommand{\sol}{\mathrm{sol}}

\newcommand{\Fock}{\mathcal{F}}
\newcommand{\Hil}{\mathfrak{H}}

\let\phi=\varphi

\newcommand{\vsigma}{{\bm{\sigma}}}
\newcommand{\vzeta}{{\bm{\zeta}}}
\newcommand{\vepsilon}{{\bm{\epsilon}}}
\newcommand{\vpi}{{\bm{\pi}}}

\newcommand{\vOmega}{{\bm{\Omega}}}

\renewcommand{\va}{\mathbf{a}}
\newcommand{\ve}{\mathbf{e}}

\newcommand{\vy}{\mathbf{y}}
\newcommand{\vc}{\mathbf{c}}
\newcommand{\vk}{\mathbf{k}}
\newcommand{\vm}{\mathbf{m}}
\newcommand{\vx}{\mathbf{x}}
\newcommand{\vj}{\mathbf{j}}
\newcommand{\vA}{{\mathbf{A}}}
\newcommand{\vB}{{\mathbf{B}}}
\newcommand{\vF}{{\mathbf{F}}}
\newcommand{\vX}{{\mathbf{X}}}
\newcommand{\vD}{{\mathbf{D}}}
\undef\vb
\newcommand{\vb}{{\mathbf{b}}}

\newcommand{\wh}[1]{\widehat{#1}}
\DeclareSymbolFont{PFwideaccents}{OMX}{yhex}{m}{n}
\DeclareMathAccent{\PFwideaccent}{\mathord}{PFwideaccents}{"62}
\newcommand{\widehatfull}[1]{\mathpalette\PFwidehatfull{#1}}
\makeatletter
\newcommand{\PFwidehatfull}[2]{%
  \begingroup
  \sbox0{$\m@th#1#2$}%
  \sbox2{$\m@th#1\PFwideaccent{\vphantom{#2}\rule{4em}{0pt}}$}%
  \mathord{\rlap{\resizebox{\wd0}{\ht2}{\usebox2}}\usebox0}%
  \endgroup}
\makeatother

\renewcommand{\epsilon}{\varepsilon}

\title[Unique continuation for the Pauli--Fierz Hamiltonian]{Unique continuation for the\\
Pauli--Fierz Hamiltonian}

\author[M.\,A. Csirik]{Mih\'aly~A.~Csirik}
\address{Department of Mathematical Sciences, Chalmers University of
Technology and the University of Gothenburg, G\"oteborg, Sweden}
\thanks{Email: \href{mailto:csirik@gmail.com}{\texttt{csirik@gmail.com}}}

\begin{document}
\begin{abstract}
We prove a unique continuation property for eigenstates of the Pauli--Fierz
Hamiltonian, viewed as functions of the electron
configuration with values in the spin--photon space.
The vector potential may lie in $L^6_\mathrm{loc}$, with
its divergence, curl, and the scalar potential in $L^p_\mathrm{loc}$ for some
$p>2$.  As an application, we prove a Hohenberg--Kohn theorem at
positive coupling strength: the spin-summed ground-state density and
combined current density determine the external potentials up to a
gauge transformation and an additive constant.
\end{abstract}
\maketitle

\section{Introduction}\label{sec-intro}

A differential operator is said to have the unique continuation property if every
solution that vanishes on an open set, or to infinite order at a point,
vanishes identically.  Carleman~\cite{carleman1939probleme} introduced
the weighted inequalities now bearing his name to prove such
results.  For Schr\"odinger operators $-\Delta+V$ on $\RR^n$,
Jerison and Kenig~\cite{jerison1985unique} proved unique continuation under the scale-invariant
condition $V\in L^{n/2}_\mathrm{loc}$; see~\cite{wolff1993recent} for a
review of such sharp results.  Singular magnetic fields and
nonsmooth coefficients were treated by Kurata~\cite{kurata1997unique}
and by Koch and Tataru~\cite{koch2001carleman}.  In density-functional
theory, the property enters through the Hohenberg--Kohn
theorem~\cite{hohenberg1964inhomogeneous}; see~\cite{lewin2023universal}
for a review.  As Lieb
emphasized~\cite{lieb1983density}, for the usual $L^p$ potentials its
proof requires that ground states do not vanish on sets of positive
measure.  For Laplace form-bounded distributional potentials, Corso
recently showed that the relevant condition is instead strict
positivity of the density quasi-everywhere, and verified it for
non-interacting systems~\cite{corso2026maximal}; see
also~\cite{corso2025torus,corso2026spinless} in one dimension.

For $N$ electrons, the unique continuation theorems
of~\cite{jerison1985unique,kurata1997unique,koch2001carleman} must be
applied on $\RR^{3N}$, where their conditions deteriorate with $N$,
although the physical potentials are sums of one- and two-body terms.
For instance, the condition of Jerison and Kenig becomes
$V\in L^{3N/2}_\mathrm{loc}$.  Garrigue~\cite{garrigue2018unique}
removed this dependence.  He proved strong unique continuation for
many-body Schr\"odinger operators whose one- and two-body potentials
lie in $L^p_\mathrm{loc}(\RR^3)$ with $p>2$, for every $N$, and
deduced the Hohenberg--Kohn theorem under the same assumptions.  In a
sequel~\cite{garrigue2020unique} he treated the many-body Pauli
Hamiltonian with external magnetic fields.  Garrigue's main tool is a
Carleman estimate with the singular weight
$-\log|\vx|+(-\log|\vx|)^{-\alpha}$ that also controls fractional
derivatives, so that each potential is estimated in its own
three-dimensional variables.  For the Hohenberg--Kohn theorem, unique
continuation from sets of positive measure suffices.  He obtains it from
an argument of de Figueiredo and Gossez~\cite{defigueiredo1992strict},
in which a local energy inequality
shows that an eigenfunction vanishes to infinite order at
every Lebesgue density point of its zero set.\looseness=-1

Quantum-electrodynamical density-functional theory (QEDFT) extends
density-functional theory to electrons coupled to quantized photons.
Rajagopal formulated a time-dependent version~\cite{rajagopal1994time}.
Ruggenthaler and coworkers developed a time-dependent Kohn--Sham
approach to relativistic quantum
electrodynamics~\cite{ruggenthaler2011time} and a hierarchy of such
theories~\cite{ruggenthaler2014quantum}.  In the
non-relativistic limit, described by the Pauli--Fierz Hamiltonian, the
basic variables are the electronic current density and the expectation
value of the quantized vector potential~\cite{ruggenthaler2014quantum}.
The ground-state theory~\cite{ruggenthaler2015ground} derives a
Hohenberg--Kohn result: the external scalar potential and a transversal
external current acting on the photon field are in one-to-one
correspondence with the electronic density and the expectation value of
the vector potential.  As in the original
argument~\cite{hohenberg1964inhomogeneous}, the step determining the
scalar potential tacitly assumes that the ground state does not vanish on a
set of positive measure.  In the dipole approximation,
Tokatly~\cite{tokatly2013time} used the electronic density, or current
density, together with the expectation values of the photon mode
coordinates.  The time-dependent
works~\cite{ruggenthaler2011time,ruggenthaler2014quantum,tokatly2013time}
derive the time-dependent analogue of the Hohenberg--Kohn theorem, the
Runge--Gross theorem, for their variables.

In this paper, we carry out Garrigue's program to derive a unique
continuation property for the Pauli--Fierz Hamiltonian.  Two new challenges arise: the eigenstates take values in the tensor product of
the electron spin space and the photon Fock space, and the quantized
vector potential is, at each point, an unbounded operator on the photon
Fock space rather than a function.  Our Carleman estimate uses
the same weight as
in~\cite{garrigue2018unique,garrigue2020unique} but keeps the covariant
kinetic energy and the
radiation Hamiltonian together in the principal part.  Under an
additional ultraviolet moment condition on the cutoff,
\cref{ucp-main-theorem} gives unique continuation for external vector potentials in
$L^6_\mathrm{loc}$, with the divergence, curl, and scalar potential in
$L^p_\mathrm{loc}$ for some $p>2$.  As an application,
\cref{thm-hk-nonvanishing} gives a Hohenberg--Kohn
theorem at positive coupling.  For this theorem we
follow~\cite{Q}: unlike in~\cite{ruggenthaler2015ground}, the photon
field carries no external current, and the basic variables are the
spin-summed density and the combined current density, including the
curl of the magnetization density.
The proofs are deferred to \cref{sec-proofs}, and the Carleman
estimates they rely on are proved in \cref{ucp-carleman}.
Counterexamples to the Hohenberg--Kohn theorem at zero coupling, both
without and with electron spin, are constructed in
\cref{app-zero-coupling}.

\medskip
\noindent\textbf{Acknowledgements.}
The author thanks T.~C. Corso for his useful comments on the
manuscript.  The author acknowledges funding from the European Research Council
through ERC-2021-STG Grant Agreement No.~101041487 \mbox{(REGAL)}.

\medskip
\noindent\textbf{Lean formalization.}
The unique continuation and Hohenberg--Kohn theorems, the Carleman
estimates in \cref{ucp-carleman}, and both zero-coupling
counterexamples in \cref{app-zero-coupling} are formalized in Lean~4,
building on the formalization of~\cite{Q}.  The formalization is in the
directory \texttt{lean} of the repository of this
paper\footnote{\url{https://github.com/xyzw/pf_ucp}}.  Standard results from
analysis and operator theory enter as common assumptions.

\enlargethispage{2\baselineskip}
\medskip
\noindent\textbf{AI disclosure.}
The large language models GPT-6 Astra (OpenAI), and Claude Opus~5.5
and Claude Fable~5.2 (Anthropic), were used throughout the development of this paper and its
Lean formalization.  The author has reviewed all content and takes full
responsibility for the contents of this paper.

\section{Setting}\label{sec-setting}

We consider $N\geq1$ spin-$1/2$ electrons.
We follow the conventions and notations of~\cite{Q},
although that paper only considers spinless electrons.
The Hilbert space and spin--photon fiber are
\[
 \HC_N=\bigwedge^N L^2(\RR^3,\CC^2)\otimes\Fock_\ph,
 \qquad \mathcal K_N=(\CC^2)^{\otimes N}\otimes\Fock_\ph,
\]
\[
 \Fock_\ph=\Fock_s(\Hil_\ph),\qquad
 \Hil_\ph=L^2(\RR^3\times\{1,2\}).
\]
We write $\Omega\in\Fock_\ph$ for the vacuum.  The dispersion relation
is $\omega(\vk)=|\vk|$, and the radiation Hamiltonian is
\[
 H_\rad=\sum_{\lambda=1,2}\int_{\RR^3}\omega(\vk)\,
        a^*(\vk,\lambda)a(\vk,\lambda)\dd\vk,
\]
where $a^*(\vk,\lambda)$ and $a(\vk,\lambda)$ create and annihilate
a photon with momentum $\vk$ and polarization $\lambda$.
To define the Hamiltonian, the photon field must be smeared out by an
ultraviolet cutoff $\wh\chi$; without it, the quantized vector
potential $\AF(\vx)$ is only an operator-valued distribution in $\vx$, not an
operator-valued function \cite{lieb2010stability}.  Throughout, the cutoff satisfies
\begin{equation}\label{cutoff-assumption}
 \wh\chi(\vk)=\ol{\wh\chi(-\vk)},\qquad
 \wh\chi\in L^2(\RR^3),\qquad
 \wh\chi/\sqrt\omega\in L^2(\RR^3).
\end{equation}
With real transverse polarization vectors $\vepsilon_\lambda$, set
\begin{equation}\label{Gx-def}
 G^\mu_\vx(\vk,\lambda)
 =\frac{\epsilon_\lambda^\mu(\vk)\wh\chi(-\vk)
                  \e^{-\i\vk\cdot\vx}}{\sqrt2},\qquad
 \Af^\mu(\vx)=a^*(G^\mu_\vx)+a(G^\mu_\vx).
\end{equation}
Here, for $f\in\Hil_\ph$, $a^*(f)$ and $a(f)$ are the photon
creation and annihilation operators on $\Fock_\ph$, as in~\cite{Q}.
The operator $a^*(f)$ is linear and $a(f)$ is antilinear in $f$, they are
mutually adjoint, and they satisfy the canonical commutation relations
\[
 [a(g),a^*(f)]=\dua{g}{f}_{\Hil_\ph},\qquad
 [a(g),a(f)]=[a^*(g),a^*(f)]=0,
 \qquad f,g\in\Hil_\ph.
\]
By \cref{cutoff-assumption}, $G^\mu_\vx\in\Hil_\ph$, so $\Af^\mu(\vx)$
is well-defined.

For fixed $e\geq0$, the covariant derivative and external Hamiltonian are
\begin{align}
 \vD_n&=-\i\grad_{\vx_n}-e\AF(\vx_n),\notag\\
 H^{v,\vA}
 &=\sum_{n=1}^N\left[\frac12\bigl(\vsigma_n\cdot(\vD_n+\vA(\vx_n))\bigr)^2
                         +v(\vx_n)\right]
   +\sum_{1\leq n<m\leq N}\frac1{|\vx_n-\vx_m|}+H_\rad.
 \label{qvA-def}
\end{align}
Here $\vsigma=(\sigma^1,\sigma^2,\sigma^3)$ is the vector of Pauli
matrices
\[
 \sigma^1=\begin{pmatrix}0&1\\1&0\end{pmatrix},\qquad
 \sigma^2=\begin{pmatrix}0&-\i\\\i&0\end{pmatrix},\qquad
 \sigma^3=\begin{pmatrix}1&0\\0&-1\end{pmatrix},
\]
and $\vsigma_n$ denotes $\vsigma$ acting on the spin factor of
the $n$th electron in $(\CC^2)^{\otimes N}$, so that
$\vsigma_n\cdot\vb=\sum_{\mu=1}^3\sigma_n^\mu b^\mu$ for
$\vb\in\CC^3$.
Moreover, we assume that the real external potentials belong to
\begin{equation}\label{potential-spaces}
 v\in L^{3/2}(\RR^3)+L^\infty(\RR^3),\qquad
 \vA\in L^3(\RR^3,\RR^3)+L^\infty(\RR^3,\RR^3).
\end{equation}
The common closed form domain $\mathcal Q_\PF$ is the
antisymmetric subspace of
\begin{equation*}
 \mathcal Q
 =H^1(\RR^{3N},\mathcal K_N)\cap\DC(H_\rad^{1/2}),
\end{equation*}
where the radiation operator acts fiberwise.  Closedness and
semiboundedness under these assumptions are established in
\cref{prop-form-domain}.
We denote by $H^{v,\vA}$ the self-adjoint operator associated with this
closed form, and write $\dua{\Phi}{H^{v,\vA}\Psi}$ for the form
evaluated at $\Phi,\Psi\in\mathcal Q_\PF$.  By the first representation
theorem \cite[Theorem~VI.2.1]{kato1995perturbation}, $\Psi\in\mathcal Q_\PF$ is an eigenstate of $H^{v,\vA}$ with
eigenvalue $E$ if and only if it satisfies the weak eigenvalue equation
\begin{equation}\label{weak-eigenvalue-equation}
 \dua{\Phi}{H^{v,\vA}\Psi}=E\dua{\Phi}{\Psi}
 \qquad\text{for all }\Phi\in\mathcal Q_\PF.
\end{equation}

We also use the spinless operator
\begin{equation*}
 H_{\mathrm{sl}}^{v,\vA}
 =\sum_{n=1}^N\left[\frac12\big(\vD_n+\vA(\vx_n)\big)^2+v(\vx_n)\right]
   +\sum_{1\leq n<m\leq N}\frac1{|\vx_n-\vx_m|}+H_\rad.
\end{equation*}
It acts on the space
$\HC_N^{\mathrm{sl}}=\bigwedge^N L^2(\RR^3)\otimes\Fock_\ph$,
with form domain $\mathcal Q_\PF^{\mathrm{sl}}$, or componentwise on spin
when explicitly stated.  As for $H^{v,\vA}$, it is realized as the
self-adjoint operator associated with its closed form, and its
eigenstates are the solutions $\Psi\in\mathcal Q_\PF^{\mathrm{sl}}$ of the
corresponding weak eigenvalue equation.  With
\[
 \vpi_n=\vD_n+\vA(\vx_n),\qquad
 \vB=\grad\times\vA,\qquad
 \BF=\grad\times\AF,
\]
the following identity of quadratic forms holds whenever
$\omega\wh\chi\in L^2(\RR^3)$ and the distributional curl satisfies
$\vB\in L^{3/2}_\mathrm{loc}(\RR^3,\RR^3)$: for every
compactly supported $\Psi\in\mathcal Q$,
\begin{equation}\label{pauli-expansion}
 \|(\vsigma_n\cdot\vpi_n)\Psi\|^2
 =\|\vpi_n\Psi\|^2
   +\dua{\Psi}{\vsigma_n\cdot\bigl(\vB(\vx_n)-e\BF(\vx_n)\bigr)\Psi}.
\end{equation}

For $\Phi,\Psi\in L^2(\RR^{3N},\mathcal K_N)$, the partial inner
product
\[
 \dua{\Phi}{\Psi}_\vx
 =\int_{\RR^{3(N-1)}}
   \dua{\Phi(\vx,\vx_2,\ldots,\vx_N)}{\Psi(\vx,\vx_2,\ldots,\vx_N)}_{\mathcal K_N}
   \dd\vx_2\cdots\dd\vx_N
\]
integrates out all electronic positions except $\vx_1=\vx$; the inner
product of $\mathcal K_N$ sums over all spin indices and integrates
out the photons.  It is defined for a.e.\ $\vx\in\RR^3$ and is
integrable in $\vx$.
For a normalized $\Psi\in\mathcal Q_\PF$, the spin-summed density,
magnetization density, and current densities are
\begin{align}
 \rho_\Psi(\vx)&=N\dua{\Psi}{\Psi}_\vx,\notag\\
 \vm_\Psi(\vx)&=N\dua{\Psi}{\vsigma_1\Psi}_\vx,\notag\\
 \vjp_\Psi(\vx)&=N\Im\dua{\Psi}{\grad_\vx\Psi}_\vx,\notag\\
 \vjph_\Psi(\vx)&=-eN\Re\dua{\Psi}{\AF(\vx)\Psi}_\vx,\notag\\
 \vjc_\Psi&=\vjp_\Psi+\vjph_\Psi+\tfrac12\grad\times\vm_\Psi,
 \notag\\
 \vj_\Psi&=\vjc_\Psi+\rho_\Psi\vA.
 \label{physical-current-def}
\end{align}
Here $\vjp_\Psi$ is the paramagnetic current density, $\vjph_\Psi$
the field-induced current density, and $\frac12\grad\times\vm_\Psi$
the magnetization current density; their sum $\vjc_\Psi$ is the
combined current density.  Finally, $\vj_\Psi$ is the physical current
density for the specified external vector potential.  In contrast,
$\vjc_\Psi$ is a function of the state alone at fixed $e,\chi$.  The bounds in \cite[Proposition~4.1]{Q} are stated
for spinless states.  Their proof regards $\vx_1\mapsto\Psi(\vx_1,\cdot)$
as an $H^1$ function with values in the Hilbert space of the remaining
variables, so it applies verbatim when this space contains the spin
factor.  They give $\rho_\Psi\in L^1\cap L^3$ and
$\vjp_\Psi,\vjph_\Psi\in L^1\cap L^{3/2}$.
Similarly, $\grad\vm_\Psi\in L^1\cap L^{3/2}$, and hence
$\vjc_\Psi\in L^1\cap L^{3/2}$.

Let $\Cc_\PF\subset\mathcal Q_\PF$ be the antisymmetric part of
$C_c^\infty(\RR^{3N},(\CC^2)^{\otimes N})$ tensored algebraically with
finite-photon vectors whose one-photon factors have bounded momentum
support.  Since the form norm of $H^{v,\vA}$ is equivalent to that of
$\mathcal Q$ (\cref{prop-form-domain}), $\Cc_\PF$ is a form core
for $H^{v,\vA}$.  For smooth potentials, expanding the square in
\cref{qvA-def} gives, as operators on $\Cc_\PF$,
\[
 \bigl(\vsigma_n\cdot(\vD_n+\vA(\vx_n))\bigr)^2
 =(\vsigma_n\cdot\vD_n)^2
  +\vD_n\cdot\vA(\vx_n)+\vA(\vx_n)\cdot\vD_n
  +|\vA(\vx_n)|^2+\vsigma_n\cdot\vB(\vx_n).
\]
Next, we take the expectation in a normalized $\Psi\in\Cc_\PF$ and sum over
$n$.  By the antisymmetry of $\Psi$, the terms linear in $\vA$ contribute
$\int_{\RR^3}\vA\cdot(\vjp_\Psi+\vjph_\Psi)$, the quadratic term
contributes $\frac12\int_{\RR^3}|\vA|^2\rho_\Psi$, and the external
Zeeman term contributes
$\frac12\int_{\RR^3}\vB\cdot\vm_\Psi
 =\frac12\int_{\RR^3}\vA\cdot(\grad\times\vm_\Psi)$
after an integration by parts.  Collecting terms gives the energy
decomposition (cf.~\cite[Section~2.3.2]{garrigue2020unique})
\begin{equation}\label{energy-decomposition}
 \dua{\Psi}{H^{v,\vA}\Psi}
 =\dua{\Psi}{H^{0,0}\Psi}
  +\dua{v+\tfrac12|\vA|^2}{\rho_\Psi}+\dua{\vA}{\vjc_\Psi}.
\end{equation}
We derived \cref{energy-decomposition} for smooth potentials and
normalized $\Psi\in\Cc_\PF$.  It extends to all potentials satisfying
\cref{potential-spaces} and all normalized $\Psi\in\mathcal Q_\PF$ by smooth
approximation of the potentials, density of $\Cc_\PF$ in the form norm,
and continuity of both sides.
We write $E(v,\vA)=\inf\spec(H^{v,\vA})$, which is finite since
$H^{v,\vA}$ is bounded from below.

\section{Main results}\label{sec-main-results}

We start by establishing a unique continuation property for the
Pauli--Fierz Hamiltonian $H^{v,\vA}$.
A state $\Psi\in\HC_N$ is said to \emph{vanish to infinite order} at
$\vX_0\in\RR^{3N}$ if, for every $m>0$,
\[
 \int_{B_r(\vX_0)}\|\Psi(\vX)\|_{\mathcal K_N}^2\dd\vX=\mathcal O(r^m)
 \qquad(r\downarrow0).
\]
This notion connects unique continuation with the positivity of
densities.  The \emph{de Figueiredo--Gossez argument}
\cite{defigueiredo1992strict} (\cref{ucp-infinite-order}) uses a local
energy estimate to show that an eigenstate vanishes to infinite order
at every Lebesgue density point of its zero set.  Hence an eigenstate
without zeros of infinite order cannot vanish on a set of positive
measure.

In order to state our result, define the collision-free configuration space via
\begin{equation}\label{ucp-regular-configurations}
 \mathcal U=\{\vX=(\vx_1,\ldots,\vx_N)\in\RR^{3N}
                      :\vx_n\ne\vx_m\text{ if }n\ne m\}.
\end{equation}
Our first main result shows that a nonzero eigenstate cannot vanish to
infinite order at any point of $\mathcal U$.

\begin{theorem}[Pauli--Fierz unique continuation]\label[theorem]{ucp-main-theorem}
Assume that
\begin{equation}\label{ucp-extra-cutoff}
 \omega\wh\chi\in L^2(\RR^3)
\end{equation}
and that, for some $p>2$,
\[
 \begin{gathered}
 v\in L^p_\mathrm{loc}(\RR^3),\qquad
 \vA\in L^6_\mathrm{loc}(\RR^3,\RR^3),\\
 \div\vA\in L^p_\mathrm{loc}(\RR^3),\qquad
 \vB\in L^p_\mathrm{loc}(\RR^3,\RR^3).
 \end{gathered}
\]
An eigenstate $\Psi$ of $H^{v,\vA}$ that vanishes to infinite order
at any point of $\mathcal U$ is identically zero.
Moreover, every nonzero eigenstate satisfies
\begin{equation*}
 \|\Psi(\vX)\|_{\mathcal K_N}>0\quad\text{for a.a. }\vX\in\RR^{3N},
 \qquad \rho_\Psi(\vx)>0\quad\text{for a.a. }\vx\in\RR^3.
\end{equation*}
The same conclusions hold without the condition on $\vB$ for the
spinless operator $H_{\mathrm{sl}}^{v,\vA}$ on $\HC_N^{\mathrm{sl}}$,
with the fiber $\mathcal K_N$ replaced by $\Fock_\ph$.
\end{theorem}
The proof in \cref{sec-ucp} follows the method of
Garrigue~\cite{garrigue2018unique,garrigue2020unique} and relies on the
covariant Carleman estimates of \cref{ucp-carleman}.

\begin{theorem}[Hohenberg--Kohn]
\label[theorem]{thm-hk-nonvanishing}
Let $e>0$, and let the cutoff satisfy $\wh\chi\not\equiv0$ and
$\omega\wh\chi\in L^2(\RR^3)$.  Let $(v_k,\vA_k)$, $k=1,2$, be two
potential pairs with $\vB_k=\grad\times\vA_k$, and suppose that,
for at least one $k$ and some $p>2$,
\begin{equation}\label{ucp-potential-regularity}
 \begin{gathered}
 v_k\in L^p_\mathrm{loc}(\RR^3),\qquad
 \vA_k\in L^6_\mathrm{loc}(\RR^3,\RR^3),\\
 \div\vA_k\in L^p_\mathrm{loc}(\RR^3),\qquad
 \vB_k\in L^p_\mathrm{loc}(\RR^3,\RR^3).
 \end{gathered}
\end{equation}

Let $\Psi_k$ be normalized ground states of $H^{v_k,\vA_k}$, with
energies $E_k=E(v_k,\vA_k)$, $k=1,2$.  Assume that there is a real
function $\theta\in W^{1,1}_\mathrm{loc}(\RR^3)$ with
$\grad\theta\in L^3(\RR^3,\RR^3)+L^\infty(\RR^3,\RR^3)$ such that
\[
 \rho_{\Psi_1}=\rho_{\Psi_2}=\rho,\qquad
 \vjc_{\Psi_2}=\vjc_{\Psi_1}+\rho\grad\theta
 \quad\text{a.e.}
\]
Assume moreover that at least one of the following holds:
$\wh\chi\ne0$ almost everywhere, or $\rho$ has an exponential moment,
that is, for some $\alpha>0$,
\begin{equation}\label{density-exponential-moment}
 \int_{\RR^3}\e^{\alpha|\vx|}\rho(\vx)\dd\vx<\infty.
\end{equation}
Then
\begin{equation}\label{hk-conclusion}
 \boxed{\vA_2=\vA_1-\grad\theta,\qquad v_1=v_2+c\quad\text{a.e.},
                         \qquad c=\frac{E_1-E_2}{N}.}
\end{equation}
The same conclusion holds without the condition on $\vB_k$ for ground
states of the spinless operators $H_{\mathrm{sl}}^{v_k,\vA_k}$, with
combined current density $\vjc_{\Psi}=\vjp_\Psi+\vjph_\Psi$.
\end{theorem}

Condition \cref{ucp-potential-regularity} includes molecular Coulomb scalar
potentials with locally $L^p$ perturbations for $p>2$.  The
singularities at the nuclei are allowed, since
$|\vx|^{-1}\in L^q_\mathrm{loc}(\RR^3)$ for every $q<3$.  In particular, the global class
$L^p(\RR^3)+L^\infty(\RR^3)$,
$p>2$, satisfies the scalar hypothesis.

The hard ultraviolet cutoff
$\wh\chi_\Lambda(\vk)=\iden_{\{|\vk|\leq\Lambda\}}/\sqrt{|\vk|}$,
$\Lambda>0$, satisfies $\omega\wh\chi_\Lambda\in L^2(\RR^3)$ but
vanishes for $|\vk|>\Lambda$, so it is covered only by the
exponential-moment alternative \cref{density-exponential-moment}.
Exponential decay below an ionization threshold is known
for Pauli--Fierz models under appropriate binding hypotheses
\cite{griesemer2004exponential}.

The vector-potential part of the conclusion, restricted to the support
of the density, does not need the local regularity
\cref{ucp-potential-regularity}. In fact, for identical combined current
densities, \cref{prop-vector-uniqueness} below gives $\rho(\vA_1-\vA_2)=0$
under the remaining hypotheses alone.  This is the only step that uses $e>0$,
through the photon variations in \cref{prop-cutoff-kernel} below. Note that the
unique continuation theorem holds for every $e\geq0$.

\begin{remark}[Regularity away from an exceptional set]
\label{rem-ucp-open-set}
The conclusion of \cref{thm-hk-nonvanishing} remains valid if the
conditions in \cref{ucp-potential-regularity} are required only on a
connected open set $\mathcal O\subset\RR^3$ with
$|\RR^3\setminus\mathcal O|=0$.  The divergence and curl conditions are then
understood in the sense of distributions on $\mathcal O$.  This extension follows
from \cref{rem-ucp-domain-extension} by the same proof.  It allows singularities
that fail the stated local integrability conditions near the excluded
set.  For example, $v(\vx)=\eta(\vx)|\vx|^{-3/2}$, with a smooth
compactly supported cutoff $\eta$ equal to one near the origin, belongs
to $L^{3/2}(\RR^3)$ and is locally bounded on
$\mathcal O=\RR^3\setminus\{0\}$, but is not in
$L^p_\mathrm{loc}(\RR^3)$ for any $p>2$.
\end{remark}

The positive coupling assumption $e>0$ cannot be dropped from
\cref{thm-hk-nonvanishing}, either without or with electron spin.
In \cref{app-zero-coupling} we construct, for $N=1$ and $e=0$,
families of bounded azimuthal vector potentials $\vA_t$, linear in $t$
and with different magnetic fields for different $t$, together with
scalar potentials $v_t$, such that every $H^{v_t,\vA_t}$ has the same
nondegenerate ground state, the hydrogen $1s$ orbital times the photon
vacuum
(\cref{prop-zero-coupling-counterexample,prop-spinful-zero-coupling-counterexample}).
Hence the density and combined current density do not depend on $t$, and
for a nonzero compact ultraviolet cutoff all hypotheses of
\cref{thm-hk-nonvanishing} other than $e>0$ hold.  The same examples
work at $e>0$ if $\wh\chi=0$.

\section{Proofs}\label{sec-proofs}

The rest of the paper is devoted to proofs. 

\subsection{Preliminaries}

For the convenience of the reader, we recall that all Hamiltonians $H^{v,\vA}$ with potentials
satisfying \cref{potential-spaces} are well defined, with a common
form domain.

\begin{proposition}[Form domain]\label[proposition]{prop-form-domain}
Let $v$ and $\vA$ satisfy \cref{potential-spaces}.  The form
\begin{align*}
 q^{v,\vA}[\Phi]
 &=\frac12\sum_{n=1}^N\|\vsigma_n\cdot\vpi_n\Phi\|^2
   +\|H_\rad^{1/2}\Phi\|^2\\
 &\quad+\int_{\RR^{3N}}\Bigl(\sum_{n=1}^Nv(\vx_n)
     +\sum_{1\leq n<m\leq N}\frac{1}{|\vx_n-\vx_m|}\Bigr)
     \|\Phi(\vX)\|_{\mathcal K_N}^2\dd\vX
\end{align*}
on $L^2(\RR^{3N},\mathcal K_N)$ is closed and bounded below on
$\mathcal Q$, and its shifted form norm is equivalent
to the norm of $H^1(\RR^{3N},\mathcal K_N)\cap\DC(H_\rad^{1/2})$.  In
particular, $H^{v,\vA}$ is the self-adjoint operator associated with
the restriction of this form to $\mathcal Q_\PF$, and there are
constants $c>0$ and $C<\infty$ such that
\begin{equation}\label{pauli-form-coercivity}
 q^{v,\vA}[\Phi]+C\|\Phi\|^2
 \geq c\Bigl(\sum_{n=1}^N\|\vpi_n\Phi\|^2
   +\|\grad_{\vX}\Phi\|^2+\|H_\rad^{1/2}\Phi\|^2\Bigr),
 \qquad \Phi\in\mathcal Q.
\end{equation}
\end{proposition}

\begin{proof}
For the spinless Hamiltonian, this is the KLMN argument of
\cite[Section~3.4]{Q}.  The same argument applies with spin.  The only
additional facts needed are the free Pauli identity
$\|\grad_{\vX}\Phi\|^2
 =\sum_n\|\vsigma_n\cdot(-\i\grad_{\vx_n})\Phi\|^2$ and
$(\vsigma_n\cdot\va)^2=|\va|^2$ for real $\va\in\RR^3$. 
\end{proof}

Next, we derive a consequence of the variational principle: if the
ground states of two
Hamiltonians share their density and combined current density, then
each is also a ground state of the other Hamiltonian.

\begin{proposition}[Common ground states]\label[proposition]{prop-common-ground-state}
Let $e\geq0$, let $(v_k,\vA_k)$, $k=1,2$, be two potential pairs, and
let $\Psi_k$ be normalized ground states of $H^{v_k,\vA_k}$ with
energies $E_k=E(v_k,\vA_k)$.  Suppose that
\[
 \rho_{\Psi_1}=\rho_{\Psi_2}=\rho,\qquad
 \vjc_{\Psi_1}=\vjc_{\Psi_2}=\vjc.
\]
Then $\Psi_1$ and $\Psi_2$ are ground states of both $H^{v_1,\vA_1}$
and $H^{v_2,\vA_2}$, and
\begin{equation*}
 E_1-E_2=\dua{v_1-v_2+\tfrac12(|\vA_1|^2-|\vA_2|^2)}{\rho}+\dua{\vA_1-\vA_2}{\vjc}.
\end{equation*}
\end{proposition}

This is the analogue of the ``HK1 theorem'' for paramagnetic CDFT
\cite[Theorem~1]{penz2023structureII}, with the combined current density
in place of the paramagnetic one; \cref{energy-decomposition} supplies
the required linearity of the energy in $(v+\frac12|\vA|^2,\vA)$. 
Note that the ground states may be degenerate.

\begin{proof}
Let
$d=\dua{v_1-v_2+\tfrac12(|\vA_1|^2-|\vA_2|^2)}{\rho}+\dua{\vA_1-\vA_2}{\vjc}$.  The variational principle
and \cref{energy-decomposition} give
\[
 E_1\leq\dua{\Psi_2}{H^{v_1,\vA_1}\Psi_2}=E_2+d,
 \qquad
 E_2\leq\dua{\Psi_1}{H^{v_2,\vA_2}\Psi_1}=E_1-d.
\]
The differences $(E_2+d)-E_1$ and $(E_1-d)-E_2$ are nonnegative and
sum to zero, so both vanish.  Hence $E_1-E_2=d$,
$\dua{\Psi_2}{H^{v_1,\vA_1}\Psi_2}=E_1$ and
$\dua{\Psi_1}{H^{v_2,\vA_2}\Psi_1}=E_2$.  Since $E_k=\inf\spec(H^{v_k,\vA_k})$, the spectral theorem shows that
every normalized $\Phi\in\mathcal Q_\PF$ with
$\dua{\Phi}{H^{v_k,\vA_k}\Phi}=E_k$ satisfies
$H^{v_k,\vA_k}\Phi=E_k\Phi$.  This proves both assertions.
\end{proof}

At positive coupling, we obtain the following restriction
on $\rho(\vA_1-\vA_2)$.

\begin{proposition}[The ambiguity left by the cutoff]
\label[proposition]{prop-cutoff-kernel}
Let $e>0$, let $(v_k,\vA_k)$, $k=1,2$, be two potential pairs, and
let $\Psi_k$ be normalized ground states of $H^{v_k,\vA_k}$ with
energies $E_k=E(v_k,\vA_k)$.  Suppose that
\[
 \rho_{\Psi_1}=\rho_{\Psi_2}=\rho,\qquad
 \vjc_{\Psi_1}=\vjc_{\Psi_2}=\vjc.
\]
Then
\begin{equation}\label{cutoff-kernel}
 \boxed{\begin{aligned}
 &\div_{\vx}\!\left[\rho(\vx)
                   \big(\vA_1(\vx)-\vA_2(\vx)\big)\right]=0,\\
 &\wh\chi(-\vk)\,
   \widehatfull{\Bigl(\rho(\cdot)
            \bigl(\vA_1(\cdot)-\vA_2(\cdot)\bigr)\Bigr)}(\vk)=0
            \quad\text{for a.a. }\vk.
 \end{aligned}}
\end{equation}
The first identity is distributional in $\vx$.
\end{proposition}

\begin{proof}
By \cref{prop-common-ground-state}, we may use the same state
$\Psi=\Psi_1$ as a ground state of both Hamiltonians.  Its density and
combined current density are $\rho$ and $\vjc$.  Let $\vj_k$ be its
physical current density \cref{physical-current-def} for the vector
potential $\vA_k$.  Then
\begin{equation*}
 \vj_1-\vj_2=\rho(\vA_1-\vA_2).
\end{equation*}
Both $\vj_k$ belong to $L^1$, since $\rho\in L^1\cap L^{3/2}$ and
$\vA_k\in L^3+L^\infty$.

For a real function $\theta$ on $\RR^3$, let $U_\theta$ denote
multiplication by $\e^{\i\sum_n\theta(\vx_n)}$.  For
$\theta\in C_c^\infty(\RR^3)$ and $t\in\RR$, the state
$U_{t\theta}\Psi$ belongs to $\mathcal Q_\PF$, and
its energy expectation in $H^{v_k,\vA_k}$ is
\[
 \dua{U_{t\theta}\Psi}{H^{v_k,\vA_k}U_{t\theta}\Psi}
 =E_k+t\int_{\RR^3}\grad\theta\cdot\vj_k
       +\frac{t^2}{2}\int_{\RR^3}\rho|\grad\theta|^2.
\]
Stationarity at $t=0$ gives $\div\vj_k=0$.  Subtraction proves
$\div\big(\rho(\vA_1-\vA_2)\big)=0$.
Here the magnetization density is unchanged by the phase, and
$\int_{\RR^3}\grad\theta\cdot\grad\times\vm_\Psi=0$.

For the photon variation, let $f\in \DC(\omega)\subset\Hil_\ph$, and set
\[
 W(tf)=\exp\big(t a^*(f)-t a(f)\big),\qquad
 c_f^\mu(\vx)=2\Re\dua{G^\mu_\vx}{f}.
\]
The Weyl field shift of \cite[Proposition~A.6(iii)]{Q} gives
$W(tf)^*\vpi_n W(tf)=\vpi_n-et\vc_f(\vx_n)$.  By
\cite[Corollary~A.8]{Q}, $W(tf)$ also preserves the common form
domain.  Apply \cref{energy-decomposition} to this shift
of the external vector potential and add the radiation-energy
variation.  The Pauli cross term includes
$-(et/2)\int_{\RR^3}\vc_f\cdot(\grad\times\vm_\Psi)$, which is already
part of the physical current density $\vj_k$.  Thus
\begin{equation}\label{weyl-energy-polynomial}
 \begin{split}
 \dua{W(tf)\Psi}{H^{v_k,\vA_k} W(tf)\Psi}
 &=E_k+t\left(2\Re\dua{\Psi}{a(\omega f)\Psi}
                            -e\int_{\RR^3}\vc_f\cdot\vj_k\right)\\
 &\quad+t^2\left(\|\sqrt\omega f\|^2
                            +\frac{e^2}{2}\int_{\RR^3}\rho|\vc_f|^2\right).
 \end{split}
\end{equation}
All terms are finite.  In particular, the annihilation expectation is
controlled by $\|\sqrt\omega f\|\|H_\rad^{1/2}\Psi\|$, and
$\vc_f$ is bounded. 

The linear coefficient in \cref{weyl-energy-polynomial} vanishes for
both $k$.  Since the photon expectation refers to the same $\Psi$,
subtraction and $e>0$ give
\begin{equation}\label{weyl-orthogonality}
 \int_{\RR^3}\rho\,\vc_f\cdot(\vA_1-\vA_2)=0\qquad(f\in \DC(\omega)).
\end{equation}
The Bochner integral
\begin{equation}\label{gb-def}
 g=\sum_{\mu=1}^3\int_{\RR^3}\rho(\vx)\,\big(A_1^\mu(\vx)-A_2^\mu(\vx)\big)
                      G^\mu_\vx\dd\vx\in\Hil_\ph
\end{equation}
exists, since $\rho(\vA_1-\vA_2)\in L^1$ and $\|G^\mu_\vx\|$ is bounded
uniformly in $\vx$.  By \cref{weyl-orthogonality},
$2\Re\dua{g}{f}=0$.  Using also $\i f$ and density of
$\DC(\omega)$ in $\Hil_\ph$ proves $g=0$.  Inserting \cref{Gx-def} into \cref{gb-def}
and evaluating the Bochner integral pointwise gives
\begin{equation*}
 g(\vk,\lambda)
 =\frac{(2\pi)^{3/2}}{\sqrt2}\wh\chi(-\vk)
                     \vepsilon_\lambda(\vk)\cdot\widehatfull{\bigl(\rho\,(\vA_1-\vA_2)\bigr)}(\vk).
\end{equation*}
Thus $\wh\chi(-\vk)P_\sol(\vk)\widehatfull{\bigl(\rho\,(\vA_1-\vA_2)\bigr)}(\vk)=0$ almost
everywhere, where $P_\sol(\vk)=\iden-\vk\otimes\vk/|\vk|^2$.  On the
other hand, $\div\big(\rho(\vA_1-\vA_2)\big)=0$ gives
$\vk\cdot\widehatfull{\bigl(\rho\,(\vA_1-\vA_2)\bigr)}(\vk)=0$.  The transverse
projection is therefore redundant for $\vk\ne0$, proving
\cref{cutoff-kernel}.
\end{proof}

The continuity equation, which gives the first identity in
\cref{cutoff-kernel}, and the photon variation, which gives
\cref{weyl-orthogonality}, have distinct roles.
Write $\vF=\rho(\vA_1-\vA_2)$ and decompose, for $\vk\ne0$,
$\wh\vF(\vk)=P_\sol(\vk)\wh\vF(\vk)
 +(\vk\otimes\vk/|\vk|^2)\wh\vF(\vk)$.
The photon variation gives
$\wh\chi(-\vk)P_\sol(\vk)\wh\vF(\vk)=0$, so it detects the transverse
part only at momenta with $\wh\chi(-\vk)\ne0$.  The continuity equation
$\div\vF=0$ gives $\vk\cdot\wh\vF(\vk)=0$, which removes the
longitudinal part at all momenta.

Under an additional condition on either the cutoff or the density,
\cref{cutoff-kernel} determines the vector potential on the support of
the density.

\begin{proposition}[Vector potential on the density support]
\label[proposition]{prop-vector-uniqueness}
Let $e>0$ and $\wh\chi\not\equiv0$, let $(v_k,\vA_k)$, $k=1,2$, be two
potential pairs, and let $\Psi_k$ be normalized ground states of
$H^{v_k,\vA_k}$.  Suppose that
\[
 \rho_{\Psi_1}=\rho_{\Psi_2}=\rho,\qquad
 \vjc_{\Psi_1}=\vjc_{\Psi_2}=\vjc,
\]
and that at least one of the following conditions holds.
\begin{enumerate}[(i)]
\item $\wh\chi(\vk)\ne0$ for almost every $\vk\in\RR^3$.
\item The density has the exponential moment
\cref{density-exponential-moment} for some $\alpha>0$.
\end{enumerate}
Then $\rho(\vA_1-\vA_2)=0$ almost everywhere.  In particular, if
$\rho>0$ almost everywhere, then $\vA_1=\vA_2$ almost everywhere.
\end{proposition}

\begin{proof}
By H\"older's inequality, $\rho(\vA_1-\vA_2)\in L^1(\RR^3,\RR^3)$.
Under condition (i), \cref{cutoff-kernel} of \cref{prop-cutoff-kernel} gives $\widehatfull{\bigl(\rho\,(\vA_1-\vA_2)\bigr)}=0$
almost everywhere.  Fourier uniqueness on $L^1$ then
gives $\rho(\vA_1-\vA_2)=0$.

Under condition (ii), write $\vA_k=\vA_{k,3}+\vA_{k,\infty}$
with $\vA_{k,3}\in L^3$ and $\vA_{k,\infty}\in L^\infty$ for $k=1,2$.
Interpolation gives
\begin{equation*}
 \|\e^{\alpha|\vx|/2}\rho\|_{3/2}
 \leq\|\e^{\alpha|\vx|}\rho\|_1^{1/2}\|\rho\|_3^{1/2}.
\end{equation*}
Consequently,
\begin{equation}\label{b-exponential-moment}
 \begin{split}
 &\int_{\RR^3}\e^{\alpha|\vx|/2}\rho(\vx)|\vA_1(\vx)-\vA_2(\vx)|\dd\vx\\
 &\quad\leq\sum_{k=1}^2\left(
       \|\vA_{k,3}\|_3\|\e^{\alpha|\vx|/2}\rho\|_{3/2}
       +\|\vA_{k,\infty}\|_\infty\|\e^{\alpha|\vx|/2}\rho\|_1
       \right)
 <\infty.
 \end{split}
\end{equation}
Since $|\e^{-\i\vzeta\cdot\vx}|\leq \e^{|\Im\vzeta||\vx|}$, the bound
\cref{b-exponential-moment} makes the Fourier integral of each
component of $\rho(\vA_1-\vA_2)$ converge absolutely and locally
uniformly on $\{\vzeta\in\CC^3:|\Im\vzeta|<\alpha/2\}$.  Each component
of $\widehatfull{\bigl(\rho\,(\vA_1-\vA_2)\bigr)}$ therefore extends
holomorphically to this strip and is real analytic on $\RR^3$.

Because $\wh\chi\not\equiv0$ in $L^2$, the set on which
$\wh\chi(-\vk)\ne0$ has positive Lebesgue measure.  Equation
\cref{cutoff-kernel} of \cref{prop-cutoff-kernel} makes every component of $\widehatfull{\bigl(\rho\,(\vA_1-\vA_2)\bigr)}$
zero almost
everywhere on this set.  A real-analytic function on a connected open
set that vanishes on a set of positive measure is identically zero.
Applying this fact to real and imaginary parts gives
$\widehatfull{\bigl(\rho\,(\vA_1-\vA_2)\bigr)}=0$ on $\RR^3$.  Fourier uniqueness gives
$\rho(\vA_1-\vA_2)=0$.  Division by $\rho$
on $\{\rho>0\}$ proves the last assertion.
\end{proof}

\begin{remark}[The role of conditions (i) and (ii)]\label{rem-cutoff-kernel}
Without condition (i) or (ii) of \cref{prop-vector-uniqueness}, the
identities \cref{cutoff-kernel} alone do not force
$\rho(\vA_1-\vA_2)=0$.  For example, the hard cutoff
$\wh\chi_\Lambda$ vanishes for $|\vk|>\Lambda$.  Let
$\eta\in C_c^\infty(\RR^3)$ be nonzero, real and even, with support in
$\{\Lambda+1<|\vk|<\Lambda+2\}$, and let $\vF$ be the vector field with
$\wh\vF(\vk)=\i\eta(\vk)\,\vk\times\ve_3$.  Then $\vF$ is a nonzero,
real, divergence-free Schwartz field with
$\wh\chi_\Lambda(-\vk)\wh\vF(\vk)=0$ for all $\vk$, so it satisfies
both identities in \cref{cutoff-kernel} in place of
$\rho(\vA_1-\vA_2)$.  Condition (i) rules this out, since it forces the
Fourier transform to vanish almost everywhere.  Condition (ii) rules it
out as well, since it makes the Fourier transform of
$\rho(\vA_1-\vA_2)$ real analytic, whereas $\wh\vF$ is nonzero with
compact support.  The example concerns only the identities
\cref{cutoff-kernel} and is not built from two Pauli--Fierz ground
states, so it is not a counterexample at positive coupling.  Removing
conditions (i) and (ii) would require information beyond
\cref{cutoff-kernel}.
\end{remark}

\subsection{Proof of \texorpdfstring{\cref{ucp-main-theorem}}{Theorem \getrefnumber{ucp-main-theorem}}}\label{sec-ucp}

The total scalar potential
\[
 V(\vX)=\sum_{n=1}^N v(\vx_n)+\sum_{1\leq n<m\leq N}\frac{1}{|\vx_n-\vx_m|}
\]
has one-body terms in $L^p_\mathrm{loc}$ and an interaction term that
is locally bounded on the collision-free configuration space
$\mathcal U$ from \cref{ucp-regular-configurations}.  The scalar perturbation argument
below treats the one-body terms in their three-dimensional variables.

We first prove continuation for the spinless Hamiltonian.  Its local
estimates will then be combined with the magnetic terms to prove
\cref{ucp-main-theorem}.  The first step of Garrigue's argument \cite{garrigue2020unique}
applies to the full spinless Pauli--Fierz form.  In particular, $H_\rad$
need not be estimated as a scalar multiplication operator in this step.
All configuration-space balls below are balls in $\RR^{3N}$.  The
form of $H_{\mathrm{sl}}^{v,\vA}$ is also defined on the full space
$L^2(\RR^{3N},\Fock_\ph)$, without antisymmetry, with form domain
$H^1(\RR^{3N},\Fock_\ph)\cap\DC(H_\rad^{1/2})$, and we denote it by
$\dua{\Phi}{H_{\mathrm{sl}}^{v,\vA}\Psi}$ as well.  This is needed because
localization by a function of $\vX$ does not preserve antisymmetry.
We also use the standard creation and annihilation operator bounds
\cite[Section~3.3]{Q},
which with \cref{cutoff-assumption} give, uniformly in $\vx$,
\begin{equation}\label{ucp-field-form-bound}
 \sum_{\mu=1}^3\|\Af^\mu(\vx)\xi\|_{\Fock_\ph}^2
 \leq C_\chi\left(\|H_\rad^{1/2}\xi\|_{\Fock_\ph}^2
                         +\|\xi\|_{\Fock_\ph}^2\right),
 \qquad \xi\in\DC(H_\rad^{1/2}).
\end{equation}

We begin with a Caccioppoli-type inequality, which bounds the local
energy of an eigenstate by its $L^2$ mass on a larger ball.  For
$\Phi\in L^2(\RR^{3N},\Fock_\ph)$, write
$u_\Phi(\vX)=\|\Phi(\vX)\|_{\Fock_\ph}$.

\begin{lemma}[Local kinetic energy estimates]\label[lemma]{ucp-caccioppoli}
Assume \cref{cutoff-assumption,potential-spaces}, and let
$\Psi\in\mathcal Q_\PF^{\mathrm{sl}}$ be an eigenstate of $H_{\mathrm{sl}}^{v,\vA}$ with eigenvalue
$E$.  There is a constant $C$ such that,
for every $\vX_0\in\RR^{3N}$ and $0<r\leq1$, the following hold.
\begin{enumerate}[(i)]
\item The local kinetic and radiation energy is bounded by the $L^2$
mass on the doubled ball,
\begin{align}
 &\int_{B_r(\vX_0)}\left(
   \sum_{n=1}^N\|\vpi_n\Psi(\vX)\|_{\Fock_\ph}^2
   +\|H_\rad^{1/2}\Psi(\vX)\|_{\Fock_\ph}^2
                      \right)\dd\vX\notag\\
 &\hspace{35mm}\leq \frac{C}{r^{2}}
       \int_{B_{2r}(\vX_0)}\|\Psi(\vX)\|_{\Fock_\ph}^2\dd\vX.
 \label{ucp-local-energy}
\end{align}
\item For every $\eta\in C_c^\infty(\RR^{3N},\RR)$ with
$0\leq\eta\leq1$, $\eta=1$ on $B_r(\vX_0)$,
$\supp\eta\subset B_{2r}(\vX_0)$, and $|\grad\eta|\leq C_1/r$,
\begin{equation}\label{ucp-local-gradient}
 \|\grad_{\vX}(\eta u_\Psi)\|^2
 \leq\frac{C}{r^2}\int_{B_{2r}(\vX_0)}u_\Psi(\vX)^2\dd\vX.
\end{equation}
\end{enumerate}
The constant $C$ depends on the Hamiltonian, $E$, and $C_1$, but not on
$r$, $\vX_0$, or $\eta$.
\end{lemma}

\begin{proof}
Let
$\mathcal Q^{\mathrm{sl}}=H^1(\RR^{3N},\Fock_\ph)\cap\DC(H_\rad^{1/2})$
be the form domain without antisymmetry, so that
$\mathcal Q_\PF^{\mathrm{sl}}$ is its antisymmetric subspace.
Permutations of the electronic coordinates act unitarily on
$\mathcal Q^{\mathrm{sl}}$ and leave the form invariant.  Averaging over permutations,
the antisymmetrizer $\Pi_-$ satisfies
$\dua{\Pi_-\Phi}{H_{\mathrm{sl}}^{v,\vA}\Psi}
 =\dua{\Phi}{H_{\mathrm{sl}}^{v,\vA}\Pi_-\Psi}$ for all
$\Phi,\Psi\in\mathcal Q^{\mathrm{sl}}$.  Since the eigenstate satisfies $\Pi_-\Psi=\Psi$,
this gives
\[
 \dua{\Phi}{H_{\mathrm{sl}}^{v,\vA}\Psi}=\dua{\Pi_-\Phi}{H_{\mathrm{sl}}^{v,\vA}\Psi},\qquad
 \dua{\Phi}{\Psi}=\dua{\Pi_-\Phi}{\Psi}.
\]
The weak eigenvalue equation on the antisymmetric subspace consequently
gives
\begin{equation}\label{ucp-full-weak-equation}
 \dua{\Phi}{H_{\mathrm{sl}}^{v,\vA}\Psi}=E\dua{\Phi}{\Psi}
\end{equation}
for every $\Phi\in\mathcal Q^{\mathrm{sl}}$.

\begin{claim}
For every $\Phi\in\mathcal Q^{\mathrm{sl}}$, $u_\Phi\in H^1(\RR^{3N})$
and
\begin{equation}\label{ucp-diamagnetic}
 |\grad_{\vX}u_\Phi(\vX)|^2
 \leq\sum_{n=1}^N\|(\vpi_n\Phi)(\vX)\|_{\Fock_\ph}^2
 \qquad\text{for a.a. }\vX\in\RR^{3N}.
\end{equation}
\end{claim}

\begin{proof}[Proof of the claim]
Fix $\epsilon>0$ and put
$u_\epsilon(\vX)=(u_\Phi(\vX)^2+\epsilon^2)^{1/2}-\epsilon$.  The map
$\xi\mapsto(\|\xi\|_{\Fock_\ph}^2+\epsilon^2)^{1/2}$ is continuously
differentiable and Lipschitz on $\Fock_\ph$, so the chain rule for
$\Phi\in H^1(\RR^{3N},\Fock_\ph)$ gives $u_\epsilon\in H^1(\RR^{3N})$ with
\[
 \partial_{x_n^\mu}u_\epsilon(\vX)
 =\frac{\Re\dua{\Phi(\vX)}{\partial_{x_n^\mu}\Phi(\vX)}_{\Fock_\ph}}
       {(u_\Phi(\vX)^2+\epsilon^2)^{1/2}}.
\]
By the definition of $\vpi_n$,
\[
 \partial_{x_n^\mu}\Phi(\vX)=\i(\pi_n^\mu\Phi)(\vX)
 -\i\bigl(A^\mu(\vx_n)-e\Af^\mu(\vx_n)\bigr)\Phi(\vX).
\]
The operator $A^\mu(\vx_n)-e\Af^\mu(\vx_n)$ is symmetric, so its
expectation in $\Phi(\vX)$ is real, and therefore
\[
 \Re\dua{\Phi(\vX)}{\partial_{x_n^\mu}\Phi(\vX)}_{\Fock_\ph}
 =\Re\dua{\Phi(\vX)}{\i(\pi_n^\mu\Phi)(\vX)}_{\Fock_\ph}
 \leq u_\Phi(\vX)\|(\pi_n^\mu\Phi)(\vX)\|_{\Fock_\ph}.
\]
These expectations are defined by \cref{cutoff-assumption}, since
$\Phi(\vX)\in\DC(H_\rad^{1/2})$ for almost every $\vX$.
Summing over $n$ and $\mu$ gives
\[
 |\grad_{\vX}u_\epsilon(\vX)|^2
 \leq\sum_{n=1}^N\|(\vpi_n\Phi)(\vX)\|_{\Fock_\ph}^2
 \qquad\text{for almost every }\vX.
\]
Since $0\leq u_\Phi(\vX)-u_\epsilon(\vX)\leq\min(u_\Phi(\vX),\epsilon)$,
$u_\epsilon\to u_\Phi$ in $L^2$ as $\epsilon\downarrow0$, while
$\grad_{\vX}u_\epsilon$ stays bounded in $L^2$.  Hence
$u_\Phi\in H^1(\RR^{3N})$, and $\grad_{\vX}u_\epsilon$ converges weakly
to $\grad_{\vX}u_\Phi$ along a subsequence.  The set of
$g\in L^2(\RR^{3N},\RR^{3N})$ with
$|g(\vX)|^2\leq\sum_n\|(\vpi_n\Phi)(\vX)\|_{\Fock_\ph}^2$ for almost
every $\vX$ is convex and closed, hence weakly closed, which proves
\cref{ucp-diamagnetic}.
\end{proof}

The assumption on $v$ and the three-dimensional Sobolev inequality,
applied separately to every electronic coordinate and then integrated
in the other coordinates, give, with $v_-=\max(-v,0)$ the negative part
of $v$, for every $\epsilon>0$,
\begin{equation}\label{ucp-negative-potential}
 \sum_{n=1}^N\int_{\RR^{3N}}v_-(\vx_n)u_\Phi(\vX)^2\dd\vX
 \leq\epsilon\|\grad_{\vX}u_\Phi\|^2
       +C_\epsilon\|\Phi\|^2.
\end{equation}
To obtain the arbitrarily small coefficient, split the $L^{3/2}$
part of $v_-$ into a bounded function and an $L^{3/2}$ function of
arbitrarily small norm.  Since the electron--electron interaction
and $H_\rad$ are nonnegative, \cref{ucp-diamagnetic,ucp-negative-potential}
with $\epsilon=1/4$ imply
\begin{equation}\label{ucp-form-lower-bound}
 \dua{\Phi}{H_{\mathrm{sl}}^{v,\vA}\Phi}\geq\frac14\sum_{n=1}^N\|\vpi_n\Phi\|^2
                +\|H_\rad^{1/2}\Phi\|^2-C_0\|\Phi\|^2.
\end{equation}

Let $\eta$ be as in (ii).  Multiplication by $\eta$ preserves
$\mathcal Q^{\mathrm{sl}}$.  Since $\eta$ is a real scalar function,
$\vpi_n(\eta\Psi)=\eta\vpi_n\Psi-\i(\grad_{\vx_n}\eta)\Psi$.  This and
the eigenvalue equation \cref{ucp-full-weak-equation}, tested with $\eta^2\Psi$, give the IMS
localization identity
\begin{equation}\label{ucp-ims}
 \begin{split}
 \dua{\eta\Psi}{H_{\mathrm{sl}}^{v,\vA}\eta\Psi}
 &=\Re\dua{\eta^2\Psi}{H_{\mathrm{sl}}^{v,\vA}\Psi}
   +\frac12\int_{\RR^{3N}}|\grad_{\vX}\eta(\vX)|^2u_\Psi(\vX)^2\dd\vX\\
 &=E\|\eta\Psi\|^2
   +\frac12\int_{\RR^{3N}}|\grad_{\vX}\eta(\vX)|^2u_\Psi(\vX)^2\dd\vX.
 \end{split}
\end{equation}
Combining \cref{ucp-form-lower-bound,ucp-ims} proves
\begin{equation}\label{ucp-localized-form-energy}
 \sum_{n=1}^N\|\vpi_n(\eta\Psi)\|^2
 +\|H_\rad^{1/2}\eta\Psi\|^2
 +\|\grad_{\vX}(\eta u_\Psi)\|^2
 \leq Cr^{-2}\int_{B_{2r}(\vX_0)}u_\Psi(\vX)^2\dd\vX.
\end{equation}
Since $\eta=1$ and $\grad\eta=0$ on $B_r(\vX_0)$, the product rule
above gives $\vpi_n(\eta\Psi)=\vpi_n\Psi$ and
$H_\rad^{1/2}(\eta\Psi)=H_\rad^{1/2}\Psi$ on $B_r(\vX_0)$.  Restricting
the norms on the left of \cref{ucp-localized-form-energy} to
$B_r(\vX_0)$ therefore proves \cref{ucp-local-energy}, and dropping
the first two terms on the left proves \cref{ucp-local-gradient}.
\end{proof}

Next, the de Figueiredo--Gossez argument \cite{defigueiredo1992strict},
in the form used by Garrigue \cite{garrigue2020unique},
turns the preceding inequality into infinite-order vanishing: an eigenstate
vanishes to infinite order at every Lebesgue density point of its zero
set.

\begin{proposition}[Zeros of infinite order, cf.\ \cite{defigueiredo1992strict}]\label[proposition]{ucp-infinite-order}
Assume \cref{cutoff-assumption,potential-spaces}, let
$\Psi\in\mathcal Q_\PF^{\mathrm{sl}}$ be an eigenstate of
$H_{\mathrm{sl}}^{v,\vA}$ with eigenvalue $E$, and put
\[
 \mathcal Z=\{\vX\in\RR^{3N}:\Psi(\vX)=0\}.
\]
At every Lebesgue density point $\vX_0$ of $\mathcal Z$, $\Psi$ vanishes to
infinite order, with the norm of $\Fock_\ph$ in place of that of
$\mathcal K_N$.
Moreover, for every $m>0$,
\begin{equation}\label{ucp-energy-flat}
 \int_{B_r(\vX_0)}\Bigl(\sum_{n=1}^N\|\vpi_n\Psi(\vX)\|_{\Fock_\ph}^2
   +\|H_\rad^{1/2}\Psi(\vX)\|_{\Fock_\ph}^2\Bigr)\dd\vX=\mathcal O(r^m)
 \qquad(r\downarrow0).
\end{equation}
In particular, if $|\mathcal Z|>0$, such a point exists.
\end{proposition}

\begin{proof}
Without loss of generality, we can translate $\vX_0$ to the
origin and set
\[
 U(r)=\int_{B_r}u_\Psi(\vX)^2\dd\vX.
\]
Then $\theta(r):=|B_r\setminus \mathcal Z|/|B_r|\to0$ as $r\downarrow0$.
Let $\eta\in C_c^\infty(\RR^{3N},\RR)$ satisfy $0\leq\eta\leq1$, $\eta=1$ on
$B_r$, $\supp\eta\subset B_{2r}$, and $|\grad\eta|\leq C_1/r$, as in
\cref{ucp-caccioppoli}(ii).  H\"older's inequality,
the Sobolev inequality on $\RR^{3N}$, and
\cref{ucp-local-gradient} of \cref{ucp-caccioppoli} give
\begin{align}
 U(r)&\leq\|\eta u_\Psi\|_2^2\notag\\
 &\leq |B_{2r}\setminus \mathcal Z|^{2/(3N)}
                  \|\eta u_\Psi\|_{6N/(3N-2)}^2\notag\\
 &\leq C|B_{2r}\setminus \mathcal Z|^{2/(3N)}
                  \|\grad_{\vX}(\eta u_\Psi)\|_2^2\notag\\
 &\leq C\theta(2r)^{2/(3N)}U(2r).
 \label{ucp-density-recurrence}
\end{align}
Fix $m>0$.  Since $\theta(s)\to0$ as $s\downarrow0$, there is
$r_m\in(0,1)$ such that $C\theta(s)^{2/(3N)}\leq2^{-m}$ whenever
$0<s\leq r_m$.  Let $k\geq1$ and apply \cref{ucp-density-recurrence}
with $r=2^{-k}r_m\leq1$.  Since $2r=2^{-k+1}r_m\leq r_m$, the choice
of $r_m$ gives
\[
 U(2^{-k}r_m)\leq2^{-m}U(2^{-k+1}r_m).
\]
Induction on $k$ yields
\[
 U(2^{-k}r_m)\leq2^{-km}U(r_m),\qquad k\geq0.
\]
Now let $0<r\leq r_m$, and choose $k\geq0$ with
$2^{-k-1}r_m<r\leq2^{-k}r_m$.  Then $2^{-k}<2r/r_m$.  Since $U$ is
nondecreasing and $U(r_m)\leq\|\Psi\|^2$,
\[
 U(r)\leq U(2^{-k}r_m)\leq2^{-km}U(r_m)
 \leq\Big(\frac{2r}{r_m}\Big)^m\|\Psi\|^2.
\]
Thus
\begin{equation}\label{ucp-U-bound}
 U(r)\leq C_mr^m\quad(0<r\leq r_m),\qquad
 C_m=\Big(\frac{2}{r_m}\Big)^m\|\Psi\|^2,
\end{equation}
so $\Psi$ vanishes to infinite order at the origin.

For \cref{ucp-energy-flat}, let $0<r\leq r_{m+2}/2$.  Then $r\leq1$
and $2r\leq r_{m+2}$, so \cref{ucp-caccioppoli} and \cref{ucp-U-bound}
with $m+2$ in place of $m$ give
\begin{align*}
 &\int_{B_r}\Bigl(\sum_{n=1}^N\|\vpi_n\Psi(\vX)\|_{\Fock_\ph}^2
   +\|H_\rad^{1/2}\Psi(\vX)\|_{\Fock_\ph}^2\Bigr)\dd\vX\\
 &\qquad\leq\frac{C}{r^2}U(2r)
 \leq\frac{C}{r^2}C_{m+2}(2r)^{m+2}
 =2^{m+2}CC_{m+2}\,r^m.
\end{align*}
The two extra powers of $r$ compensate for the factor $r^{-2}$ in
\cref{ucp-caccioppoli}.

Finally, by the Lebesgue density theorem, almost every point of $\mathcal Z$
is a Lebesgue density point of $\mathcal Z$, that is, a point $\vX_0$ with
$|B_r(\vX_0)\setminus \mathcal Z|/|B_r(\vX_0)|\to0$ as $r\downarrow0$.  If
$|\mathcal Z|>0$, the set of such points has positive measure, and in
particular it is nonempty.
\end{proof}

The Carleman estimates in \cref{ucp-carleman} are proved for smooth
functions.  To apply them
to a localized eigenstate, we show that the localized state lies in the
operator domain of $-\Delta_{\vX}+H_\rad$ and can be approximated there
by smooth functions.
This step uses the first ultraviolet moment
\cref{ucp-extra-cutoff}, which includes the hard ultraviolet cutoff.

We repeatedly use the moment inequality: for a nonnegative self-adjoint
operator $A$ and $0\leq\theta\leq1$,
\begin{equation}\label{moment-inequality}
 \|A^\theta u\|\leq\|Au\|^\theta\|u\|^{1-\theta},
 \qquad u\in\DC(A).
\end{equation}
It follows from the spectral theorem and H\"older's inequality for the
spectral measure of $u$.  For $\theta<1$, Young's inequality turns it
into $\|A^\theta u\|\leq\epsilon\|Au\|+C_{\epsilon,\theta}\|u\|$ for
every $\epsilon>0$.

\begin{lemma}[Local operator domain]\label[lemma]{ucp-local-operator-domain}
Assume \cref{cutoff-assumption,potential-spaces,ucp-extra-cutoff}.
Let $\mathcal O\subset\mathcal U$ be open, with $\mathcal U$ as in
\cref{ucp-regular-configurations}, and let $\mathcal O_1\subset\RR^3$ be the
union of its electronic coordinate projections.  Suppose that, for
some $p>2$,
\[
 v\in L^p_\mathrm{loc}(\mathcal O_1),\qquad
 \vA\in L^6_\mathrm{loc}(\mathcal O_1,\RR^3),\qquad
 \div\vA\in L^p_\mathrm{loc}(\mathcal O_1),
\]
where the divergence is distributional.
If $\Psi$ is an
eigenstate of $H_{\mathrm{sl}}^{v,\vA}$ with eigenvalue $E$, then for every
$\eta\in C_c^\infty(\mathcal O)$,
\begin{equation}\label{ucp-local-free-domain}
 \eta\Psi\in\DC(-\Delta_{\vX})\cap\DC(H_\rad),
\end{equation}
and
\begin{equation}\label{ucp-localized-operator-equation}
 (H_{\mathrm{sl}}^{v,\vA}-E)(\eta\Psi)
 =-\i\sum_{n=1}^N(\grad_{\vx_n}\eta)\cdot\vpi_n\Psi
                         -\frac12(\Delta_{\vX}\eta)\Psi.
\end{equation}
Here $H_{\mathrm{sl}}^{v,\vA}$ acts on
$u\in\DC(-\Delta_{\vX})\cap\DC(H_\rad)$ with compact support in
$\mathcal O$ by
\begin{equation}\label{ucp-local-expression}
 H_{\mathrm{sl}}^{v,\vA}u
 =T^0u+\sum_{n=1}^N\Bigl(\vA(\vx_n)\cdot\vD_nu
   -\frac{\i}{2}(\div\vA)(\vx_n)u+\frac12|\vA(\vx_n)|^2u\Bigr)+Vu,
\end{equation}
where $T^0=\frac12\sum_{n=1}^N\vD_n^2+H_\rad$, $V$ is the total
scalar potential, and each term belongs to $L^2(\RR^{3N},\Fock_\ph)$.
\end{lemma}

\begin{proof}
Fix $\eta\in C_c^\infty(\mathcal O)$.  Since $\supp\eta$ is compact,
there is an open set $\mathcal V\supset\supp\eta$ with compact closure
in $\mathcal O$.  Choose $\zeta\in C_c^\infty(\mathcal V)$ equal to one
near $\supp\eta$.  We replace $H_{\mathrm{sl}}^{v,\vA}$ by an operator
$\widetilde H_{\mathrm{sl}}$ on $L^2(\RR^{3N},\Fock_\ph)$ whose
potentials are cut off by $\zeta$.  It agrees with
$H_{\mathrm{sl}}^{v,\vA}$ near $\supp\eta$, and its perturbations are
relatively bounded with respect to $-\Delta_{\vX}+H_\rad$.  We show
in Steps~1--3 that
$\DC(\widetilde H_{\mathrm{sl}})=\DC(-\Delta_{\vX}+H_\rad)$, with
equivalent graph norms, and in Step~4 that
$\eta\Psi\in\DC(\widetilde H_{\mathrm{sl}})$ and
\begin{equation}\label{ucp-localized-tilde-equation}
 \widetilde H_{\mathrm{sl}}(\eta\Psi)
 =E\eta\Psi-\i\sum_{n=1}^N(\grad_{\vx_n}\eta)\cdot\vpi_n\Psi
                         -\frac12(\Delta_{\vX}\eta)\Psi.
\end{equation}
Since $\DC(-\Delta_{\vX}+H_\rad)=\DC(-\Delta_{\vX})\cap\DC(H_\rad)$,
these two facts prove \cref{ucp-local-free-domain}.  By Step~1,
$\widetilde H_{\mathrm{sl}}u$ is given by \cref{ucp-local-expression}
for $u\in\DC(\widetilde H_{\mathrm{sl}})$ supported where $\zeta=1$, so
\cref{ucp-localized-tilde-equation} also proves
\cref{ucp-localized-operator-equation}.  Since $\eta$ is
arbitrary, Steps~1--3, applied with a cutoff equal to one near the
support of a given $u$, also show that each term of
\cref{ucp-local-expression} belongs to $L^2$.

\noindent\textbf{Step 1: The localized Hamiltonian.}
Let $K_n\subset\mathcal O_1$ be the compact projection of
$\ol{\mathcal V}$ onto the coordinate $\vx_n$, and put
$v_n=v\iden_{K_n}\in L^p(\RR^3)$.  Since $\ol{\mathcal V}$ is a compact
subset of $\mathcal U$, the electron--electron interaction is bounded
on $\mathcal V$.  Hence the total scalar potential $V$ from the
beginning of this subsection satisfies
\begin{equation*}
 V(\vX)=\sum_{n=1}^N v_n(\vx_n)+W(\vX),\qquad
 W\in L^\infty(\mathcal V),
\end{equation*}
for $\vX\in\mathcal V$.  Let $\widetilde V$ be the extension of
$\zeta V$ by zero.  It is a real multiplication operator, equal to $V$
near $\supp\eta$.  For
$n=1,\ldots,N$, let $P_n\subset\mathcal O_1$ be the compact projection
of $\supp\zeta$ onto the coordinate $\vx_n$.  Since
$\vA\in L^6_\mathrm{loc}(\mathcal O_1,\RR^3)$, we have
$\iden_{P_n}\vA\in L^6(\RR^3,\RR^3)$.  Put
\[
 \widetilde\vA_n(\vX)=\zeta(\vX)\vA(\vx_n),\qquad n=1,\ldots,N,
\]
with $\widetilde\vA_n(\vX)=0$ for $\vX\notin\supp\zeta$.  Then
$\widetilde\vA_n(\vX)=\vA(\vx_n)$ for $\vX$ near $\supp\eta$, and
$\widetilde\vA_n(\vX)=\zeta(\vX)(\iden_{P_n}\vA)(\vx_n)$ for all $\vX$.  The
distributional product rule gives
\[
 \grad_{\vx_n}\cdot\widetilde\vA_n
 =\zeta\,(\div\vA)(\vx_n)+\grad_{\vx_n}\zeta\cdot\vA(\vx_n).
\]
The first term is bounded in modulus by a one-body $L^p$ multiplier,
and the second by a one-body $L^6$ multiplier.  Also
$|\widetilde\vA_n|^2$ is bounded in modulus by a one-body $L^3$
multiplier.  The localized Hamiltonian is
\begin{equation}\label{ucp-localized-hamiltonian}
 \widetilde H_{\mathrm{sl}}
 =\frac12\sum_{n=1}^N\bigl(\vD_n+\widetilde\vA_n\bigr)^2+H_\rad+\widetilde V
 =T^0+\widetilde R+\widetilde V,
\end{equation}
with $T^0$ as in \cref{ucp-local-expression}, and
\[
 \widetilde R
 =\sum_{n=1}^N\Bigl(\widetilde\vA_n\cdot\vD_n
   -\frac{\i}{2}\grad_{\vx_n}\cdot\widetilde\vA_n
   +\frac12|\widetilde\vA_n|^2\Bigr).
\]
Note that $\widetilde H_{\mathrm{sl}}$ is not of the form
$H_{\mathrm{sl}}^{v',\vA'}$, since
$\widetilde\vA_n$ and $\widetilde V$ depend on all electronic coordinates.
Steps~2 and~3 realize \cref{ucp-localized-hamiltonian} as a
self-adjoint operator.

\noindent\textbf{Step 2: The free operator $T^0$.}
Theorem 6 of \cite{hasler2008selfadjoint} gives
$\DC(T^0)=\DC(-\Delta_{\vX}+H_\rad)$ and states that any operator core
for $-\Delta_{\vX}+H_\rad$ is an operator core for $T^0$.  Theorem 7
there shows that $T^0$, defined through its closed form, acts on this
domain by the expanded expression $\frac12\sum_n\vD_n^2+H_\rad$.  To
check the hypotheses of both theorems,
write $\Hil_\omega$ for the one-photon space equipped with the norm
\[
 \|g\|_{\Hil_\omega}^2=\|g\|^2+\|g/\sqrt\omega\|^2.
\]
Our coupling functions lie uniformly in $\Hil_\omega$.  Since
$\sqrt\omega\wh\chi\in L^2$ follows from
$\wh\chi,\omega\wh\chi\in L^2$, the extra
assumption \cref{ucp-extra-cutoff} puts $\omega G^\mu_\vx$ and all
first electronic derivatives of $G^\mu_\vx$ uniformly in this space.
Moreover, $\sum_{\mu=1}^3\partial_{x^\mu}G^\mu_\vx=0$, since
$\vk\cdot\vepsilon_\lambda(\vk)=0$; that is, $\div\AF=0$.
These are Hypothesis (G) of Theorem 6 and the hypotheses of Theorem 7
in that reference.  The different positive coefficient of the electronic
kinetic energy is accommodated by a linear rescaling of electronic
coordinates.  The coupling strength is arbitrary.  In particular,
$T^0$ and $-\Delta_{\vX}+H_\rad$ have equivalent graph norms.  Put
\begin{equation}\label{ucp-free-comparison-operator}
 L=-\Delta_{\vX}+H_\rad+1.
\end{equation}

\noindent\textbf{Step 3: Relative bounds and the domain of
$\widetilde H_{\mathrm{sl}}$.}
Let $b\in L^q(\RR^3)$ with $q>2$
and put $s=3/q$.  Then
the three-dimensional Sobolev inequality in the $n$th electronic
coordinate gives
\begin{equation}\label{ucp-coordinate-multiplier}
 \begin{split}
 \|b(\vx_n)\Phi\|_2
 &\leq C_q\|b\|_q
                 \|(-\Delta_{\vx_n})^{s/2}\Phi\|_2\\
 &\leq C_q\|b\|_q
                 \|(-\Delta_{\vX})^{s/2}\Phi\|_2.
 \end{split}
\end{equation}
The Sobolev inequality holds for Hilbert-space-valued functions.  We
apply it with values in $\Fock_\ph$ at fixed values of the remaining
electronic coordinates and then integrate over these coordinates.  For $q=6$, apply this estimate
to electronic derivatives and to $(H_\rad+1)^{1/2}\Phi$.
The electronic and radiation spectral multipliers commute, and
\[
 \|(-\Delta_{\vx_n})^{1/4}(H_\rad+1)^{1/2}\Phi\|_2
 \leq\|L^{3/4}\Phi\|_2.
\]
Together with the pointwise field bound
\cref{ucp-field-form-bound}, this yields, for
$\vb\in L^6(\RR^3,\RR^3)$,
\begin{equation}\label{ucp-vector-multiplier}
 \|\vb(\vx_n)\cdot\vD_n\Phi\|_2
 \leq C\|\vb\|_6\|L^{3/4}\Phi\|_2.
\end{equation}
Since $\widetilde\vA_n=\zeta\,(\iden_{P_n}\vA)(\vx_n)$ and $\zeta$ is
bounded, \cref{ucp-vector-multiplier} with $\vb=\iden_{P_n}\vA$ bounds
$\widetilde\vA_n\cdot\vD_n$.  The scalar terms are bounded by $\|L^{s/2}\Phi\|_2$ with
$s/2<3/4$, and the first-order term by $\|L^{3/4}\Phi\|_2$.  Since
these powers of $L$ are below one, the moment inequality
\cref{moment-inequality} gives, for every $\epsilon>0$,
\[
 \|\widetilde R\Phi\|_2+\|\widetilde V\Phi\|_2
 \leq\epsilon\|L\Phi\|_2
                         +C_\epsilon\|\Phi\|_2.
\]
By Step~2, both perturbations are infinitesimally
bounded with respect to $T^0$.  The distributional product rule, justified first by
smooth approximation and then by these bounds, shows that
$\widetilde R$ is symmetric on $\DC(T^0)$ and agrees there with the
difference of the kinetic forms.  The Kato--Rellich theorem
\cite[Theorem~V.4.3]{kato1995perturbation} identifies
$T^0+\widetilde R+\widetilde V$ with the form realization of
\cref{ucp-localized-hamiltonian} and gives
\begin{equation}\label{ucp-truncated-domain}
 \DC(\widetilde H_{\mathrm{sl}})=\DC(-\Delta_{\vX}+H_\rad),
\end{equation}
again with equivalent graph norms and the same operator cores.

\noindent\textbf{Step 4: The localized eigenvalue equation.}
Since $\zeta=1$ near $\supp\eta$, the form of
$\widetilde H_{\mathrm{sl}}$ agrees with that of
$H_{\mathrm{sl}}^{v,\vA}$ on pairs $(\Phi,\eta\Psi)$ with
$\Phi\in\mathcal Q^{\mathrm{sl}}$.  The weak eigenvalue equation,
extended to full-space tests as in \cref{ucp-full-weak-equation}, is
tested with $\eta\Phi$.  The product rule
$\vpi_n(\eta\Psi)=\eta\vpi_n\Psi-\i(\grad_{\vx_n}\eta)\Psi$ then gives
\cref{ucp-localized-tilde-equation} in the form sense.  The
right-hand side of \cref{ucp-localized-tilde-equation} is in $L^2$,
since $\Psi$ is in the form domain.  The first representation
theorem for closed forms \cite[Theorem~VI.2.1]{kato1995perturbation}
puts $\eta\Psi$ in $\DC(\widetilde H_{\mathrm{sl}})$.
\end{proof}

Let $\Fock_{\mathrm{fin}}\subset\Fock_\ph$ be the span of $\Omega$
and of the vectors $a^*(h_1)\cdots a^*(h_m)\Omega$ with $m\geq1$
and $h_1,\ldots,h_m\in\Hil_\ph$ of bounded support, and let
$\mathcal C\subset L^2(\RR^{3N},\Fock_\ph)$ be the span of the functions
$\vX\mapsto\phi(\vX)\xi$ with $\phi\in C_c^\infty(\RR^{3N})$ and
$\xi\in\Fock_{\mathrm{fin}}$.

\begin{lemma}[Approximation with prescribed support]\label[lemma]{ucp-core-lemma}
Let $w$ be a compactly supported element of
$\DC(-\Delta_{\vX})\cap\DC(H_\rad)$, and let $\mathcal G\subset\RR^{3N}$ be an open neighborhood of $\supp w$.
Then there are $w_j\in\mathcal C$ with $\supp w_j\subset\mathcal G$ and
\begin{equation}\label{ucp-core-approximation}
 \|w_j-w\|+\|\Delta_{\vX}(w_j-w)\|+\|H_\rad(w_j-w)\|\to0
 \qquad(j\to\infty).
\end{equation}
The same holds for $w\in L^2(\RR^{3N},\mathcal K_N)$, with
$\mathcal C$ replaced by $\mathcal C\otimes(\CC^2)^{\otimes N}$.
\end{lemma}

\begin{proof}
Let $L=-\Delta_{\vX}+H_\rad+1$ as in
\cref{ucp-free-comparison-operator}.  Since
$-\Delta_{\vX}$ and $H_\rad$ are commuting nonnegative self-adjoint
operators, $\DC(L)=\DC(-\Delta_{\vX})\cap\DC(H_\rad)$, and the graph
norm of $L$ is equivalent to the norm in \cref{ucp-core-approximation}.
The set $\mathcal C$ is the algebraic tensor product of the core
$C_c^\infty(\RR^{3N})$ of $-\Delta_{\vX}$ and the core
$\Fock_{\mathrm{fin}}$ of $H_\rad$, and hence an operator core for
$L$.  Choose $\Phi_j\in\mathcal C$ with $\Phi_j\to w$ in the graph
norm of $L$, and let $\zeta\in C_c^\infty(\mathcal G)$ be equal to one
near $\supp w$.  Then $w_j=\zeta\Phi_j\in\mathcal C$,
$\supp w_j\subset\mathcal G$, and
\[
 Lw_j=\zeta L\Phi_j-2\grad_{\vX}\zeta\cdot\grad_{\vX}\Phi_j
      -(\Delta_{\vX}\zeta)\Phi_j.
\]
Since $\|\grad_{\vX}\Phi\|^2\leq\|\Phi\|\|L\Phi\|$, also
$\grad_{\vX}\Phi_j\to\grad_{\vX}w$, and therefore $w_j\to\zeta w=w$ in
the graph norm of $L$.  This proves \cref{ucp-core-approximation}.
The version with spin follows
by applying this argument to each spin component, since
$-\Delta_{\vX}$ and $H_\rad$ act componentwise on
$(\CC^2)^{\otimes N}$.
\end{proof}

\begin{proof}[Proof of \cref{ucp-main-theorem} in the spinless case]
Suppose first that $\Psi$ vanishes to infinite order at
$\vX_0\in\mathcal U$.  Without loss of generality, we can translate this
point to the origin.  Fix $\alpha>0$, and let
\[
 r=|\vX|,\qquad t=-\log r,\qquad
 f(t)=t+t^{-\alpha},\qquad
 \beta(t)=\ddot f(t)=\alpha(\alpha+1)t^{-\alpha-2}.
\]
Choose $R>0$ with $\ol{B_R}\subset\mathcal U$.  On $B_R$ the
electron--electron interaction is bounded, so $V-E$ is a sum of
one-body $L^p$ terms and a bounded term.  Since
$H_{\mathrm{sl}}^{v,\vA}-E=T^{\vA}+(V-E)$ with $T^{\vA}$ as in
\cref{ucp-carleman}, \cref{ucp-scalar-perturbation} applies with
$W=V-E$.  After decreasing $R$, it gives $\tau_0\geq1$ and $C<\infty$
such that, for $\tau\geq\tau_0$ and every $u$ in the local operator
domain of $H_{\mathrm{sl}}^{v,\vA}$ with compact support in
$B_R\setminus\{0\}$,
\begin{equation}\label{ucp-spinless-carleman}
 \tau^3\int_{B_R}\beta(t)\e^{2\tau f(t)}r^{-2}\|u(\vX)\|_{\Fock_\ph}^2\dd\vX
 \leq C\int_{B_R}\e^{2\tau f(t)}r^2
   \|(H_{\mathrm{sl}}^{v,\vA}-E)u(\vX)\|_{\Fock_\ph}^2\dd\vX.
\end{equation}
This is \cref{ucp-scalar-carleman} of \cref{ucp-scalar-perturbation}, keeping only
the first term on its left-hand side.  By \cref{ucp-caccioppoli}, the covariant derivatives of
$\Psi$ also vanish to infinite order at the origin, in the sense that
$\int_{B_s}\sum_n\|\vpi_n\Psi(\vX)\|_{\Fock_\ph}^2\dd\vX=\mathcal O(s^m)$ as
$s\downarrow0$ for every $m>0$.

Choose $\zeta\in C_c^\infty(B_R)$ with $\zeta=1$ on $B_{R/2}$,
and choose a radial cutoff $\eta_\epsilon$ equal to zero on
$B_\epsilon$ and one outside $B_{2\epsilon}$, with
$|\grad\eta_\epsilon|\leq C/\epsilon$ and
$|\Delta\eta_\epsilon|\leq C/\epsilon^2$.
For $0<\epsilon<R/4$, the localized vector
$u_\epsilon=\zeta\eta_\epsilon\Psi$ belongs to the local operator
domain by \cref{ucp-local-operator-domain}, and
\cref{ucp-localized-operator-equation} of \cref{ucp-local-operator-domain} with $\zeta\eta_\epsilon$ in
place of $\eta$ gives
\[
 (H_{\mathrm{sl}}^{v,\vA}-E)u_\epsilon
 =-\i\sum_{n=1}^N\grad_{\vx_n}(\zeta\eta_\epsilon)\cdot\vpi_n\Psi
  -\frac12\Delta_{\vX}(\zeta\eta_\epsilon)\Psi.
\]
Inserting $u=u_\epsilon$ in \cref{ucp-spinless-carleman} therefore gives
\begin{equation}\label{ucp-cutoff-carleman}
 \begin{split}
 &\tau^3\int_{B_R}\beta(t)\e^{2\tau f(t)}r^{-2}\|u_\epsilon(\vX)\|_{\Fock_\ph}^2\dd\vX\\
 &\quad\leq C\int_{B_R}\e^{2\tau f(t)}r^2
   \Bigl\|\sum_{n=1}^N\grad_{\vx_n}(\zeta\eta_\epsilon)\cdot\vpi_n\Psi(\vX)
     -\frac{\i}{2}\Delta_{\vX}(\zeta\eta_\epsilon)\Psi(\vX)\Bigr\|_{\Fock_\ph}^2\dd\vX.
 \end{split}
\end{equation}
The terms with derivatives of $\eta_\epsilon$ are supported in
$B_{2\epsilon}\setminus B_\epsilon$, where $f(t)\leq f(-\log\epsilon)$.
For fixed $\tau$, their contribution to the right-hand side of
\cref{ucp-cutoff-carleman} is at most
\[
 C \e^{2\tau f(-\log\epsilon)}
 \left(\epsilon^{-2}\int_{B_{2\epsilon}}\|\Psi(\vX)\|_{\Fock_\ph}^2\dd\vX
             +\int_{B_{2\epsilon}}\sum_{n=1}^N\|\vpi_n\Psi(\vX)\|_{\Fock_\ph}^2\dd\vX\right).
\]
Since $\e^{2\tau f(-\log\epsilon)}=\mathcal O_\tau(\epsilon^{-2\tau})$,
infinite-order vanishing makes this expression tend to zero.
The terms with derivatives of $\zeta$ are supported where
$R/2\leq r<R$ and do not depend on $\epsilon$.
Taking $\epsilon\downarrow0$ and using Fatou's lemma on the left
gives
\begin{equation}\label{ucp-outer-error}
 \tau^3\int_{B_{R/2}}\beta(t)\e^{2\tau f(t)}r^{-2}
                         \|\Psi(\vX)\|_{\Fock_\ph}^2\dd\vX
 \leq C_\Psi \e^{2\tau f(-\log(R/2))},
\end{equation}
where $C_\Psi$ is independent of $\tau$.
For $r<R/4$, the weight $f(-\log r)$ is at least
$f(-\log(R/4))>f(-\log(R/2))$, and
$\beta(-\log r)r^{-2}$ has a positive lower bound.
Letting $\tau\to\infty$ in \cref{ucp-outer-error} proves
$\Psi=0$ on $B_{R/4}$.

This local vanishing propagates throughout $\mathcal U$.
The set $\mathcal U$ is open, connected and of full measure.
Indeed, paths in $\RR^3$ can avoid any
finite set of points.  Moving the electron positions one at a time,
through distinct auxiliary positions if necessary, therefore connects
any two configurations in $\mathcal U$.  Its complement in
$\RR^{3N}$ consists of diagonals of codimension three.
To justify the propagation, let $\mathcal Z_\mathrm{open}$ be the set of points
having a neighborhood on which $\Psi=0$ almost everywhere.
It is nonempty and open.  On each sufficiently small ball compactly
contained in $\mathcal U$, the preceding argument gives a vanishing
radius that is uniform in the center.  If a point belongs to the
relative closure of $\mathcal Z_\mathrm{open}$, choose a point of
$\mathcal Z_\mathrm{open}$ closer than half this uniform radius.
Infinite-order vanishing at that interior point then supplies a zero
neighborhood of the closure point.  Thus $\mathcal Z_\mathrm{open}$ is also
relatively closed.  Connectedness of $\mathcal U$ implies
$\mathcal Z_\mathrm{open}=\mathcal U$.  Since its complement has measure zero,
$\Psi=0$ almost everywhere on the full configuration space $\RR^{3N}$.

Now suppose that the configuration-space zero set of a nonzero
$\Psi$ has positive measure.  It has a Lebesgue density point in
$\mathcal U$, because $\mathcal U$ has full measure.
Proposition~\ref{ucp-infinite-order} gives infinite-order vanishing
there, contradicting the conclusion just proved.  This establishes
the nonvanishing assertion of \cref{ucp-main-theorem} for the
spinless Hamiltonian.
If $\rho_\Psi$ vanished on a set $\Sigma\subset\RR^3$ of positive measure,
then by the definition of $\rho_\Psi$ and Tonelli's theorem, $\Psi$ would
vanish almost everywhere on $\Sigma\times\RR^{3(N-1)}$, a set of positive
measure, contradicting the nonvanishing just proved.  This gives the
density-positivity assertion.
\end{proof}

With spin, the local kinetic energy estimates take the following form.

\begin{lemma}[Local kinetic energy estimates with spin]\label[lemma]{ucp-caccioppoli-spin}
Assume \cref{cutoff-assumption,potential-spaces}, and let
$\Psi\in\mathcal Q_\PF$ be an eigenstate of $H^{v,\vA}$ with
eigenvalue $E$.  There is a constant $C$ such that, for every
$\vX_0\in\RR^{3N}$ and $0<r\leq1$, the following hold.
\begin{enumerate}[(i)]
\item The local kinetic and radiation energy is bounded by the $L^2$
mass on the doubled ball,
\begin{equation}\label{ucp-local-energy-spin}
 \int_{B_r(\vX_0)}\Bigl(\sum_{n=1}^N\|\vpi_n\Psi(\vX)\|_{\mathcal K_N}^2
   +\|H_\rad^{1/2}\Psi(\vX)\|_{\mathcal K_N}^2\Bigr)\dd\vX
 \leq\frac{C}{r^2}\int_{B_{2r}(\vX_0)}\|\Psi(\vX)\|_{\mathcal K_N}^2\dd\vX.
\end{equation}
\item For every $\eta$ as in \cref{ucp-caccioppoli}(ii),
\begin{equation}\label{ucp-local-gradient-spin}
 \|\grad_{\vX}(\eta\Psi)\|^2
 \leq\frac{C}{r^2}\int_{B_{2r}(\vX_0)}\|\Psi(\vX)\|_{\mathcal K_N}^2\dd\vX.
\end{equation}
\end{enumerate}
The constant $C$ depends on the Hamiltonian, $E$, and $C_1$, but not on
$r$, $\vX_0$, or $\eta$.
\end{lemma}

\begin{proof}
Simultaneous permutations of the position and spin variables act
unitarily on $\mathcal Q$ and leave $q^{v,\vA}$ invariant.
As in the proof of \cref{ucp-caccioppoli}, averaging over permutations
and $\Pi_-\Psi=\Psi$ show that $\Psi$ satisfies
\cref{weak-eigenvalue-equation} for every $\Phi\in\mathcal Q$,
not only for antisymmetric $\Phi$.  Let $\eta$ be as in
\cref{ucp-caccioppoli}(ii).  The product rule and
\[
 \|(\vsigma_n\cdot\grad_{\vx_n}\eta(\vX))\Psi(\vX)\|_{\mathcal K_N}^2
 =|\grad_{\vx_n}\eta(\vX)|^2\|\Psi(\vX)\|_{\mathcal K_N}^2
\]
give the IMS localization identity for the Pauli form
\[
 q^{v,\vA}[\eta\Psi]
 =E\|\eta\Psi\|^2
       +\frac12\int_{\RR^{3N}}|\grad_{\vX}\eta(\vX)|^2
                              \|\Psi(\vX)\|_{\mathcal K_N}^2\dd\vX.
\]
Applying \cref{pauli-form-coercivity} of \cref{prop-form-domain} to $\eta\Psi$ and using $r\leq1$
and $|\grad\eta|\leq C_1/r$, we obtain
\begin{equation*}
 \sum_{n=1}^N\|\vpi_n(\eta\Psi)\|^2
 +\|H_\rad^{1/2}\eta\Psi\|^2
 +\|\grad_{\vX}(\eta\Psi)\|^2
 \leq\frac{C}{r^2}\int_{B_{2r}(\vX_0)}\|\Psi(\vX)\|_{\mathcal K_N}^2\dd\vX.
\end{equation*}
Restricting the norms on the left to $B_r(\vX_0)$, as in the proof of
\cref{ucp-caccioppoli}, proves \cref{ucp-local-energy-spin}, and
dropping the first two terms on the left proves
\cref{ucp-local-gradient-spin}.
\end{proof}

With spin, we split off the Zeeman terms.  Set
\[
 S_{\mathrm{ext}}(\vX)=\frac12\sum_{n=1}^N\vsigma_n\cdot\vB(\vx_n),
 \qquad
 S_{\mathrm{ph}}(\vX)=-\frac e2\sum_{n=1}^N\vsigma_n\cdot\BF(\vx_n).
\]
Thus, by \cref{pauli-expansion},
$H^{v,\vA}=H_{\mathrm{sl}}^{v,\vA}+S_{\mathrm{ext}}+S_{\mathrm{ph}}$
in the sense of forms on compactly supported functions,
where the spinless operator on the right acts componentwise on spin.

The quantized magnetic field has coupling functions
\[
 K_\vx^\mu(\vk,\lambda)
 =-\frac{\i}{\sqrt2}(\vk\times\vepsilon_\lambda(\vk))^\mu
       \wh\chi(-\vk)\e^{-\i\vk\cdot\vx},
 \qquad
 \Bf^\mu(\vx)=a^*(K_\vx^\mu)+a(K_\vx^\mu).
\]
Put $M_{1/2}=\|\sqrt\omega\wh\chi\|_2$ and
$M_1=\|\omega\wh\chi\|_2$.  Cauchy--Schwarz gives
$M_{1/2}^2\leq\|\wh\chi\|_2M_1$.
The orthonormal transverse polarizations give, uniformly in $\vx$,
\[
 \sum_\mu\|K_\vx^\mu/\sqrt\omega\|_{\Hil_\ph}^2=M_{1/2}^2,
 \qquad
 \sum_\mu\|K_\vx^\mu\|_{\Hil_\ph}^2=M_1^2.
\]
The creation and annihilation bounds \cite[Section~3.3]{Q} therefore
yield
\begin{equation}\label{pauli-magnetic-field-bound}
 \|\BF(\vx)\xi\|^2
 \leq 8M_{1/2}^2\|H_\rad^{1/2}\xi\|^2
                      +2M_1^2\|\xi\|^2,
 \qquad \xi\in\DC(H_\rad^{1/2}).
\end{equation}
The same bound holds componentwise on the spin space.
In particular, $S_{\mathrm{ph}}$ is bounded from the radiation form
domain to the fiber, uniformly in $\vX$.  This uses only the first
ultraviolet moment \cref{ucp-extra-cutoff}.

The local operator domain of \cref{ucp-local-operator-domain} carries
over to the Hamiltonian with spin.

\begin{lemma}[Local operator domain with spin]\label[lemma]{ucp-local-operator-domain-spin}
Assume \cref{cutoff-assumption,potential-spaces,ucp-extra-cutoff}, let $\mathcal O$, $\mathcal O_1$, $p$, $v$ and
$\vA$ be as in \cref{ucp-local-operator-domain}, and suppose in
addition that $\vB\in L^p_\mathrm{loc}(\mathcal O_1,\RR^3)$.  If
$\Psi\in\mathcal Q_\PF$ is an eigenstate of $H^{v,\vA}$ with eigenvalue
$E$, then for every $\eta\in C_c^\infty(\mathcal O)$,
$\eta\Psi\in\DC(-\Delta_{\vX})\cap\DC(H_\rad)$ and
\begin{equation}\label{pauli-localized-equation}
 (H^{v,\vA}-E)(\eta\Psi)
 =-\i\sum_{n=1}^N(\grad_{\vx_n}\eta)\cdot\vpi_n\Psi
                              -\frac12(\Delta_{\vX}\eta)\Psi.
\end{equation}
Here $H^{v,\vA}u=H_{\mathrm{sl}}^{v,\vA}u+S_{\mathrm{ext}}u
+S_{\mathrm{ph}}u$ for $u\in\DC(-\Delta_{\vX})\cap\DC(H_\rad)$ with
compact support in $\mathcal O$, with $H_{\mathrm{sl}}^{v,\vA}u$ given
by \cref{ucp-local-expression} componentwise on the spin space, and
each term belongs to $L^2(\RR^{3N},\mathcal K_N)$.
\end{lemma}

\begin{proof}
Fix $\eta\in C_c^\infty(\mathcal O)$, and let $\mathcal V\supset\supp\eta$
be an open set with compact closure in $\mathcal O$, as in the proof of
\cref{ucp-local-operator-domain}.
On $\mathcal V$, take the local realization
$\widetilde H_{\mathrm{sl}}$ from \cref{ucp-local-operator-domain} and let it
act componentwise on the spin space.  Choose a real cutoff of $S_{\mathrm{ext}}$
that agrees with $S_{\mathrm{ext}}$ near $\supp\eta$, and denote it by
$\widetilde S_{\mathrm{ext}}$.  It can be
bounded in operator norm by a finite sum of one-body $L^p$
multipliers, since $\vB\in L^p_{\mathrm{loc}}$.
For $L=-\Delta_{\vX}+H_\rad+1$, the estimate
\cref{ucp-coordinate-multiplier} and the magnetic-field bound give
\[
 \|\widetilde S_{\mathrm{ext}}\Phi\|_2
       \leq C\|L^{3/(2p)}\Phi\|_2,
 \qquad
 \|S_{\mathrm{ph}}\Phi\|_2
       \leq C\|L^{1/2}\Phi\|_2.
\]
Both operators are infinitesimally bounded relative to
$\widetilde H_{\mathrm{sl}}$, whose graph norm is equivalent to that of $L$.
The Kato--Rellich theorem \cite[Theorem~V.4.3]{kato1995perturbation}
thus gives
\begin{equation}\label{pauli-local-domain}
 \DC(\widetilde H)
 =\DC(-\Delta_{\vX})\cap\DC(H_\rad),
 \qquad
 \widetilde H
 =\widetilde H_{\mathrm{sl}}+\widetilde S_{\mathrm{ext}}+S_{\mathrm{ph}},
\end{equation}
with equivalent graph norms and the operator core
$\mathcal C\otimes(\CC^2)^{\otimes N}$.  Both added terms
also have infinitesimal form bounds on $\mathcal Q$.
For $\widetilde S_{\mathrm{ext}}$, use the compactly supported
$L^p$ coefficients in the $L^{3/2}$ Sobolev form bound.
For $S_{\mathrm{ph}}$, use \cref{pauli-magnetic-field-bound}.
Thus this operator is associated with the corresponding closed form
sum.  The form identity \cref{pauli-expansion}
with the external field holds on this neighborhood.  It follows first on
$\mathcal C\otimes(\CC^2)^{\otimes N}$ by integration by parts and then on localized form vectors by
continuity, using \cref{potential-spaces}, the
Sobolev inequality, and $\vB\in L^p_{\mathrm{loc}}\subset
L^{3/2}_{\mathrm{loc}}$.
Thus $\widetilde H$ agrees locally with the form
realization of $H^{v,\vA}$.  The independently chosen
truncation of $S_{\mathrm{ext}}$ need not be the curl of the truncated
vector potential outside that neighborhood.

Both Zeeman terms commute with scalar electronic cutoffs.  Therefore
the localized weak equation is still
\[
 \widetilde H(\eta\Psi)
 =E\eta\Psi-\i\sum_{n=1}^N(\grad_{\vx_n}\eta)\cdot\vpi_n\Psi
                              -\frac12(\Delta_{\vX}\eta)\Psi.
\]
Its right-hand side belongs to $L^2$, since $\vpi_n\Psi\in L^2$ by
\cref{pauli-form-coercivity} of \cref{prop-form-domain}.
The first representation theorem
\cite[Theorem~VI.2.1]{kato1995perturbation} and \cref{pauli-local-domain}
give $\eta\Psi\in\DC(-\Delta_{\vX})\cap\DC(H_\rad)$.  For $u$ in this
domain supported where the cutoffs of $\widetilde H_{\mathrm{sl}}$ and
$\widetilde S_{\mathrm{ext}}$ equal one, $\widetilde H u$ is the
expression for $H^{v,\vA}u$ in the statement, by Step~1 of the proof
of \cref{ucp-local-operator-domain} and the choice of
$\widetilde S_{\mathrm{ext}}$.  Taking $u=\eta\Psi$ proves
\cref{pauli-localized-equation}.  The terms $S_{\mathrm{ext}}u$ and
$S_{\mathrm{ph}}u$ belong to $L^2$ by the bounds above.
\end{proof}

\begin{proof}[Proof of \cref{ucp-main-theorem}]
All norms in this proof use the spin--photon fiber
$\mathcal K_N=(\CC^2)^{\otimes N}\otimes\Fock_\ph$.
The Sobolev chain rule gives, for almost every $\vX\in\RR^{3N}$, the
pointwise bound
$|\grad_{\vX}(\eta(\vX)\|\Psi(\vX)\|_{\mathcal K_N})|\leq\|\grad_{\vX}(\eta(\vX)\Psi(\vX))\|_{\mathcal K_N}$, so
\cref{ucp-local-gradient-spin} of \cref{ucp-caccioppoli-spin} replaces \cref{ucp-local-gradient} of \cref{ucp-caccioppoli} in
the proof of \cref{ucp-infinite-order}.  Hence that proof, including
the density recurrence \cref{ucp-density-recurrence}, applies to the
total spinor norm, and $\Psi$ vanishes to infinite order at every
density point of its zero set.  Moreover, by
\cref{ucp-caccioppoli-spin}(i), infinite-order vanishing of $\Psi$ at
a point implies that
$\int_{B_s}\sum_n\|\vpi_n\Psi(\vX)\|_{\mathcal K_N}^2\dd\vX=\mathcal O(s^m)$
as $s\downarrow0$ for every $m>0$.

Now let $\Psi$ vanish to infinite order at a point of $\mathcal U$,
translated to the origin.  Choose $R>0$ with
$\ol{B_R}\subset\mathcal U$.  On $B_R$, the total scalar potential $V$
is the sum of the one-body terms $v(\vx_n)$ and the bounded
electron--electron interaction.  Let
$H^{V,\vA}=T^{\vA}+V+S_{\mathrm{ext}}+S_{\mathrm{ph}}$.  We apply
\cref{ucp-pauli-carleman} with $\vX_0=0$, $v_n=v$, $V_b$ the
electron--electron interaction and $E$ the eigenvalue, and decrease
$R$ so that $R\leq r_0$.  Keeping only the first term on the left of
\cref{pauli-carleman} of \cref{ucp-pauli-carleman}, we obtain $\tau_0\geq1$ and $C<\infty$ such
that, for $\tau\geq\tau_0$ and every
$u\in\DC(-\Delta_{\vX})\cap\DC(H_\rad)$ with compact support in
$B_R\setminus\{0\}$,
\[
 \tau^3\int_{B_R}\beta(t)\e^{2\tau f(t)}r^{-2}
        \|u(\vX)\|_{\mathcal K_N}^2\dd\vX
 \leq C\int_{B_R}\e^{2\tau f(t)}r^2
   \|(H^{V,\vA}-E)u(\vX)\|_{\mathcal K_N}^2\dd\vX.
\]
Let $\zeta$, $\eta_\epsilon$ and $u_\epsilon=\zeta\eta_\epsilon\Psi$ be
as in the spinless case.  By \cref{ucp-local-expression} of \cref{ucp-local-operator-domain},
$H^{v,\vA}u=H^{V,\vA}u$ for $u\in\DC(-\Delta_{\vX})\cap\DC(H_\rad)$
with compact support in $B_R$.  Thus
\cref{ucp-local-operator-domain-spin}, applied with $\mathcal O=B_R$
and $\eta=\zeta\eta_\epsilon$, gives
$u_\epsilon\in\DC(-\Delta_{\vX})\cap\DC(H_\rad)$ and
\[
 (H^{V,\vA}-E)u_\epsilon
 =-\i\sum_{n=1}^N\grad_{\vx_n}(\zeta\eta_\epsilon)\cdot\vpi_n\Psi
  -\frac12\Delta_{\vX}(\zeta\eta_\epsilon)\Psi.
\]
Inserting $u=u_\epsilon$ in the Carleman estimate above gives, for
$0<\epsilon<R/4$,
\[
 \begin{split}
 &\tau^3\int_{B_R}\beta(t)\e^{2\tau f(t)}r^{-2}
        \|u_\epsilon(\vX)\|_{\mathcal K_N}^2\dd\vX\\
 &\quad\leq C\int_{B_R}\e^{2\tau f(t)}r^2
   \Bigl\|\sum_{n=1}^N\grad_{\vx_n}(\zeta\eta_\epsilon)\cdot\vpi_n\Psi(\vX)
     -\frac{\i}{2}\Delta_{\vX}(\zeta\eta_\epsilon)\Psi(\vX)\Bigr\|_{\mathcal K_N}^2\dd\vX,
 \end{split}
\]
which is \cref{ucp-cutoff-carleman} with norms in $\mathcal K_N$.  As
in the spinless case, the infinite-order local energy bounds make the
inner-cutoff error vanish as $\epsilon\downarrow0$, and Fatou's lemma
on the left gives
\[
 \tau^3\int_{B_{R/2}}\beta(t)\e^{2\tau f(t)}r^{-2}
                         \|\Psi(\vX)\|_{\mathcal K_N}^2\dd\vX
 \leq C_\Psi \e^{2\tau f(-\log(R/2))},
\]
where $C_\Psi$ is independent of $\tau$.  Letting $\tau\to\infty$
proves $\Psi=0$ on $B_{R/4}$.  This vanishing propagates
through the connected collision-free configuration space $\mathcal U$.
If the zero set of a nonzero spinor had positive measure, the
density-point argument would produce infinite-order vanishing at
a point of $\mathcal U$, a contradiction.  Positivity almost everywhere
of the spin-summed density follows as in the spinless case: if
$\rho_\Psi$ vanished on a set $\Sigma\subset\RR^3$ of positive measure, then by
Tonelli's theorem $\Psi$ would vanish almost everywhere on
$\Sigma\times\RR^{3(N-1)}$.
\end{proof}

\begin{remark}[Unique continuation off an exceptional set]
\label{rem-ucp-domain-extension}
The local integrability assumptions of \cref{ucp-main-theorem} may
instead be imposed on a connected open set $\mathcal O\subset\RR^3$
with $|\RR^3\setminus\mathcal O|=0$.  The same conclusions hold with
$\mathcal U$ replaced by $\mathcal U\cap\mathcal O^N$ and the
divergence and curl understood in the sense of distributions on $\mathcal O$.
For the spinless version the curl condition is omitted.
This configuration space is still connected, since paths in a
connected open subset of $\RR^3$ can avoid finitely many points,
and it has full measure.  The local regularity of $v$ and $\vA$ enters
the proof only through
\cref{ucp-local-operator-domain,ucp-local-operator-domain-spin} and the
Carleman estimates of \cref{ucp-carleman}.  Their constants depend only
on local norms of the potentials and are uniform on compact subsets of
$\mathcal U\cap\mathcal O^N$.  Hence the proof applies unchanged with
$\mathcal U$ replaced by $\mathcal U\cap\mathcal O^N$.
\end{remark}

\subsection{Proof of \texorpdfstring{\cref{thm-hk-nonvanishing}}{Theorem \getrefnumber{thm-hk-nonvanishing}}}

First consider the case $\theta=0$, in which
$\rho_{\Psi_1}=\rho_{\Psi_2}=\rho$ and $\vjc_{\Psi_1}=\vjc_{\Psi_2}=\vjc$.  The
general case is reduced to this one below by a gauge transformation.  By \cref{prop-common-ground-state},
$\Psi_1$ is a common ground state.
We apply \cref{ucp-main-theorem} to the potential pair satisfying
\cref{ucp-potential-regularity}.  It gives
\begin{equation}\label{ucp-joint-nonvanishing}
 \|\Psi_1(\vX)\|_{\mathcal K_N}>0
 \quad\text{for a.a. }\vX\in\RR^{3N},\qquad
 \rho(\vx)>0\quad\text{for a.a. }\vx\in\RR^3.
\end{equation}
\Cref{prop-vector-uniqueness} now gives $\vA_1=\vA_2$.
Subtracting the weak eigenvalue equations cancels the kinetic,
interaction, and radiation forms, leaving
\begin{equation}\label{scalar-multiplier-equation}
 M(\vX)\Psi_1(\vX)=0,\qquad
 M(\vX)=\sum_{n=1}^N\big(v_1(\vx_n)-v_2(\vx_n)\big)-(E_1-E_2).
\end{equation}
The equation $M\Psi_1=0$ holds in the sense of distributions.  The weak eigenvalue equations only
provide antisymmetric test functions $\Phi\in\mathcal Q_\PF$.  Since the
multiplier $M$ is symmetric under permutations, $M\Psi_1$ is antisymmetric, so
$\dua{\Phi}{M\Psi_1}=\dua{\Pi_-\Phi}{M\Psi_1}$ for every test function
$\Phi$.  Hence \cref{scalar-multiplier-equation} holds against all test
functions.  The products
$\big(v_1(\vx_n)-v_2(\vx_n)\big)\Psi_1(\vX)$ are locally Bochner integrable.  Indeed, on each
bounded configuration-space set $K$, Cauchy--Schwarz gives
\[
 \begin{aligned}
 &\int_K |v_1(\vx_n)-v_2(\vx_n)|\|\Psi_1(\vX)\|_{\mathcal K_N}\dd\vX\\
 &\quad\leq\left(\int_K|v_1(\vx_n)-v_2(\vx_n)|\|\Psi_1(\vX)\|_{\mathcal K_N}^2\dd\vX\right)^{1/2}\\
 &\qquad\times\left(\int_K|v_1(\vx_n)-v_2(\vx_n)|\dd\vX\right)^{1/2}<\infty,
 \end{aligned}
\]
by scalar form boundedness and local integrability of $v_1-v_2$.
Consequently the multiplier equation holds almost everywhere.
Using \cref{ucp-joint-nonvanishing}, it reduces to
\[
 \sum_{n=1}^N \big(v_1(\vx_n)-v_2(\vx_n)\big)=E_1-E_2
 \qquad\text{for a.a. }\vX=(\vx_1,\ldots,\vx_N)\in\RR^{3N}.
\]
Write $f=v_1-v_2\in L^1_\mathrm{loc}(\RR^3)$ and fix a ball $B\subset\RR^3$.
Averaging the identity over $(\vx_2,\ldots,\vx_N)\in B^{N-1}$ and using
Fubini's theorem gives
\[
 f(\vx_1)=E_1-E_2-\frac{N-1}{|B|}\int_B f(\vy)\dd\vy
 \qquad\text{for a.a. }\vx_1\in\RR^3.
\]
Thus $f=c$ almost everywhere for a constant $c$, and substituting this
into the identity gives $Nc=E_1-E_2$.

Now let $\theta$ be any function satisfying the hypotheses of the
theorem, so that
\[
 \rho_{\Psi_1}=\rho_{\Psi_2}=\rho,\qquad
 \vjc_{\Psi_2}=\vjc_{\Psi_1}+\rho\grad\theta.
\]
As in the proof of \cref{prop-cutoff-kernel}, $U_\theta$ denotes
multiplication by $\e^{\i\sum_n\theta(\vx_n)}$.
The weak chain rule and the multiplier bounds for $L^3+L^\infty$
show that $U_\theta$ maps the fermionic form domain onto itself.
The covariant derivatives satisfy
\[
 \bigl(\vD_n+\vA(\vx_n)-\grad\theta(\vx_n)\bigr)U_\theta\Psi
   =U_\theta\bigl(\vD_n+\vA(\vx_n)\bigr)\Psi.
\]
The phase also commutes with the Pauli matrices, so this identity
holds after contraction with $\vsigma_n$.
Since $U_\theta$ commutes with the scalar, interaction, and radiation
terms, the closed forms give
\[
 H^{v,\vA-\grad\theta}
    =U_\theta H^{v,\vA}U_\theta^{-1}.
\]
Thus gauge changes preserve ground-state energies and transform the
data by $\rho_{U_\theta\Psi}=\rho_\Psi$ and
$\vjc_{U_\theta\Psi}=\vjc_\Psi+\rho_\Psi\grad\theta$.
The magnetization density is unchanged by the phase.

We now reduce to the case $\theta=0$ by gauge transforming one of the
two pairs.  The pair satisfying \cref{ucp-potential-regularity} is left
unchanged, since that condition is not gauge invariant:
$\div(\vA_k\pm\grad\theta)=\div\vA_k\pm\Delta\theta$ need not lie in
$L^p_\mathrm{loc}$.

Suppose first that the first potential pair satisfies
\cref{ucp-potential-regularity}.  We apply the case $\theta=0$ to the
pairs $(v_1,\vA_1)$ and $(v_2,\vA_2+\grad\theta)$, with the states
$\Psi_1$ and $U_{-\theta}\Psi_2$.  Its hypotheses hold for these data.
First, since $\grad\theta\in L^3+L^\infty$, the pair
$(v_2,\vA_2+\grad\theta)$ satisfies \cref{potential-spaces}.  Second, the
conjugation identity above with $-\theta$ in place of $\theta$ gives
$H^{v_2,\vA_2+\grad\theta}=U_{-\theta}H^{v_2,\vA_2}U_{-\theta}^{-1}$.  Hence
$U_{-\theta}\Psi_2$ is a normalized ground state of
$H^{v_2,\vA_2+\grad\theta}$, and its energy is
$E(v_2,\vA_2+\grad\theta)=E_2$.  Third, by the transformation rule above,
\[
 \rho_{U_{-\theta}\Psi_2}=\rho=\rho_{\Psi_1},\qquad
 \vjc_{U_{-\theta}\Psi_2}=\vjc_{\Psi_2}-\rho\grad\theta=\vjc_{\Psi_1},
\]
so the two states have the same density and combined current density.
Fourth, the first pair is unchanged and still satisfies
\cref{ucp-potential-regularity}.  Finally, $e$, $\wh\chi$, and the common
density $\rho$ are unchanged, so whichever cutoff alternative holds for
the original data also holds for the new data.  The case $\theta=0$
therefore gives $\vA_1=\vA_2+\grad\theta$ and $v_1-v_2=(E_1-E_2)/N$
almost everywhere.

If instead the second pair supplies the local regularity, we apply the
case $\theta=0$ to the pairs $(v_1,\vA_1-\grad\theta)$ and
$(v_2,\vA_2)$, with the states $U_\theta\Psi_1$ and $\Psi_2$.  The same
checks apply.  The pair $(v_1,\vA_1-\grad\theta)$ satisfies
\cref{potential-spaces}, and $U_\theta\Psi_1$ is a normalized ground state
of $H^{v_1,\vA_1-\grad\theta}=U_\theta H^{v_1,\vA_1}U_\theta^{-1}$ with
energy $E_1$.  Moreover,
\[
 \rho_{U_\theta\Psi_1}=\rho=\rho_{\Psi_2},\qquad
 \vjc_{U_\theta\Psi_1}=\vjc_{\Psi_1}+\rho\grad\theta=\vjc_{\Psi_2}.
\]
The unchanged second pair satisfies \cref{ucp-potential-regularity},
and the cutoff alternative is preserved.  The case $\theta=0$ gives
$\vA_1-\grad\theta=\vA_2$ and the same scalar constant.  This proves
\cref{hk-conclusion}.

\appendix
\section{Covariant Pauli--Fierz Carleman estimates}\label[appendix]{ucp-carleman}

Define the auxiliary spinless operator on
$L^2(\RR^{3N},\Fock_\ph)$ by
\[
 T^{\vA}=\frac12\sum_{n=1}^N\vpi_n^2+H_\rad
 =\frac12\sum_{n=1}^N\big(\vD_n+\vA(\vx_n)\big)^2+H_\rad.
\]
We first prove a Carleman estimate for $T^0$, keeping the covariant
derivatives and the radiation energy in the principal operator, so
that the left-hand side controls $\vD_nu$ rather than $\grad u$.  This
covariant form is essential for two reasons.  First, $\AF(\vx)$ is an
unbounded operator on $\Fock_\ph$, so the first-order term
$e\AF(\vx_n)\cdot\grad_{\vx_n}$ cannot be treated as a perturbation of
a Carleman estimate for the Laplacian. It is controlled only through
$H_\rad^{1/2}$.  Second, in a local radial gauge the commutators in the
Carleman computation involve the quantized field only through its
curvature, the quantized magnetic field.  This is controlled by the
first ultraviolet moment \cref{ucp-extra-cutoff} and is small near the
center of the weight.  Fractional Carleman estimates in the mixed
electronic and radiation norms of $L=-\Delta_{\vX}+H_\rad+1$ then
permit the external vector potential in $L^6_\mathrm{loc}$ as a
first-order perturbation.

For an electronic configuration $\vX_0$ and $\alpha>0$, we use the
weights
\[
 r=|\vX-\vX_0|,\qquad t=-\log r,\qquad
 f(t)=t+t^{-\alpha},\qquad
 \beta(t)=\ddot f(t)=\alpha(\alpha+1)t^{-\alpha-2}.
\]
For $0<r_0<1$, $\tau>0$ and $u\in\DC(T^0)$ with compact support in
$B_{r_0}(\vX_0)\setminus\{\vX_0\}$, put
\[
 \mathcal T^0(u)=\int_{B_{r_0}(\vX_0)}\e^{2\tau f(t)}r^2
                    \|T^0u(\vX)\|_{\Fock_\ph}^2\dd\vX.
\]
We suppress the dependence of $\mathcal T^0(u)$ on $\vX_0$, $\alpha$,
$r_0$ and $\tau$.

\begin{lemma}[A Carleman estimate with $\vA=0$]
\label[lemma]{lem-pf-free-carleman}
Assume \cref{cutoff-assumption} and
$\omega\wh\chi\in L^2(\RR^3)$.  Fix an electronic configuration
$\vX_0$ and $\alpha>0$, with $r$, $t$, $f$ and $\beta$ as above.
There are $r_0\in(0,\e^{-2})$, $\tau_0\geq1$ and $C<\infty$
such that, for $\tau\geq\tau_0$ and every $u\in\DC(T^0)$ with
compact support in $B_{r_0}(\vX_0)\setminus\{\vX_0\}$,
\begin{align}
 &\tau^3\int_{B_{r_0}(\vX_0)}\beta(t)\e^{2\tau f(t)}r^{-2}
                        \|u(\vX)\|_{\Fock_\ph}^2\dd\vX
   \nonumber\\
 &\quad+\tau\int_{B_{r_0}(\vX_0)}\beta(t)\e^{2\tau f(t)}
                        \sum_{n=1}^N\|\vD_nu(\vX)\|_{\Fock_\ph}^2\dd\vX
   \nonumber\\
 &\quad+\tau\int_{B_{r_0}(\vX_0)}\e^{2\tau f(t)}
                       \|H_\rad^{1/2}u(\vX)\|_{\Fock_\ph}^2\dd\vX
  \leq C\mathcal T^0(u).
 \label{pf-free-carleman}
\end{align}
The constants are independent of $\vX_0$, and the same estimate holds
on every smaller ball with the same $C$ and $\tau_0$.
\end{lemma}

\begin{proof}
The proof has three steps.  We first prove \cref{pf-free-carleman} by
a computation in a radial gauge, assuming that $\wh\chi$ has compact
support and that $u\in\mathcal C$, with $\mathcal C$ as defined before
\cref{ucp-core-lemma}.  The resulting constants depend only on the first ultraviolet
moment.  We then remove the compact support of $\wh\chi$ by an
ultraviolet limit, and finally extend the estimate to all of
$\DC(T^0)$ by an approximation argument.

\noindent\textbf{Step 1: Compactly supported cutoff.}
Without loss of generality, we translate $\vX_0$ to the origin.
Translating the $n$th electron by $(\vX_0)_n$ multiplies
$G^\mu_{\vx_n}$ by the phase $\e^{-\i\vk\cdot(\vX_0)_n}$, which leaves
all properties used below unchanged, and we suppress it from the
notation.  We write $\vx_n=(x_n^1,x_n^2,x_n^3)$ and
$\varPhi(h)=a^*(h)+a(h)$, so that $\Af^\mu(\vx)=\varPhi(G^\mu_\vx)$.
The cutoff assumptions give uniform bounds for every spatial
derivative of $G^\mu_\vx$ in
\[
 \Hil_\omega
 =\{h\in\Hil_\ph:h/\sqrt\omega\in\Hil_\ph\},
 \qquad
 \|h\|_\omega^2=\|h\|^2+\|h/\sqrt\omega\|^2,
\]
and the same holds after multiplying these derivatives by $\omega$.
The reality assumption and the real, even transverse polarization
projector imply
\[
 \Im\dua{G^\mu_\vx}{G^\nu_\vy}=0
 \qquad(\vx,\vy\in\RR^3).
\]
Consequently all the fields $\Af^\mu(\vx)$ strongly commute
\cite[Proposition~A.6(iii)]{Q}.
Their spatial derivatives have the same commutation property.

Define the radial line integral and its unitary exponential by
\begin{align*}
 \Theta(\vX)&=-e\sum_{n=1}^N\int_0^1\vx_n\cdot\AF(s\vx_n)\dd s
            =\varPhi(h_\vX),
 \qquad U_\vX=\e^{\i\Theta(\vX)},\\
 h_\vX&=-e\sum_{n=1}^N\sum_{\mu=1}^3\int_0^1x_n^\mu G^\mu_{s\vx_n}\dd s.
\end{align*}
These formulas may first be read on vectors with finitely many photons.
Strong commutativity, the field bounds and the Weyl relations then
give their closed-operator and form interpretations.
Write $U$ for the unitary $(Uu)(\vX)=U_\vX u(\vX)$ on
$L^2(\RR^{3N},\Fock_\ph)$.  Since $U_\vX$ commutes with all the fields,
conjugation gives
\[
 U\vD_nU^*=-\i\grad_{\vx_n}+\vbf_n,
\]
where
\[
 \vbf_n(\vX)=-e\AF(\vx_n)-\grad_{\vx_n}\Theta(\vX).
\]
The operator-valued connection $\vbf=(\vbf_1,\ldots,\vbf_N)$ is
$-e\AF$ in radial gauge, that is, it has no radial component:
$\vx_n\cdot\vbf_n(\vX)=0$.  This follows from the classical
radial-gauge identity: for a smooth vector field $\vA$ on $\RR^3$,
differentiation under the integral and the fundamental theorem of
calculus in $s$ give
\[
 \vA(\vx)-\grad_\vx\int_0^1\vx\cdot\vA(s\vx)\dd s
 =-\int_0^1s\,\vx\times(\grad\times\vA)(s\vx)\dd s.
\]
The identity is real-linear in $\vA$, so it also holds for the
$\Hil_\ph$-valued field $\vx\mapsto(G^1_\vx,G^2_\vx,G^3_\vx)$.
Applying $\varPhi$, with $\BF=\grad\times\AF$, gives
\[
 \vbf_n(\vX)=e\int_0^1s\,\vx_n\times\BF(s\vx_n)\dd s.
\]

Let $\vX=r\vOmega$ with $\vOmega=(\vOmega_1,\ldots,\vOmega_N)\in\mathbb S^{3N-1}$, and recall
$t=-\log r$, so that $r=\e^{-t}$ and $\partial_t=-r\partial_r$.  Put
$t_0=-\log r_0>2$.  In $(t,\vOmega)$-coordinates,
$B_{r_0}\setminus\{0\}$ is the cylinder
$(t_0,\infty)\times\mathbb S^{3N-1}$.  For each $t>t_0$,
define the angular connection, an operator-valued one-form on the unit
sphere $\mathbb S^{3N-1}$ acting on $\Fock_\ph$-valued functions, by
\[
 \mathfrak a(t,\vOmega)[\vzeta]=r\,\vzeta\cdot\vbf(r\vOmega),
 \qquad \vzeta=(\vzeta_1,\ldots,\vzeta_N)\in T_\vOmega\mathbb S^{3N-1}.
\]
Since $\vX\cdot\vbf(\vX)=0$, the vector $\vbf(r\vOmega)$ is tangent to the
sphere, so $\mathfrak a$ determines $\vbf$ completely.
Differentiating the radial integral in $t$, with $\vOmega$ fixed, gives
\begin{equation}\label{pf-angular-derivative}
 \dot{\mathfrak a}(t,\vOmega)[\vzeta]
 =er^2\sum_{n=1}^N(\vOmega_n\times\vzeta_n)\cdot\BF(r\vOmega_n).
\end{equation}
Let $\nabla_{\mathbb S}$ denote the spherical gradient on
$\mathbb S^{3N-1}$, acting on $\Fock_\ph$-valued functions.  Set
\[
 \nabla_{\mathbb S,\mathfrak a}
       =\nabla_{\mathbb S}+\i\mathfrak a,
 \qquad K(t)=\nabla_{\mathbb S,\mathfrak a}^*
                         \nabla_{\mathbb S,\mathfrak a},
 \qquad
 \Upsilon(t,\vOmega)=2U_{r\vOmega}H_\rad U_{r\vOmega}^*.
\]
Since $\vX\cdot\vbf(\vX)=0$, the gauge-transformed covariant Laplacian has
no connection in the radial direction, so
\[
 (-\i\grad_{\vX}+\vbf)^2
 =-\partial_r^2-\frac{3N-1}{r}\partial_r+\frac{1}{r^2}K(t).
\]
Since $UT^0U^*=\frac12(-\i\grad_\vX+\vbf)^2+UH_\rad U^*$, multiplying
by $2r^2$ and using $\partial_t=-r\partial_r$ gives
\begin{equation}\label{pf-polar-form}
 2r^2\,UT^0U^*=-\partial_t^2+2\gamma\,\partial_t+K(t)+r^2\Upsilon(t),
 \qquad \gamma=\frac{3N-2}{2}.
\end{equation}
Here $\Upsilon(t)=\Upsilon(t,\cdot)$ is a nonnegative multiplication operator on the angular
Hilbert space $L^2(\mathbb S^{3N-1},\Fock_\ph)$, with the surface
measure $\dd\vOmega$.  For $\Xi$ in this space we write
\[
 \Upsilon(t)[\Xi]=2\int_{\mathbb S^{3N-1}}
   \|H_\rad^{1/2}U_{r\vOmega}^*\Xi(\vOmega)\|_{\Fock_\ph}^2\dd\vOmega,
 \qquad
 K(t)[\Xi]=\int_{\mathbb S^{3N-1}}
   \|\nabla_{\mathbb S,\mathfrak a}\Xi(\vOmega)\|^2\dd\vOmega
\]
for the corresponding quadratic forms, where the integrand of $K(t)[\Xi]$
is the norm of a tangent-vector-valued function with values in
$\Fock_\ph$.  By the Weyl relations and $\sqrt\omega\wh\chi\in L^2$,
$U_\vX$ preserves $\DC(H_\rad^{1/2})$, and by the field bound
$\mathfrak a$ is bounded by $(H_\rad+1)^{1/2}$.  Hence the form domain
of $K(t)+\Upsilon(t)$ is
\[
 \mathcal Q_{\mathbb S}
 =H^1(\mathbb S^{3N-1},\Fock_\ph)\cap\DC(H_\rad^{1/2}),
\]
with $H_\rad$ acting pointwise on the sphere, independently of $t>t_0$.

We write $\dot K(t)[\Xi]$ for the derivative of $t\mapsto K(t)[\Xi]$ with $\Xi$
fixed.

\begin{claim}
There is $C<\infty$, independent of $t$ and $\Xi$, such that for
$t>t_0$ and $\Xi\in\mathcal Q_{\mathbb S}$,
\begin{equation}\label{pf-angular-form-derivative}
 |\dot K(t)[\Xi]|
 \leq Cr^2K(t)[\Xi]^{1/2}
                     \big(\Upsilon(t)[\Xi]+\|\Xi\|^2\big)^{1/2}.
\end{equation}
\end{claim}

\begin{proof}[Proof of the claim]
In
\[
 K(t)[\Xi]=\int_{\mathbb S^{3N-1}}
   \|\nabla_{\mathbb S}\Xi(\vOmega)+\i\mathfrak a(t,\vOmega)\Xi(\vOmega)\|^2\dd\vOmega
\]
the term $\nabla_{\mathbb S}\Xi$ does not depend on $t$, so only the
connection is differentiated, and
\[
 \dot K(t)[\Xi]=2\Re\dua{\nabla_{\mathbb S,\mathfrak a}\Xi}
 {\i\dot{\mathfrak a}\Xi}.
\]
By the Cauchy--Schwarz inequality,
$|\dot K(t)[\Xi]|\leq2K(t)[\Xi]^{1/2}\|\dot{\mathfrak a}\Xi\|$, so it
suffices to show
\begin{equation}\label{pf-angular-connection-derivative}
 \|\dot{\mathfrak a}\Xi\|
 \leq Cr^2\big(\Upsilon(t)[\Xi]+\|\Xi\|^2\big)^{1/2}.
\end{equation}
Let $\vzeta\in T_\vOmega\mathbb S^{3N-1}$ and $\xi\in\Fock_\ph$.  By \cref{pf-angular-derivative} and
$\sum_{n=1}^N|\vOmega_n||\vzeta_n|\leq|\vOmega||\vzeta|=|\vzeta|$,
\[
 \|\dot{\mathfrak a}(t,\vOmega)[\vzeta]\xi\|
 \leq er^2|\vzeta|\max_n\|\BF(r\vOmega_n)\xi\|,
\]
where $\|\BF(\vx)\xi\|^2=\sum_\lambda\|\Bf^\lambda(\vx)\xi\|^2$.  The
form factors of $\BF(\vx)$ are first derivatives of $G^\mu_\vx$, which
are uniformly bounded in $\Hil_\omega$, so the field bound gives
$\|\BF(\vx)\eta\|\leq C(\|H_\rad^{1/2}\eta\|^2+\|\eta\|^2)^{1/2}$
uniformly in $\vx$.  Since $\BF(\vx)$ commutes with $U_{r\vOmega}$,
applying this to $\eta=U_{r\vOmega}^*\xi$ gives
\[
 \|\BF(r\vOmega_n)\xi\|=\|\BF(r\vOmega_n)U_{r\vOmega}^*\xi\|
 \leq C\big(\|H_\rad^{1/2}U_{r\vOmega}^*\xi\|^2+\|\xi\|^2\big)^{1/2}.
\]
Taking $\xi=\Xi(\vOmega)$, summing over an orthonormal basis of
$T_\vOmega\mathbb S^{3N-1}$ and integrating over the sphere gives
\cref{pf-angular-connection-derivative}.
\end{proof}

Similarly, we write $\dot\Upsilon(t)[\Xi]$ for the derivative of
$t\mapsto\Upsilon(t)[\Xi]$ with $\Xi$ fixed.

\begin{claim}
There is $C<\infty$, independent of $t$ and $\Xi$, such that for
$t>t_0$ and $\Xi\in\mathcal Q_{\mathbb S}$,
\begin{equation}\label{pf-radiation-derivative}
 |\dot\Upsilon(t)[\Xi]|\leq Cr\,\Upsilon(t)[\Xi]^{1/2}\|\Xi\|.
\end{equation}
\end{claim}

\begin{proof}[Proof of the claim]
Write $\varPi(h)=\i(a^*(h)-a(h))$, so that
$[\varPhi(h),H_\rad]=\i\varPi(\omega h)$.  Since
$U_{r\vOmega}=\e^{\i\varPhi(h_{r\vOmega})}$ and $\varPhi(h_{r\vOmega})$ commutes with
$\varPhi(\dot h_{r\vOmega})$,
\[
 \dot\Upsilon(t,\vOmega)
 =2\i\,U_{r\vOmega}\big[\varPhi(\dot h_{r\vOmega}),H_\rad\big]U_{r\vOmega}^*
 =-2U_{r\vOmega}\varPi(\omega\,\dot h_{r\vOmega})U_{r\vOmega}^*,
\]
where
\[
 \dot h_{r\vOmega}
       =er\sum_{n=1}^N\sum_{\mu=1}^3\Omega_n^\mu G^\mu_{r\vOmega_n}.
\]
Since $\sqrt\omega\wh\chi\in L^2$, we have
$\|\sqrt\omega\,\dot h_{r\vOmega}\|\leq Cr$.  With
$\xi=U_{r\vOmega}^*\Xi(\vOmega)$, the bound
$|\dua{\xi}{\varPi(g)\xi}|\leq2\|g/\sqrt\omega\|\|H_\rad^{1/2}\xi\|\|\xi\|$
gives
\[
 |\dua{\Xi(\vOmega)}{\dot\Upsilon(t,\vOmega)\Xi(\vOmega)}|
 \leq Cr\|H_\rad^{1/2}U_{r\vOmega}^*\Xi(\vOmega)\|\|\Xi(\vOmega)\|.
\]
Integrating over the sphere and using the Cauchy--Schwarz inequality
gives \cref{pf-radiation-derivative}.
\end{proof}

Thus $\dot\Upsilon$ is controlled by the transformed radiation energy
itself, so $H_\rad$ can be kept in the principal part of the operator
$Q$ in \cref{pf-cylinder-operator} below.

Conjugating \cref{pf-polar-form} by $r^\gamma=\e^{-\gamma t}$ removes the
first-order term:
\begin{equation}\label{pf-cylinder-operator}
 2r^{\gamma+2}UT^0u=Q(r^\gamma Uu),
 \qquad
 Q=-\partial_t^2+K(t)+\gamma^2+r^2\Upsilon(t).
\end{equation}
Set
\[
 z(t,\vOmega)=\e^{\tau f(t)}r^\gamma U_{r\vOmega}u(r\vOmega),\qquad
 z(t)=z(t,\cdot)\in L^2(\mathbb S^{3N-1},\Fock_\ph).
\]
Then $z$ has compact support in the cylinder
$(t_0,\infty)\times\mathbb S^{3N-1}$.  We write $\|z(t)\|$ for the norm in $L^2(\mathbb S^{3N-1},\Fock_\ph)$ and
$\|z\|$ for the norm in $L^2((t_0,\infty)\times\mathbb S^{3N-1},\Fock_\ph)$
with measure $\dd t\dd\vOmega$, so that
$\|z\|^2=\int_{t_0}^\infty\|z(t)\|^2\dd t$.  The operators $K(t)$, $\gamma^2$
and $r^2\Upsilon(t)$ act at fixed $t$ and commute with multiplication by
$\e^{\tau f(t)}$, while
\[
 \e^{\tau f}(-\partial_t^2)\e^{-\tau f}=-(\partial_t-\tau\dot f)^2
 =-\partial_t^2-\tau^2(\dot f)^2+2\tau\dot f\partial_t+\tau\ddot f.
\]
Hence
\[
 \e^{\tau f}Q\e^{-\tau f}=S+B,
 \qquad
 S=Q-\tau^2(\dot f)^2,
 \qquad B=2\tau\dot f\partial_t+\tau\ddot f.
\]
By \cref{pf-cylinder-operator}, $(S+B)z=2\e^{\tau f}r^{\gamma+2}UT^0u$.
Since $\wh\chi$ has compact support, all moments $\omega^k\wh\chi$ are
finite, and the Weyl relations show that $z$, $Sz$ and $Bz$ are smooth,
compactly supported in $(t_0,\infty)\times\mathbb S^{3N-1}$, and take
values in $\DC(H_\rad^k)$ for all $k$, together with all field moments.
On such vectors, integration by parts in $t$ shows that $S$ is
symmetric and $B$ is skew symmetric, so
$\|(S+B)z\|^2=\|Sz\|^2+\|Bz\|^2+\dua{z}{[S,B]z}$.  Since $B$ involves
only $\partial_t$ and functions of $t$, the operators $K$ and $\Upsilon$
enter $[S,B]$ only through $\dot K$ and $\partial_t(r^2\Upsilon)$, which are
controlled by the claims above; no commutator of $K$ and $\Upsilon$ is
needed.  Computing $[S,B]$
term by term, with $r=\e^{-t}$ so that $\partial_t(r^2)=-2r^2$ and with
$\beta=\ddot f>0$, gives the
 identity
\begin{align}
 \|(S+B)z\|^2
 ={}&\|Sz\|^2+\|Bz\|^2
      +4\tau\int_{t_0}^\infty\beta(t)\|\partial_tz(t)\|^2\dd t
       \nonumber\\
 &+4\tau^3\int_{t_0}^\infty(\dot f(t))^2\beta(t)\|z(t)\|^2\dd t
      -\tau\int_{t_0}^\infty f^{(4)}(t)\|z(t)\|^2\dd t
       \nonumber\\
 &-2\tau\int_{t_0}^\infty\dot f(t)\dot K(t)[z(t)]\dd t
      +4\tau\int_{t_0}^\infty\dot f(t)r^2\Upsilon(t)[z(t)]\dd t
       \nonumber\\
 &-2\tau\int_{t_0}^\infty\dot f(t)r^2\dot\Upsilon(t)[z(t)]\dd t.
 \label{pf-cylinder-commutator}
\end{align}

It remains to absorb the three terms in \cref{pf-cylinder-commutator}
that may be negative, those containing $f^{(4)}$, $\dot K$ and $\dot\Upsilon$,
into the nonnegative terms.
For $t\geq2$, $\dot f(t)=1-\alpha t^{-\alpha-1}$ lies between
$2/(\alpha+2)$ and $1$, because $t^{\alpha+1}\geq2^{\alpha+1}\geq\alpha+2$
by Bernoulli's inequality, and $|f^{(4)}|\leq C\beta$.
For any fixed $\epsilon>0$, \cref{pf-angular-form-derivative} gives
\begin{align*}
 \tau\int_{t_0}^\infty|\dot K(t)[z(t)]|\dd t
 &\leq\epsilon\tau\int_{t_0}^\infty\beta(t)K(t)[z(t)]\dd t\\
 &\quad+C_\epsilon\tau\int_{t_0}^\infty\frac{r^4}{\beta(t)}
                         \big(\Upsilon(t)[z(t)]+\|z(t)\|^2\big)\dd t.
\end{align*}
Similarly, \cref{pf-radiation-derivative} gives
\[
 \tau\int_{t_0}^\infty r^2|\dot\Upsilon(t)[z(t)]|\dd t
 \leq\epsilon\tau\int_{t_0}^\infty r^2\Upsilon(t)[z(t)]\dd t
        +C_\epsilon\tau\int_{t_0}^\infty r^4\|z(t)\|^2\dd t.
\]
Since $K$ enters \cref{pf-cylinder-commutator} only through $\dot K$, none
of the nonnegative terms there can absorb
$\epsilon\tau\int_{t_0}^\infty\beta(t)K(t)[z(t)]\dd t$.  Instead, such a
term is controlled by $\|Sz\|^2$.

\begin{claim}
There is $C<\infty$, independent of $\tau\geq1$ and $z$, such that
\begin{equation}\label{pf-angular-multiplier}
 \tau\int_{t_0}^\infty\beta(t)K(t)[z(t)]\dd t
 \leq C\left(\|Sz\|^2+
             \tau^3\int_{t_0}^\infty\beta(t)\|z(t)\|^2\dd t\right).
\end{equation}
\end{claim}

\begin{proof}[Proof of the claim]
Integrating by parts in $t$,
\begin{align*}
 \Re\dua{Sz}{\beta z}
 ={}&\int_{t_0}^\infty\beta(t)\big(\|\partial_tz(t)\|^2+K(t)[z(t)]
                   +r^2\Upsilon(t)[z(t)]+\gamma^2\|z(t)\|^2\big)\dd t\\
    &-\tau^2\int_{t_0}^\infty(\dot f(t))^2\beta(t)\|z(t)\|^2\dd t
       -\frac12\int_{t_0}^\infty\ddot\beta(t)\|z(t)\|^2\dd t.
\end{align*}
Keeping the $K$ term, discarding the other nonnegative terms and
multiplying by $\tau$ gives
\begin{align*}
 \tau\int_{t_0}^\infty\beta(t)K(t)[z(t)]\dd t
 &\leq\tau\Re\dua{Sz}{\beta z}
   +\tau^3\int_{t_0}^\infty(\dot f(t))^2\beta(t)\|z(t)\|^2\dd t\\
 &\quad+\frac\tau2\int_{t_0}^\infty|\ddot\beta(t)|\|z(t)\|^2\dd t.
\end{align*}
By Young's inequality,
\[
 \tau\Re\dua{Sz}{\beta z}
 \leq\frac12\|Sz\|^2
   +\frac{\tau^2}2\int_{t_0}^\infty\beta(t)^2\|z(t)\|^2\dd t.
\]
For $t\geq2$, $\dot f$ and $\beta$ are bounded and $|\ddot\beta|\leq C\beta$.
Combining the two displays, we get for $\tau\geq1$
\[
 \tau\int_{t_0}^\infty\beta(t)K(t)[z(t)]\dd t
 \leq\frac12\|Sz\|^2+C\tau^3\int_{t_0}^\infty\beta(t)\|z(t)\|^2\dd t,
\]
which is \cref{pf-angular-multiplier}.
\end{proof}

Choose $\epsilon$ small enough that
\[
 \epsilon\tau\int_{t_0}^\infty\beta(t)K(t)[z(t)]\dd t
 \qquad\text{and}\qquad
 \epsilon\tau\int_{t_0}^\infty r^2\Upsilon(t)[z(t)]\dd t
\]
are absorbed, the first by $\|Sz\|^2$ and the $\tau^3$ term of
\cref{pf-cylinder-commutator} using \cref{pf-angular-multiplier}, and
the second by the $\Upsilon$ term of \cref{pf-cylinder-commutator}.  The
remaining errors are
\[
 C_\epsilon\tau\int_{t_0}^\infty\Big(\frac{r^4}{\beta(t)}
   \big(\Upsilon(t)[z(t)]+\|z(t)\|^2\big)+r^4\|z(t)\|^2\Big)\dd t.
\]
Since
\[
 \sup_{0<r<r_0}\frac{r^2}{\beta(-\log r)}\longrightarrow0
 \qquad\text{as }r_0\downarrow0,
\]
choosing $r_0$ small absorbs them into the same $\Upsilon$ and $\tau^3$
terms.  Finally, since $|f^{(4)}|\leq C\beta$, choosing $\tau_0$ large
absorbs the $f^{(4)}$ term.
Together with \cref{pf-cylinder-commutator,pf-angular-multiplier},
these choices prove
\begin{align}
 &\tau^3\int_{t_0}^\infty\beta(t)\|z(t)\|^2\dd t
  +\tau\int_{t_0}^\infty\beta(t)\big(\|\partial_tz(t)\|^2+K(t)[z(t)]\big)\dd t
   \nonumber\\
 &\quad+\tau\int_{t_0}^\infty r^2\Upsilon(t)[z(t)]\dd t
  \leq C\|(S+B)z\|^2.
 \label{pf-cylinder-carleman}
\end{align}
The radial derivative satisfies
\[
 \e^{\tau f}r^{\gamma+1}\partial_r(Uu)
       =-\partial_tz+(\tau\dot f-\gamma)z,
\]
and since $\dot f$ is bounded, the contribution of $(\tau\dot f-\gamma)z$ is
controlled by the first term in \cref{pf-cylinder-carleman}.
Since $\vbf$ is tangential, the radial component of
$(-\i\grad_\vX+\vbf)(Uu)$ is $-\i\partial_r(Uu)$ and its tangential
component is $-\i r^{-1}\nabla_{\mathbb S,\mathfrak a}(Uu)$.  These are
orthogonal, so $U\vD_nU^*=-\i\grad_{\vx_n}+\vbf_n$ gives
\[
 \sum_{n=1}^N\|U\vD_nu\|^2
 =\|\partial_r(Uu)\|^2+r^{-2}\|\nabla_{\mathbb S,\mathfrak a}(Uu)\|^2.
\]
Together with the angular form $K(t)[z(t)]$ and
$\dd\vX=r^{3N}\dd t\dd\vOmega$, this gives the second term on the
left of \cref{pf-free-carleman},
\[
 \tau\int_{B_{r_0}}\beta(t)\e^{2\tau f(t)}\sum_{n=1}^N\|\vD_nu(\vX)\|_{\Fock_\ph}^2\dd\vX.
\]
The other terms transform as
\begin{align*}
 \|(S+B)z\|^2
       &=4\mathcal T^0(u),\\
 \int_{t_0}^\infty r^2\Upsilon(t)[z(t)]\dd t
       &=2\int_{B_{r_0}}\e^{2\tau f(t)}\|H_\rad^{1/2}u(\vX)\|_{\Fock_\ph}^2\dd \vX.
\end{align*}

\noindent\textbf{Step 2: Ultraviolet limit.}
The compact-support restriction on the cutoff can now be removed.
Set $\wh\chi^{\le\Lambda}=\wh\chi\iden_{\{\omega\leq\Lambda\}}$.
The constants in the preceding estimates use only the norms of
$G^\mu_\vx$, $\grad_\vx G^\mu_\vx$ and $\omega G^\mu_\vx$ in $\Hil_\omega$.
They are therefore uniform in $\Lambda$ under the assumption
$\omega\wh\chi\in L^2$.
The limit also uses the following bound for products of two fields.

\begin{claim}
There is $C<\infty$ such that, for all $h,g\in\Hil_\omega$,
\begin{equation}\label{pf-two-field-bound}
 \|\varPhi(h)\varPhi(g)(H_\rad+1)^{-1}\|
 \leq C\|h\|_\omega\|g\|_\omega.
\end{equation}
\end{claim}

\begin{proof}[Proof of the claim]
We write $\varPhi(h)\varPhi(g)$ as a sum of four products of creation
and annihilation operators.  On the $M$-photon sector, $H_\rad$ is
multiplication by $\sum_i\omega(\vk_i)$, so
$\sum_{i\neq j}\omega(\vk_i)\omega(\vk_j)\leq H_\rad^2$ there.  The
Cauchy--Schwarz inequality in the two annihilated photon variables
therefore gives
\[
 \|a(h)a(g)\xi\|\leq\|h/\sqrt\omega\|\,\|g/\sqrt\omega\|\,\|H_\rad\xi\|,
 \qquad \xi\in\DC(H_\rad).
\]
The other three products reduce to this estimate and to the
annihilation bound of \cite[Section~3.3]{Q} by the canonical
commutation relations $a(h)a^*(g)=a^*(g)a(h)+\dua{h}{g}$ and
$\|a^*(g)\xi\|^2=\|a(g)\xi\|^2+\|g\|^2\|\xi\|^2$.
\end{proof}

For each fixed $u\in\mathcal C$,
\[
 T^{0,\le\Lambda}u\longrightarrow T^0u,
 \qquad \vD_n^{\le\Lambda}u\longrightarrow\vD_nu
 \quad\text{in }L^2.
\]
To prove this, we expand the covariant squares.  Since $\div\AF=0$, as
noted in Step~2 of the proof of \cref{ucp-local-operator-domain},
\[
 \vD_n^2u=-\Delta_{\vx_n}u+2\i e\AF(\vx_n)\cdot\grad_{\vx_n}u
          +e^2|\AF(\vx_n)|^2u,
\]
and the same identity holds for the cutoff $\wh\chi^{\le\Lambda}$.
Put $G^{\mu,>\Lambda}_\vx=G^\mu_\vx\iden_{\{\omega>\Lambda\}}$, so that
$\Af^\mu(\vx)-\Af^{\mu,\le\Lambda}(\vx)=\varPhi(G^{\mu,>\Lambda}_\vx)$.
Hence the components of $\vD_n^{\le\Lambda}u-\vD_nu$ are
$e\varPhi(G^{\mu,>\Lambda}_{\vx_n})u$.  The difference
$T^{0,\le\Lambda}u-T^0u$ is a sum of terms
$\varPhi(G^{\mu,>\Lambda}_{\vx_n})\partial_{x_n^\mu}u$ and of products of
two fields applied to $u$, each with a factor
$\varPhi(G^{\mu,>\Lambda}_{\vx_n})$.  The creation and annihilation bounds
of \cite[Section~3.3]{Q} give
$\|\varPhi(h)\xi\|\leq C\|h\|_\omega\|(H_\rad+1)^{1/2}\xi\|$.  Together
with \cref{pf-two-field-bound}, this gives, with $C$ independent of
$\Lambda$,
\[
 \begin{split}
 &\|\vD_n^{\le\Lambda}u-\vD_nu\|+\|T^{0,\le\Lambda}u-T^0u\|\\
 &\qquad\leq C\sup_{\mu,\vx}\|G^{\mu,>\Lambda}_\vx\|_\omega
   \bigl(\|(H_\rad+1)u\|+\|(H_\rad+1)^{1/2}\grad_\vX u\|\bigr).
 \end{split}
\]
The norms of $u$ on the right are finite, since
$\Fock_{\mathrm{fin}}\subset\DC(H_\rad)$.  Finally,
$\|G^{\mu,>\Lambda}_\vx\|_\omega$ does not depend on $\vx$, and it tends to
zero by \cref{cutoff-assumption} and dominated convergence.
Thus, for each such $u$, both sides of \cref{pf-free-carleman} for the
cutoff $\wh\chi^{\le\Lambda}$ converge to those for $\wh\chi$, so
\cref{pf-free-carleman} holds for all such $u$.  The gauge
transformation $U$ depends on $\Lambda$ but does not appear in
\cref{pf-free-carleman}, so it plays no role in this limit.

\noindent\textbf{Step 3: Approximation argument.}
Step~2 of the proof of \cref{ucp-local-operator-domain} uses only
\cref{cutoff-assumption,ucp-extra-cutoff}.  It gives
$\DC(T^0)=\DC(-\Delta_{\vX})\cap\DC(H_\rad)$ and the equivalence of the
graph norms of $T^0$ and $L=-\Delta_{\vX}+H_\rad+1$.  Hence, by
\cref{ucp-core-lemma}, an annularly supported
$u\in\DC(T^0)$ is a limit in the graph norm of $T^0$ of functions in
$\mathcal C$ supported in a slightly larger annulus.  Graph convergence implies form
convergence, and the weights are bounded on this annulus.
Consequently all terms in \cref{pf-free-carleman} pass to the
limit.  After the ultraviolet limit the approximation is performed in
the original gauge, so no higher ultraviolet moments are needed for
the final cutoff.
The bounds on the form factors are uniform in the center.  Once the
radius and the lower bound on $\tau$ have been fixed, restriction to
a smaller ball preserves the estimate with the same constants.

\end{proof}

We use a dyadic decomposition of the punctured ball.  With $r$,
$\beta$, $\vX_0$ and $r_0$ as in \cref{lem-pf-free-carleman}, let, for $j\geq0$,
\[
 R_j=2^{-j}r_0,\qquad \beta_j=\beta(-\log R_j),\qquad
 \mathcal A_j=\{R_j/2<r<2R_j\}\cap B_{r_0}(\vX_0),
\]
and choose smooth radial functions $\eta_j\geq0$ with
$\supp\eta_j\subset\{R_j/2<r<2R_j\}$, $|\grad\eta_j|\leq C/R_j$,
$|\Delta\eta_j|\leq C/R_j^2$ and $\sum_{j\geq0}\eta_j^2=1$ on
$0<r<r_0$.  As before, let
\[
 L=-\Delta_{\vX}+H_\rad+1.
\]
The fractional-derivative step below plays a role similar to that of
the fractional Carleman estimate in \cite{garrigue2020unique}.

\begin{lemma}[A fractional Carleman estimate on dyadic annuli]
\label[lemma]{ucp-fractional-carleman}
Under the assumptions of \cref{lem-pf-free-carleman},
for $\tau\geq\tau_0$ and $0\leq s\leq2$,
\begin{equation}\label{ucp-dyadic-fractional}
 \tau^{3-2s}\sum_{j\geq0}\beta_jR_j^{2s-2}
       \|L^{s/2}(\eta_j\e^{\tau f}u)\|^2
 \leq C_s \mathcal T^0(u).
\end{equation}
Here $u\in\DC(T^0)$ has compact support in
$B_{r_0}(\vX_0)\setminus\{\vX_0\}$, as in \cref{lem-pf-free-carleman}, and the
norms are on
$L^2(\RR^{3N},\Fock_\ph)$ after extension by zero.  The same estimate
holds, with the same constant, when $\eta_j$ is replaced by any smooth
radial family $\zeta_j$ supported in fixed enlargements of the annuli
$\{R_j/2<r<2R_j\}$, with $|\zeta_j|\leq C$,
$|\grad\zeta_j|\leq C/R_j$ and $|\Delta\zeta_j|\leq C/R_j^2$.  The
constant $C_s$ is independent of $\vX_0$ and does not increase when
$r_0$ is decreased.
\end{lemma}

\begin{proof}
We apply \cref{lem-pf-free-carleman} on the dyadic annuli.  Its
zero-order term gives the endpoint $s=0$, its covariant-gradient term
together with the graph-norm equivalence for $T^0$ gives the endpoint
$s=2$, and the moment inequality \cref{moment-inequality} gives the
intermediate values.
By the graph-norm equivalence for $T^0$ from the proof of
\cref{ucp-local-operator-domain},
\begin{equation}\label{ucp-graph-second-derivative}
 \|Lw\|^2\leq C\big(\|T^0w\|^2+\|w\|^2\big),\qquad w\in\DC(T^0).
\end{equation}
Multiplication by the smooth annular weight $\eta_j\e^{\tau f}$ preserves
this operator domain.  On its support, $r\asymp R_j$ and
$\beta(t)\asymp\beta_j$, with constants independent of $j$.
The zero-order term of \cref{pf-free-carleman} of \cref{lem-pf-free-carleman} gives
\begin{equation}\label{ucp-dyadic-zero-order}
 \tau^3\sum_{j\geq0}\beta_jR_j^{-2}\|\eta_j\e^{\tau f}u\|^2\leq C \mathcal T^0(u).
\end{equation}
The radial derivatives of $f$ give
\[
 |\grad(\eta_j\e^{\tau f})|\leq C\tau R_j^{-1}\e^{\tau f(t)},\qquad
 |\Delta(\eta_j\e^{\tau f})|\leq C\tau^2R_j^{-2}\e^{\tau f(t)}
\]
on $\mathcal A_j$.
The exponential weight is kept at its actual radial argument.
The product rule
\[
 T^0(\eta_j\e^{\tau f}u)=\eta_j\e^{\tau f}T^0u
   -\i\grad(\eta_j\e^{\tau f})\cdot\vD u
   -\tfrac12\Delta(\eta_j\e^{\tau f})u
\]
and \cref{ucp-graph-second-derivative} yield
\[
 \begin{split}
 \|L(\eta_j\e^{\tau f}u)\|^2
 &\leq C\int_{\mathcal A_j}\e^{2\tau f(t)}
  \bigl(\|T^0u(\vX)\|_{\Fock_\ph}^2+\tau^2R_j^{-2}\|\vD u(\vX)\|_{\Fock_\ph}^2\\
 &\qquad+\tau^4R_j^{-4}\|u(\vX)\|_{\Fock_\ph}^2\bigr)\dd\vX.
 \end{split}
\]
The term $\|\eta_j\e^{\tau f}u\|^2$ has been included in the last term, using
$R_j<1$ and $\tau\geq1$.  Multiplying by $\beta_jR_j^2/\tau$,
summing over the annuli, and applying \cref{pf-free-carleman} of \cref{lem-pf-free-carleman} gives
\begin{equation}\label{ucp-dyadic-second-order}
 \frac{1}{\tau}\sum_{j\geq0}\beta_jR_j^2\|L(\eta_j\e^{\tau f}u)\|^2\leq C \mathcal T^0(u).
\end{equation}
For $0<s<2$, the moment inequality \cref{moment-inequality} with
$\theta=s/2$ gives
\[
 \|L^{s/2}(\eta_j\e^{\tau f}u)\|^2\leq\|L(\eta_j\e^{\tau f}u)\|^s\|\eta_j\e^{\tau f}u\|^{2-s}.
\]
H\"older's inequality in $j$ combines
\cref{ucp-dyadic-zero-order,ucp-dyadic-second-order}, since
\[
 \left(\frac{\beta_jR_j^2}{\tau}\right)^{s/2}
 \left(\tau^3\beta_jR_j^{-2}\right)^{1-s/2}
 =\tau^{3-2s}\beta_jR_j^{2s-2}.
\]
The endpoints are the two preceding estimates.  Every annularly
supported $u$ meets only finitely many cutoffs, and all steps apply
directly in the operator domain.  The proof uses only the bounded
overlap and the derivative bounds of the cutoffs, and the constants
come from \cref{pf-free-carleman} of \cref{lem-pf-free-carleman} and \cref{ucp-graph-second-derivative}, which
are uniform in $\vX_0$ and $r_0$.
\end{proof}

We now add the external vector potential.  Analogously to
$\mathcal T^0(u)$, for $u$ in the local operator domain of $T^{\vA}$
with compact support in $B_{r_0}(\vX_0)\setminus\{\vX_0\}$, put
\[
 \mathcal T^{\vA}(u)=\int_{B_{r_0}(\vX_0)}\e^{2\tau f(t)}r^2
                       \|T^{\vA}u(\vX)\|_{\Fock_\ph}^2\dd\vX.
\]
The mixed estimates of \cref{ucp-fractional-carleman} control the
first-order term $\vA\cdot\vD$ with $\vA\in L^6_\mathrm{loc}$.
Therefore \cref{lem-pf-free-carleman} carries over from $T^0$ to
$T^{\vA}$, with $\vpi_n$ in place of $\vD_n$ and $\mathcal T^{\vA}(u)$
in place of $\mathcal T^0(u)$, on a sufficiently small ball.

\begin{lemma}[A Carleman estimate with external vector potential]
\label[lemma]{lem-pf-covariant-carleman}
Assume \cref{cutoff-assumption} and
$\omega\wh\chi\in L^2(\RR^3)$.  Fix $\vX_0$ and suppose near each
$(\vX_0)_n$ that
\[
 \vA\in L^6_\mathrm{loc},\qquad
 \div\vA\in L^p_\mathrm{loc},\qquad p>2,
\]
where the divergence is distributional.  With the weights of
\cref{lem-pf-free-carleman}, there are $r_0>0$, $\tau_0\geq1$ and
$C<\infty$ such that, for every $\tau\geq\tau_0$ and every $u$ in
the local operator domain of $T^{\vA}$ with compact support in
$B_{r_0}(\vX_0)\setminus\{\vX_0\}$,
\begin{equation}\label{pf-covariant-carleman}
 \begin{aligned}
 \int_{B_{r_0}(\vX_0)}\e^{2\tau f(t)}\Bigl(
   &\tau^3\beta(t)r^{-2}\|u(\vX)\|_{\Fock_\ph}^2
   +\tau\beta(t)\sum_{n=1}^N\|\vpi_nu(\vX)\|_{\Fock_\ph}^2\\
   &+\tau\|H_\rad^{1/2}u(\vX)\|_{\Fock_\ph}^2\Bigr)\dd\vX
   \leq C \mathcal T^{\vA}(u).
 \end{aligned}
\end{equation}
Moreover, the mixed estimate \cref{ucp-dyadic-fractional} holds with
$\mathcal T^{\vA}(u)$ in place of $\mathcal T^0(u)$ on the right.  These choices are uniform
on compact sets with uniform local $L^6$ bounds on $\vA$ and
$L^p$ bounds on $\div\vA$.
\end{lemma}

\begin{proof}
Write $T^{\vA}=T^0+R_\vA$, where $R_\vA$ consists of the first-order
term $\sum_{n=1}^N\vA(\vx_n)\cdot\vD_n$ and a zero-order multiplier
$q_\vA$.  Using the
mixed estimate \cref{ucp-dyadic-fractional} of \cref{ucp-fractional-carleman}, we show that
\[
 \int_{B_{r_0}(\vX_0)}\e^{2\tau f(t)}r^2\|R_\vA u(\vX)\|_{\Fock_\ph}^2\dd\vX
 \leq\tfrac14\mathcal T^0(u)
\]
for $r_0$ small and $\tau$ large: the first-order part is small when
$r_0$ is small, and the zero-order part is small when $\tau$ is large.
Since $T^0u=T^{\vA}u-R_\vA u$, this gives
$\mathcal T^0(u)\leq4\mathcal T^{\vA}(u)$, so \cref{lem-pf-free-carleman}
holds with $\mathcal T^{\vA}(u)$ on the right.
Finally, we replace $\vD_n$ by $\vpi_n=\vD_n+\vA(\vx_n)$, using
\cref{ucp-dyadic-fractional} of \cref{ucp-fractional-carleman} once more.

\noindent\textbf{Step 1: Decomposition of $T^{\vA}$.}
Fix neighborhoods of the projected center on which the local norms
of $\vA$ and $\div\vA$ are finite, and write
\begin{align*}
 M_A^2&=\sum_{n=1}^N\|\vA\|_{L^6(B_R((\vX_0)_n))}^2,\\
 M_d^2&=\sum_{n=1}^N\|\div\vA\|_{L^p(B_R((\vX_0)_n))}^2
\end{align*}
for a sufficiently small fixed $R>0$.  We will decrease $r_0<R$
while using these fixed upper bounds.  Steps~1--3 of the proof of
\cref{ucp-local-operator-domain} involve no eigenstate, and they use
the collision-free condition only to bound the electron--electron
interaction.  Applied without scalar potential on a neighborhood of
$\vX_0$, they show that $T^{\vA}$ and $T^0$ have the same local
operator domain, the compactly supported elements of
$\DC(-\Delta_{\vX})\cap\DC(H_\rad)$.  On annularly supported functions
in this domain,
\begin{equation*}
 \begin{aligned}
 T^{\vA}&=T^0+R_\vA,\qquad
 R_\vA=\sum_{n=1}^N\vA(\vx_n)\cdot\vD_n+q_\vA,\\
 q_\vA&=\frac12\sum_{n=1}^N\big(|\vA(\vx_n)|^2
        -\i(\div\vA)(\vx_n)\big).
 \end{aligned}
\end{equation*}

\noindent\textbf{Step 2: The first-order term.}
This step uses only the free estimates.
The Sobolev inequality and \cref{ucp-field-form-bound} give
\begin{equation}\label{ucp-endpoint-multipliers}
 \begin{split}
 \sum_{n=1}^N\|\vA(\vx_n)\cdot\vD_nz\|^2
       &\leq C M_A^2\|L^{3/4}z\|^2,\\
 \sum_{n=1}^N\|\vA(\vx_n)z\|^2
       &\leq C M_A^2\|L^{1/4}z\|^2
 \end{split}
\end{equation}
for $z$ supported in the configuration ball.  For the field part of
the first inequality, the pointwise field bound is followed by
the Sobolev inequality in $\vx_n$ applied to $(H_\rad+1)^{1/2}z$.
The commuting electronic and photon spectral operators satisfy
\[
 \|(-\Delta_{\vx_n})^{1/4}(H_\rad+1)^{1/2}z\|
       \leq\|L^{3/4}z\|.
\]
The ordinary derivative part is bounded by the same right-hand side.
Thus no spatial derivative of the field operator is taken in
\cref{ucp-endpoint-multipliers}.

Let $\eta_j$ be the cutoffs introduced before
\cref{ucp-fractional-carleman}, and choose radial cutoffs $\zeta_j$
equal to one near $\supp\eta_j$, supported in fixed enlargements of
its annulus, with derivative bounds of orders
$R_j^{-1}$ and $R_j^{-2}$.  Set
$z_j=\zeta_j\e^{\tau f(t)}u$ and $\phi(\vX)=f(-\log r)$.
On $\supp\eta_j$,
\[
 \eta_j\e^{\tau f(t)}\vD u
   =\eta_j\big(\vD z_j+\i\tau\grad\phi\,z_j\big),\qquad
 |\grad\phi|\leq C/R_j.
\]
For $a>0$, put
\[
 \sigma_a(r_0)=\sup_{0<r\leq r_0}
                \frac{r^a}{\beta(-\log r)},
\]
which tends to zero as $r_0\downarrow0$.
The identity $\sum_{j\geq0}\eta_j^2=1$, \cref{ucp-endpoint-multipliers}, and
\cref{ucp-dyadic-fractional} of \cref{ucp-fractional-carleman}, with $\zeta_j$ in place of $\eta_j$,
therefore give
\begin{align}
 &\int_{B_{r_0}(\vX_0)}\e^{2\tau f(t)}r^2
                       \Big\|\sum_{n=1}^N\vA(\vx_n)\cdot\vD_nu(\vX)\Big\|_{\Fock_\ph}^2\dd\vX\notag\\
 &\quad\leq C M_A^2\sum_{j\geq0}\left(
               R_j^2\|L^{3/4}z_j\|^2
                    +\tau^2\|L^{1/4}z_j\|^2\right)\notag\\
 &\quad\leq C M_A^2\sigma_1(r_0)\sum_{j\geq0}\left(
               \beta_jR_j\|L^{3/4}z_j\|^2
                    +\tau^2\beta_jR_j^{-1}\|L^{1/4}z_j\|^2\right)\notag\\
 &\quad\leq C M_A^2\sigma_1(r_0)\mathcal T^0(u).
 \label{ucp-endpoint-first-order}
\end{align}
The second inequality uses $R_j\leq\sigma_1(r_0)\beta_j$, and the last
is \cref{ucp-dyadic-fractional} of \cref{ucp-fractional-carleman} with $s=3/2$ and $s=1/2$.  The
constants in this inequality are independent of $r_0$.

\noindent\textbf{Step 3: The zero-order term.}
The two parts of $q_\vA$ are one-body multipliers in $L^3$ and
$L^p$, respectively.  For a one-body multiplier $b\in L^q$,
$q>2$, the Sobolev inequality and the commuting spectral
operators give
\[
 \|b(\vx_n)\eta_j\e^{\tau f}u\|
 \leq C_q\|b\|_q\|(-\Delta_{\vx_n})^{3/(2q)}(\eta_j\e^{\tau f}u)\|
 \leq C_q\|b\|_q\|L^{3/(2q)}(\eta_j\e^{\tau f}u)\|.
\]
Applying \cref{ucp-dyadic-fractional} of \cref{ucp-fractional-carleman} with $s=1$ and $s=3/p$
yields directly
\begin{align}
 &\int_{B_{r_0}(\vX_0)}\e^{2\tau f(t)}r^2\|q_\vA(\vX)u(\vX)\|_{\Fock_\ph}^2\dd\vX\notag\\
 &\quad\leq C\left(
   M_A^4\sigma_2(r_0)\tau^{-1}
   +M_d^2\sigma_{4-6/p}(r_0)\tau^{-(3-6/p)}
         \right)\mathcal T^0(u).
 \label{ucp-endpoint-zero-order}
\end{align}
Only the moduli of the multipliers enter, so the imaginary divergence
term causes no change.

\noindent\textbf{Step 4: Absorption.}
Since
$\|R_\vA u\|^2\leq2\big\|\sum_{n=1}^N\vA(\vx_n)\cdot\vD_nu\big\|^2
 +2\|q_\vA u\|^2$,
\cref{ucp-endpoint-first-order,ucp-endpoint-zero-order} give
\begin{align*}
 &\int_{B_{r_0}(\vX_0)}\e^{2\tau f(t)}r^2\|R_\vA u(\vX)\|_{\Fock_\ph}^2\dd\vX\\
 &\quad\leq C\left(M_A^2\sigma_1(r_0)+M_A^4\sigma_2(r_0)\tau^{-1}
   +M_d^2\sigma_{4-6/p}(r_0)\tau^{-(3-6/p)}\right)\mathcal T^0(u).
\end{align*}
Choose $r_0$ so small that $CM_A^2\sigma_1(r_0)\leq\frac18$.  This is
possible because $\sigma_1(r_0)\to0$ as $r_0\downarrow0$, while $C$
and $M_A$ do not grow when $r_0$ is decreased: $C$ comes from
\cref{lem-pf-free-carleman,ucp-fractional-carleman}, which hold on
every smaller ball with the same constants, and $M_A$ is a norm on the
fixed balls $B_R((\vX_0)_n)$.  Next increase $\tau_0$ so that the
remaining two terms are at most $\frac18$, using $3-6/p>0$ and that
$\sigma_a(r_0)$ does not increase as $r_0$ decreases.  Then
\[
 \int_{B_{r_0}(\vX_0)}\e^{2\tau f(t)}r^2\|R_\vA u(\vX)\|_{\Fock_\ph}^2\dd\vX
 \leq\tfrac14\mathcal T^0(u).
\]
Since $T^0u=T^{\vA}u-R_\vA u$, this gives
$\mathcal T^0(u)\leq2\mathcal T^{\vA}(u)+\frac12\mathcal T^0(u)$, that is,
\begin{equation}\label{ucp-external-source-control}
 \mathcal T^0(u)\leq4\mathcal T^{\vA}(u).
\end{equation}
This proves the assertion about the mixed estimate and bounds the
left-hand side of \cref{pf-free-carleman} of \cref{lem-pf-free-carleman} by $C \mathcal T^{\vA}(u)$.

\noindent\textbf{Step 5: Replacing $\vD_n$ by $\vpi_n$.}
To replace $\vD_n$ by $\vpi_n$, use the second estimate in
\cref{ucp-endpoint-multipliers} and
\cref{ucp-dyadic-fractional} of \cref{ucp-fractional-carleman} with $s=1/2$ to obtain
\[
 \tau\int_{B_{r_0}(\vX_0)}\beta(t)\e^{2\tau f(t)}
                     \sum_{n=1}^N\|\vA(\vx_n)u(\vX)\|_{\Fock_\ph}^2\dd\vX
 \leq C M_A^2\frac{r_0}{\tau}\mathcal T^0(u).
\]
Together with
$\|\vpi_nu\|^2\leq2\|\vD_nu\|^2+2\|\vA(\vx_n)u\|^2$ and \cref{ucp-external-source-control}, this gives
\cref{pf-covariant-carleman}.  The fixed local norm bounds permit all
choices uniformly on compact sets.  The endpoint $L^6$ assumption is
used through smallness of the radius, without requiring smallness
of the external field itself.
\end{proof}

\begin{lemma}[A Carleman estimate with scalar potential]
\label[lemma]{ucp-scalar-perturbation}
Under the assumptions of \cref{lem-pf-covariant-carleman}, suppose on
$B_{r_0}(\vX_0)$ that, with the same $p$,
\[
 \begin{gathered}
 W(\vX)=\sum_{n=1}^Nv_n(\vx_n)+W_b(\vX),\\
 v_n\in L^p(B_{r_0}((\vX_0)_n)),\qquad
 W_b\in L^\infty(B_{r_0}(\vX_0)).
 \end{gathered}
\]
Then, after increasing $\tau_0$, for
every $\tau\geq\tau_0$ and every $u$ in the local operator domain of
$T^{\vA}+W$ with compact support in
$B_{r_0}(\vX_0)\setminus\{\vX_0\}$,
\begin{equation}\label{ucp-scalar-carleman}
 \begin{aligned}
 \int_{B_{r_0}(\vX_0)}\e^{2\tau f(t)}\Bigl(
   &\tau^3\beta(t)r^{-2}\|u(\vX)\|_{\Fock_\ph}^2
   +\tau\beta(t)\sum_{n=1}^N\|\vpi_nu(\vX)\|_{\Fock_\ph}^2\\
   &+\tau\|H_\rad^{1/2}u(\vX)\|_{\Fock_\ph}^2\Bigr)\dd\vX\\
   \leq C&\int_{B_{r_0}(\vX_0)}\e^{2\tau f(t)}r^2
        \|T^{\vA}u(\vX)+W(\vX)u(\vX)\|_{\Fock_\ph}^2\dd\vX.
 \end{aligned}
\end{equation}
The constants are uniform on compact sets under the local bounds of
\cref{lem-pf-covariant-carleman}, the indicated $L^p$ norms, and
$\|W_b\|_\infty$.
\end{lemma}

\begin{proof}
Put $s=3/p<3/2$, $M_p^2=\sum_{n=1}^N\|v_n\|_p^2$, and
$M_b=\|W_b\|_\infty$.  Extend each $v_n$ by zero outside its
three-dimensional ball.  The Sobolev
inequality, as in \cref{ucp-coordinate-multiplier}, gives
\begin{align*}
 \|v_n(\vx_n)\eta_j\e^{\tau f}u\|
 &\leq C\|v_n\|_p\|(-\Delta_{\vx_n})^{s/2}(\eta_j\e^{\tau f}u)\|\\
 &\leq C\|v_n\|_p\|(-\Delta_{\vX})^{s/2}(\eta_j\e^{\tau f}u)\|.
\end{align*}
The second inequality follows from the corresponding Fourier
multipliers.  The identity
$\sum_{j\geq0}\eta_j^2=1$ and $r\asymp R_j$ on $\supp\eta_j$ give
\begin{align*}
 \mathcal W(u)&:=\int_{B_{r_0}(\vX_0)}\e^{2\tau f(t)}r^2
                       \|W(\vX)u(\vX)\|_{\Fock_\ph}^2\dd\vX\\
 &\leq C M_p^2\sum_{j\geq0}R_j^2
                   \|(-\Delta_{\vX})^{s/2}(\eta_j\e^{\tau f}u)\|^2
       +C M_b^2\sum_{j\geq0}R_j^2\|\eta_j\e^{\tau f}u\|^2.
\end{align*}
By the joint spectral calculus,
$\|(-\Delta_{\vX})^{s/2}\Phi\|\leq\|L^{s/2}\Phi\|$.  Using \cref{ucp-dyadic-fractional} of \cref{ucp-fractional-carleman} with
$s=3/p$ and $s=0$, followed by \cref{ucp-external-source-control},
we obtain, with $\sigma_a(r_0)$ as in the proof of
\cref{lem-pf-covariant-carleman},
\begin{equation}\label{ucp-scalar-smallness}
 \mathcal W(u)\leq C\left(
       M_p^2\sigma_{4-2s}(r_0)\tau^{-(3-2s)}
                 +M_b^2\sigma_4(r_0)\tau^{-3}\right)\mathcal T^{\vA}(u).
\end{equation}
Both powers of $\tau$ decay because $p>2$.

The scalar multiplier is infinitesimally bounded relative to the
free electronic operator by \cref{ucp-coordinate-multiplier} and the
moment inequality \cref{moment-inequality}.  Thus, as in Step~3 of the
proof of \cref{ucp-local-operator-domain}, which involves no
eigenstate, $T^{\vA}+W$ and $T^{\vA}$ have the same local operator
domain.  All terms above are consequently defined for the vectors
in the statement.  Writing $T^{\vA}u=(T^{\vA}u+Wu)-Wu$ gives
\[
 \mathcal T^{\vA}(u)\leq2\int_{B_{r_0}(\vX_0)}\e^{2\tau f(t)}r^2
                       \|T^{\vA}u(\vX)+W(\vX)u(\vX)\|_{\Fock_\ph}^2\dd\vX
                       +2\mathcal W(u).
\]
Increase $\tau_0$ so that, for $\tau\geq\tau_0$,
\[
 2C\left(M_p^2\sigma_{4-2s}(r_0)\tau^{-(3-2s)}
        +M_b^2\sigma_4(r_0)\tau^{-3}\right)\leq\frac12,
\]
with $C$ from \cref{ucp-scalar-smallness}.  Then
$2\mathcal W(u)\leq\frac12\mathcal T^{\vA}(u)$, and since
$\mathcal T^{\vA}(u)<\infty$, subtracting this term gives
\[
 \mathcal T^{\vA}(u)\leq4\int_{B_{r_0}(\vX_0)}\e^{2\tau f(t)}r^2
                       \|T^{\vA}u(\vX)+W(\vX)u(\vX)\|_{\Fock_\ph}^2\dd\vX.
\]
Inserting this bound on the right of \cref{pf-covariant-carleman} of \cref{lem-pf-covariant-carleman}
gives \cref{ucp-scalar-carleman}.  All estimates use
only the stated local norms, which gives uniformity in the center.
\end{proof}

Finally, we include electron spin.  On
$L^2(\RR^{3N},(\CC^2)^{\otimes N}\otimes\Fock_\ph)$, let $T^{\vA}$ and
$L$ act componentwise and set
\[
 \begin{split}
 V(\vX)&=\sum_{n=1}^Nv_n(\vx_n)+V_b(\vX),\\
 S_{\mathrm{ext}}(\vX)&=\frac12\sum_{n=1}^N
                    \vsigma_n\cdot\vB(\vx_n),\\
 S_{\mathrm{ph}}(\vX)&=-\frac e2\sum_{n=1}^N
                    \vsigma_n\cdot\BF(\vx_n),\\
 H^{V,\vA}&=T^{\vA}+V+S_{\mathrm{ext}}+S_{\mathrm{ph}},
 \end{split}
\]
where $\vB=\grad\times\vA$ in the sense of
distributions.  By the Pauli identity \cref{pauli-expansion},
\[
 H^{V,\vA}=\frac12\sum_{n=1}^N(\vsigma_n\cdot\vpi_n)^2+H_\rad+V
\]
in the sense of quadratic forms.  The Zeeman terms $S_{\mathrm{ext}}$
and $S_{\mathrm{ph}}$ are separated because they are treated as
perturbations of $T^{\vA}$.

\begin{lemma}[A Carleman estimate with spin]
\label[lemma]{ucp-pauli-carleman}
In addition to the hypotheses of \cref{lem-pf-covariant-carleman},
suppose, with the same $p$, that on the configuration ball under
consideration
\[
 v_n\in L^p(B_{r_0}((\vX_0)_n)),\qquad
 V_b\in L^\infty(B_{r_0}(\vX_0)),\qquad
 \vB\in L^p(B_{r_0}((\vX_0)_n),\RR^3)
\]
for every $n$.  Let $E\in\CC$ be fixed.  Then, after increasing
$\tau_0$, for every $\tau\geq\tau_0$ and every $u$ in the local operator
domain of $H^{V,\vA}$ with compact support in
$B_{r_0}(\vX_0)\setminus\{\vX_0\}$,
\begin{equation}\label{pauli-carleman}
 \begin{aligned}
 \int_{B_{r_0}(\vX_0)}\e^{2\tau f(t)}\Bigl(
   &\tau^3\beta(t)r^{-2}\|u(\vX)\|_{\mathcal K_N}^2
   +\tau\beta(t)\sum_{n=1}^N\|\vpi_nu(\vX)\|_{\mathcal K_N}^2\\
   &+\tau\|H_\rad^{1/2}u(\vX)\|_{\mathcal K_N}^2\Bigr)\dd\vX\\
   \leq C&\int_{B_{r_0}(\vX_0)}\e^{2\tau f(t)}r^2
        \|(H^{V,\vA}-E)u(\vX)\|_{\mathcal K_N}^2\dd\vX,
 \end{aligned}
\end{equation}
where $\mathcal K_N=(\CC^2)^{\otimes N}\otimes\Fock_\ph$.
The constants are uniform on compact sets under
the local bounds in \cref{lem-pf-covariant-carleman}, the displayed
$L^p$ bounds, and a bound on $\|V_b-E\|_\infty$.
\end{lemma}

\begin{proof}
Expand $u$ in an orthonormal basis of $(\CC^2)^{\otimes N}$.  Each
component is an $\Fock_\ph$-valued function, and $T^{\vA}$, $\vpi_n$,
$H_\rad^{1/2}$ and $L$ act on each component separately.  The squared
$\mathcal K_N$-norms in \cref{pf-covariant-carleman} of \cref{lem-pf-covariant-carleman}
and \cref{ucp-dyadic-fractional} of \cref{ucp-fractional-carleman}
are the sums of the squared $\Fock_\ph$-norms of the components.
Applying \cref{lem-pf-covariant-carleman} to each component and
summing, we obtain \cref{pf-covariant-carleman} of \cref{lem-pf-covariant-carleman} and the mixed estimate
\cref{ucp-dyadic-fractional} of \cref{ucp-fractional-carleman}, with $\mathcal T^{\vA}(u)$ on the right,
for $\mathcal K_N$-valued $u$, with the same constants.  Here
$\mathcal T^{\vA}(u)$ is the source integral of
\cref{lem-pf-covariant-carleman}, with norms in $\mathcal K_N$.

It remains to replace $T^{\vA}$ by $H^{V,\vA}-E$.  The difference
$V+S_{\mathrm{ext}}+S_{\mathrm{ph}}-E$ is a perturbation of the
same kind as in \cref{ucp-scalar-perturbation}.  We show that its
weighted $L^2$ norm is at most $\delta(\tau)\mathcal T^{\vA}(u)$ with
$\delta(\tau)\to0$ as $\tau\to\infty$, and then absorb it.  The
classical part $V+S_{\mathrm{ext}}-E$ is handled as in
\cref{ucp-scalar-perturbation}, using the operator norms of its matrix
coefficients.  The quantized Zeeman term $S_{\mathrm{ph}}$ is
controlled by the radiation and zero-order terms of
\cref{pf-covariant-carleman} of \cref{lem-pf-covariant-carleman}.

Put
\[
 \begin{split}
 s&=3/p<3/2,\qquad M_b=\|V_b-E\|_\infty,\\
 M_p^2&=\sum_{n=1}^N\left(
       \|v_n\|_{L^p(B_{r_0}((\vX_0)_n))}^2
       +\tfrac14\|\vB\|_{L^p(B_{r_0}((\vX_0)_n))}^2\right),
 \end{split}
\]
with $\sigma_a(r_0)$ as in the proof of \cref{lem-pf-covariant-carleman}.
The Pauli identity gives
$\|\vsigma_n\cdot\vB(\vx_n)\|_{\mathrm{op}}=|\vB(\vx_n)|$.
Thus the Sobolev inequality gives
\[
 \big\|\big(v_n(\vx_n)+\tfrac12
                    \vsigma_n\cdot\vB(\vx_n)\big)\eta_j\e^{\tau f}u\big\|
 \leq C\big(\|v_n\|_p+\tfrac12\|\vB\|_p\big)
                                      \|L^{s/2}(\eta_j\e^{\tau f}u)\|.
\]
Here the three-dimensional coefficients are extended by zero outside
their respective balls.  The identity $\sum_{j\geq0}\eta_j^2=1$ and the mixed
estimates with exponents $s$ and $0$ now give, exactly as in
\cref{ucp-scalar-smallness},
\begin{equation}\label{pauli-classical-smallness}
 \begin{aligned}
 &\int_{B_{r_0}(\vX_0)}\e^{2\tau f(t)}r^2\|(V+S_{\mathrm{ext}}-E)u(\vX)\|_{\mathcal K_N}^2\dd\vX\\
 &\quad\leq C\left(
 M_p^2\sigma_{4-2s}(r_0)\tau^{-(3-2s)}
       +M_b^2\sigma_4(r_0)\tau^{-3}\right)\mathcal T^{\vA}(u).
 \end{aligned}
\end{equation}
This argument uses the operator norm of each matrix coefficient and
does not require simultaneous diagonalization of the Pauli matrices.

For the quantized term, \cref{pauli-magnetic-field-bound} and the
boundedness of the Pauli matrices give a constant $C_{\mathrm{ph}}$, depending
only on $e$, $N$ and the cutoff norms, such that
\[
 \|S_{\mathrm{ph}}(\vX)\xi\|^2
 \leq C_{\mathrm{ph}}\big(\|H_\rad^{1/2}\xi\|^2+\|\xi\|^2\big)
 \qquad\big(\xi\in\DC(H_\rad^{1/2})\big),
\]
uniformly in $\vX$.  The radiation term of
\cref{pf-covariant-carleman} of \cref{lem-pf-covariant-carleman} has no factor $\beta$.
Its radiation and zero-order terms therefore imply
\begin{align}
 &\int_{B_{r_0}(\vX_0)}\e^{2\tau f(t)}r^2\|S_{\mathrm{ph}}u(\vX)\|_{\mathcal K_N}^2\dd\vX
       \notag\\
 &\quad\leq C_{\mathrm{ph}}r_0^2\int_{B_{r_0}(\vX_0)}\e^{2\tau f(t)}
                                  \|H_\rad^{1/2}u(\vX)\|_{\mathcal K_N}^2\dd\vX
       \notag\\
 &\qquad+C_{\mathrm{ph}}\sigma_4(r_0)\int_{B_{r_0}(\vX_0)}\e^{2\tau f(t)}
                                  \beta(t)r^{-2}\|u(\vX)\|_{\mathcal K_N}^2\dd\vX
       \notag\\
 &\quad\leq CC_{\mathrm{ph}}\left(\frac{r_0^2}{\tau}
                   +\frac{\sigma_4(r_0)}{\tau^3}\right)\mathcal T^{\vA}(u).
 \label{pauli-quantized-smallness}
\end{align}
In particular, no derivative of $\BF$ occurs in this perturbation
argument.  It is carried out in the original gauge, after the
covariant estimate has been proved, and needs only the existing
moment assumption $\omega\wh\chi\in L^2$.

As in the proof of \cref{ucp-local-operator-domain-spin}, the local
domains used here are $\DC(-\Delta_{\vX})\cap\DC(H_\rad)$.  Alternatively, the classical matrix
estimate above and the moment inequality \cref{moment-inequality}
give infinitesimal operator
boundedness relative to $L$.  The same conclusion for $S_{\mathrm{ph}}$
follows from its $H_\rad^{1/2}$ bound.  Hence all the preceding terms
are defined for the vectors in the statement.

Let $R_E=V+S_{\mathrm{ext}}+S_{\mathrm{ph}}-E$.  Combining
\cref{pauli-classical-smallness,pauli-quantized-smallness} yields
\[
 \int_{B_{r_0}(\vX_0)}\e^{2\tau f(t)}r^2\|R_Eu(\vX)\|_{\mathcal K_N}^2\dd\vX
 \leq\delta(\tau)\mathcal T^{\vA}(u),
\]
where
\[
 \delta(\tau)=C\left(
 M_p^2\sigma_{4-2s}(r_0)\tau^{-(3-2s)}
 +M_b^2\sigma_4(r_0)\tau^{-3}
 +C_{\mathrm{ph}}\frac{r_0^2}{\tau}
 +C_{\mathrm{ph}}\frac{\sigma_4(r_0)}{\tau^3}\right)
 \longrightarrow0.
\]
Since $T^{\vA}u=(H^{V,\vA}-E)u-R_Eu$, we obtain
\[
 \mathcal T^{\vA}(u)\leq2\int_{B_{r_0}(\vX_0)}\e^{2\tau f(t)}r^2
        \|(H^{V,\vA}-E)u(\vX)\|_{\mathcal K_N}^2\dd\vX+2\delta(\tau)\mathcal T^{\vA}(u).
\]
Choose $\tau_0$ so that $2\delta(\tau)\leq1/2$ for
$\tau\geq\tau_0$.  Since $\mathcal T^{\vA}(u)<\infty$, subtracting the
last term gives
\[
 \mathcal T^{\vA}(u)\leq4\int_{B_{r_0}(\vX_0)}\e^{2\tau f(t)}r^2
        \|(H^{V,\vA}-E)u(\vX)\|_{\mathcal K_N}^2\dd\vX.
\]
Substitution into the componentwise version of
\cref{pf-covariant-carleman} of \cref{lem-pf-covariant-carleman} proves \cref{pauli-carleman}.  The construction uses only the stated local norms,
which also proves uniformity in the center.
\end{proof}

\section{Counterexamples at zero coupling}\label[appendix]{app-zero-coupling}

We show that the assumption $e>0$ in \cref{thm-hk-nonvanishing} cannot
be dropped, either without or with electron spin, even with a
nonzero compact ultraviolet cutoff, exponential localization, and
nondegenerate ground states.  The examples stay within the potential
spaces \cref{potential-spaces}.  

\begin{proposition}[A spinless common ground state for different magnetic fields]
\label[proposition]{prop-zero-coupling-counterexample}
Let $N=1$, $e=0$, and $Z>0$.  The cutoff may be any fixed cutoff
satisfying \cref{cutoff-assumption}.  Write $r=|\vx|$.  For
$-1<t<1$, set
\begin{equation*}
 \va(\vx)=\frac{(-x_2,x_1,0)}{1+r^2},\qquad
 \vA_t=t\va,\qquad
 v_t(\vx)=-\frac{Z}{r}-\frac{t^2}{2}|\va(\vx)|^2.
\end{equation*}
Every $H_{\mathrm{sl}}^{v_t,\vA_t}$ on $\HC_1^{\mathrm{sl}}$ has the same unique ground state, up to phase,
\begin{equation*}
 \Psi_Z=\phi_Z\otimes\Omega,\qquad
 \phi_Z(\vx)=\left(\frac{Z^3}{\pi}\right)^{1/2}\e^{-Zr},
 \qquad E_t=-\frac{Z^2}{2}.
\end{equation*}
Its density and spinless combined current density are independent of $t$,
\begin{equation}\label{counterexample-data}
 \rho(\vx)=\frac{Z^3}{\pi}\e^{-2Zr},\qquad \vjc=0.
\end{equation}
For $t\ne s$, the vector potentials $\vA_t$ and $\vA_s$ have different
magnetic fields and do not differ by a gradient.
\end{proposition}

The transformed scalar potential is the same throughout this family,
$u_t=v_t+|\vA_t|^2/2=-Z/r$.  Each pair satisfies
\cref{ucp-potential-regularity} for every $2<p<3$, and the density has
an exponential moment for every $\alpha<2Z$.  Choosing a nonzero
compact ultraviolet cutoff ensures $\omega\wh\chi\in L^2$.
Thus only the positive coupling hypothesis of the spinless version
of \cref{thm-hk-nonvanishing} is missing.
The same example works at $e>0$ if $\wh\chi=0$, when $\AF=0$.

\begin{proof}[Proof of \cref{prop-zero-coupling-counterexample}]
We have $|\va|\leq r/(1+r^2)\leq1/2$,
$\div\va=0$, and $\va\cdot\vx=0$.  Thus $\vA_t\in L^\infty$,
while $v_t\in L^{3/2}+L^\infty$ by splitting the Coulomb potential
at $r=1$.  Since $e=0$, expanding the square gives
\begin{equation*}
 H_{\mathrm{sl}}^{v_t,\vA_t}=h_t+H_\rad,\qquad
 h_t=-\frac12\Delta-\frac Zr+\frac{t}{1+r^2}L_z,
 \qquad L_z=-\i(x_1\partial_{x_2}-x_2\partial_{x_1}).
\end{equation*}
The radial function $\phi_Z$ has $L_z\phi_Z=0$.  It remains to check
that it is the ground state for every $|t|<1$.

We expand an electronic wavefunction $\psi$ in its form domain in
spherical harmonics,
\[
 \psi(\vx)=\sum_{l=0}^\infty\sum_{m=-l}^l
                  r^{-1}u_{lm}(r)Y_{lm}(\vx/r),\qquad
 u_{lm}(r)=r\int_{\mathbb S^2}\ol{Y_{lm}(\vOmega)}\,\psi(r\vOmega)\dd\vOmega.
\]
Here $Y_{lm}$ are the spherical harmonics.  They form an orthonormal
basis of $L^2(\mathbb S^2)$ and satisfy
$-\Delta_{\mathbb S^2}Y_{lm}=l(l+1)Y_{lm}$ and $L_zY_{lm}=mY_{lm}$.
The $u_{lm}$ are the radial coefficients, so that $\|\psi\|^2=\sum_{l,m}\int_0^\infty|u_{lm}(r)|^2\dd r$.
Let $w(r)=1/r-Z$ be the logarithmic derivative of the hydrogen $1s$
radial function $r\e^{-Zr}$.  For a radial coefficient $u$, completing
the square and integrating by parts give
\begin{equation}\label{radial-factorization}
 \begin{split}
 \mathcal E_Z[u]
 &:=\frac12\int_0^\infty|u'|^2\dd r
       -Z\int_0^\infty\frac{|u|^2}{r}\dd r
       +\frac{Z^2}{2}\int_0^\infty|u|^2\dd r\\
 &=\frac12\int_0^\infty
             |u'-wu|^2\dd r
 \geq0.
 \end{split}
\end{equation}
This follows by integration by parts on a radial core and extension
to the form domain.  In the $(l,m)$ channel, $-\frac12\Delta$ acts on
$u_{lm}$ as $-\frac12\frac{\dd^2}{\dd r^2}+\frac{l(l+1)}{2r^2}$, and
$(1+r^2)^{-1}L_z$ as multiplication by $m/(1+r^2)$.  Hence
\begin{equation}\label{channel-decomposition}
 \Bigl\langle\psi,\Bigl(h_t+\frac{Z^2}{2}\Bigr)\psi\Bigr\rangle
 =\sum_{l=0}^\infty\sum_{m=-l}^l\Bigl(\mathcal E_Z[u_{lm}]
   +\int_0^\infty\Bigl(\frac{l(l+1)}{2r^2}+\frac{tm}{1+r^2}\Bigr)
     |u_{lm}(r)|^2\dd r\Bigr).
\end{equation}
The weight in the last integral satisfies
\begin{equation}\label{angular-lower-bound}
 \frac{l(l+1)}{2r^2}+\frac{tm}{1+r^2}
 \geq(1-|t|)\frac{l(l+1)}{2r^2}\geq0,
\end{equation}
because $1/(1+r^2)\leq1/r^2$ and $|m|\leq l\leq l(l+1)/2$ for
$l\geq1$, while both sides vanish for $l=0$.  By
\cref{radial-factorization,angular-lower-bound}, every term in
\cref{channel-decomposition} is nonnegative, so $h_t\geq-Z^2/2$.  For
$|t|<1$, the weight in \cref{angular-lower-bound} is positive when
$l\geq1$, so equality forces every $l\geq1$ channel to vanish.  In the remaining channel, $\mathcal E_Z[u]=0$ forces
$u(r)=C r \e^{-Zr}$.  The electronic ground-state space is therefore
one dimensional and spanned by $\phi_Z$.

Since $H_\rad\geq0$ and $\ker H_\rad=\CC\Omega$, the full
ground-state space is spanned by $\Psi_Z$.  It is not separated
from the rest of the spectrum by a positive gap: since
$\omega(\vk)=|\vk|$, we have $\spec(H_\rad)=[0,\infty)$, so
$[E_t,\infty)\subset\spec(H_{\mathrm{sl}}^{v_t,\vA_t})$.  The orbital is real, and at $e=0$ the field-induced current density
vanishes, proving \cref{counterexample-data}.  Finally, $\grad\times\va$ is
clearly not identically zero, so different $t$ give different magnetic
fields.
\end{proof}

To obtain a spinful counterpart, we replace the spherical denominator
of $\va$ by a cylindrical one.  Its magnetic field then has a fixed
direction, and a scalar correction cancels the Zeeman term in the
spin-down sector.

\begin{proposition}[A spinful common ground state for different magnetic fields]
\label[proposition]{prop-spinful-zero-coupling-counterexample}
Let $N=1$, $e=0$, and $Z>0$.  The cutoff may be any fixed cutoff
satisfying \cref{cutoff-assumption}.  Write
$r=|\vx|$ and $s=(x_1^2+x_2^2)^{1/2}$.  For $0<t<1/2$, set
\begin{equation}\label{spinful-counterexample-potentials}
 \va_\perp(\vx)=\frac{(-x_2,x_1,0)}{1+s^2},\qquad
 \vA_t=t\va_\perp,\qquad
 v_t(\vx)=-\frac Zr-\frac{t^2}{2}|\va_\perp(\vx)|^2
                       +\frac{t}{(1+s^2)^2}.
\end{equation}
Every $H^{v_t,\vA_t}$ on $\HC_1$ has the same unique ground state,
up to phase,
\begin{equation}\label{spinful-counterexample-state}
 \Psi_Z^\downarrow
 =\begin{pmatrix}0\\\phi_Z\end{pmatrix}\otimes\Omega,\qquad
 \phi_Z(\vx)=\left(\frac{Z^3}{\pi}\right)^{1/2}\e^{-Zr},\qquad
 E_t=-\frac{Z^2}{2}.
\end{equation}
Its density, magnetization density, and combined current density are
independent of $t$ and satisfy, almost everywhere,
\begin{equation}\label{spinful-counterexample-data}
 \rho(\vx)=\frac{Z^3}{\pi}\e^{-2Zr},\qquad
 \vm=-\rho\ve_3,\qquad
 \vjc=-\frac12\grad\times(\rho\ve_3)
       =\frac{Z\rho}{r}(x_2,-x_1,0).
\end{equation}
The magnetic fields are
\begin{equation}\label{spinful-counterexample-field}
 \vB_t=\grad\times\vA_t=\frac{2t}{(1+s^2)^2}\ve_3.
\end{equation}
Thus distinct parameters give vector potentials that do not differ
by a gradient.
\end{proposition}

Here the transformed scalar potential is
$u_t=v_t+|\vA_t|^2/2=-Z/r+t/(1+s^2)^2$.
The additional term compensates the spin-down Zeeman energy
$-t/(1+s^2)^2$.  The potentials belong to
\cref{potential-spaces} and satisfy \cref{ucp-potential-regularity}
for every $2<p<3$, and the density has an exponential moment for
every $\alpha<2Z$.  With a nonzero compact ultraviolet cutoff,
only the positive coupling hypothesis of the spinful
\cref{thm-hk-nonvanishing} is missing.  This example also works at
$e>0$ if $\wh\chi=0$.  At $t=0$ the same state is still a ground
state, but the ground-state space is two dimensional because of spin.

\begin{proof}[Proof of \cref{prop-spinful-zero-coupling-counterexample}]
Set $f(s)=(1+s^2)^{-1}$ and
$h_Z=-\Delta/2-Z/r$.  The vector field $\va_\perp$ is smooth,
$|\va_\perp|=s/(1+s^2)\leq1/2$, and
$\div\va_\perp=0$.  Direct differentiation gives
\cref{spinful-counterexample-field}.  Hence the vector potential
and its curl are bounded, and the scalar potential is Coulomb plus
a bounded function.  This proves the stated global admissibility
and local regularity.

The Pauli identity and the scalar correction in
\cref{spinful-counterexample-potentials} give the form decomposition
\begin{equation*}
 H^{v_t,\vA_t}=h_t^{\mathrm{spin}}+H_\rad,\qquad
 h_t^{\mathrm{spin}}
 =\bigl(h_Z+tf(s)L_z\bigr)\iden_{\CC^2}
                        +2tf(s)^2P_+,
\end{equation*}
where $P_+=(\iden_{\CC^2}+\sigma^3)/2$ and
$L_z=-\i(x_1\partial_{x_2}-x_2\partial_{x_1})$.
Since $L_z\phi_Z=0$ and $P_+(0,\phi_Z)^{\mathsf T}=0$,
the displayed spinor is an eigenstate with energy $-Z^2/2$.

For a scalar $\psi\in H^1(\RR^3)$, define the nonnegative form
\begin{equation*}
 Q_Z[\psi]
 =\Bigl\langle\psi,\Bigl(h_Z+\frac{Z^2}{2}\Bigr)\psi\Bigr\rangle
 =\frac12\int_{\RR^3}
       \left|\grad\psi(\vx)+Z\frac{\vx}{r}\psi(\vx)\right|^2\dd\vx.
\end{equation*}
The identity follows by integration by parts from
$\div(\vx/r)=2/r$ in the sense of distributions, first on a smooth compact core
and then on $H^1$ by form continuity.  Moreover,
$Q_Z[\psi]=0$ precisely when $\psi$ is a scalar multiple of
$\phi_Z$, since equality implies
$\grad(\e^{Zr}\psi(\vx))=0$ locally.

In cylindrical coordinates $(s,\theta,z)$, $L_z=-\i\partial_\theta$.
Expansion in the angular Fourier modes $\e^{\i m\theta}$ and
$|m|\leq m^2$ for $m\in\ZZ$ yield
\begin{equation*}
 \left|\dua{\psi}{f(s)L_z\psi}\right|
 \leq\int_{\RR^3}f(s)|\partial_\theta\psi(\vx)|^2\dd\vx
 \leq\int_{\RR^3}\frac{|\partial_\theta\psi(\vx)|^2}{s^2}\dd\vx
 \leq2Q_Z[\psi].
\end{equation*}
For the last inequality, note that
$s^{-1}\partial_\theta\psi(\vx)$ is the azimuthal component of
$\grad\psi(\vx)+Z(\vx/r)\psi(\vx)$.
These inequalities hold first on the smooth core and extend to
$H^1$, since $f(s)L_z=-\i\va_\perp\cdot\grad$ has bounded
coefficients.

Consequently, for a spinor
$\psi=(\psi_+,\psi_-)^{\mathsf T}\in H^1(\RR^3,\CC^2)$,
\begin{equation*}
 \begin{split}
 \Bigl\langle\psi,
       \Bigl(h_t^{\mathrm{spin}}+\frac{Z^2}{2}\Bigr)\psi\Bigr\rangle
 &\geq(1-2t)\bigl(Q_Z[\psi_+]+Q_Z[\psi_-]\bigr)
             +2t\int_{\RR^3}f(s)^2|\psi_+(\vx)|^2\dd\vx\\
 &\geq0.
 \end{split}
\end{equation*}
Since $0<t<1/2$ and $f>0$, equality forces $\psi_+(\vx)=0$ and
$\psi_-(\vx)=C\phi_Z(\vx)$.  Thus the electronic ground state is unique
up to phase.  As $H_\rad\geq0$ and
$\ker H_\rad=\CC\Omega$, the same is true of the full ground state
\cref{spinful-counterexample-state}.

The orbital is real and $e=0$, so $\vjp=\vjph=0$.
The spin-down polarization gives $\vm(\vx)=-\rho(\vx)\ve_3$, and
$\grad\rho(\vx)=-2Z\rho(\vx)\vx/r$ gives \cref{spinful-counterexample-data}.
The exponential moments follow directly from the displayed density.
Finally, \cref{spinful-counterexample-field} shows that distinct
parameters give different curls, which excludes even a
distributional gradient as their difference.
\end{proof}

\enlargethispage{3\baselineskip}

\end{document}